\documentclass[aps,prd,onecolumn,eqsecnum,amsmath,nofootinbib,preprintnumbers]{revtex4}%

\usepackage[dvipsnames]{xcolor}
\usepackage{color,graphicx,float,subfigure,xcolor}
\usepackage{amsfonts,amssymb,theorem,mathrsfs,times}
\usepackage{bm}
\usepackage{multirow}
\usepackage{mathtools}
\usepackage{amsfonts,amssymb,theorem,mathrsfs}
\usepackage{dsfont}
\usepackage{setspace}
\usepackage{amsmath}
\usepackage{graphicx}
\usepackage[colorlinks,linkcolor=blue,anchorcolor=blue,citecolor=blue]{hyperref}   
\usepackage{ulem}   
\usepackage[english]{babel}

{\theorembodyfont{\upshape}
	}
{\theorembodyfont{\upshape}
	}
{\theorembodyfont{\upshape}
	}
{\theorembodyfont{\upshape}
	}
{\theorembodyfont{\upshape}
	}
{\theorembodyfont{\upshape}
	}
\addtocounter{MaxMatrixCols}{10}
\newcommand{\dalm}{\kern1pt\vbox{\hrule height 0.9pt\hbox{\vrule width
			0.9pt\hskip 2.5pt\vbox{\vskip 5.5pt}\hskip 3pt\vrule width
			0.3pt}\hrule height 0.3pt}\kern1pt}

\begin{document}
\thispagestyle{empty}
\preprint{\hfill {\small {ICTS-USTC/PCFT-24-58}}}
\title{The (in)stability of quasinormal modes of Boulware-Deser-Wheeler black hole in the hyperboloidal framework}

\author{Li-Ming Cao$^{a\, ,b}$\footnote{e-mail address: caolm@ustc.edu.cn}}

\author{Liang-Bi Wu$^{c\, ,d}$\footnote{e-mail address: liangbi@mail.ustc.edu.cn}}

\author{Yu-Sen Zhou$^a$\footnote{e-mail address: zhou\_ys@mail.ustc.edu.cn (corresponding author)}}


\affiliation{${}^a$Interdisciplinary Center for Theoretical Study and Department of Modern Physics,\\
    University of Science and Technology of China, Hefei, Anhui 230026, China}

\affiliation{${}^b$Peng Huanwu Center for Fundamental Theory, Hefei, Anhui 230026, China}

\affiliation{${}^c$School of Fundamental Physics and Mathematical Sciences, Hangzhou Institute for Advanced Study, UCAS, Hangzhou 310024, China}

\affiliation{${}^d$University of Chinese Academy of Sciences, Beijing 100049, China}

\date{\today}

\begin{abstract}
    We study the quasinormal modes of Boulware-Deser-Wheeler black hole in Einstein-Gauss-Bonnet gravity theory within the hyperboloidal framework. The effective potentials for the test Klein-Gordon field and gravitational perturbations of scalar, vector, and tensor types are thoroughly investigated and put into several typical classes. The effective potentials for the gravitational perturbations have more diverse behaviors than those in general relativity, such as double peaks, the existence of the negative region adjacent to or far away from the event horizon, etc. These lead to the existence of unstable modes ($\text{Im} \omega<0$), and the presence of gravitational wave echoes. These rich phenomenons are inherent in Einstein-Gauss-Bonnet theory, rather than artificially introduced by hand. What's more, the (in)stability of quasinormal modes is studied in frequency domain and time domain, respectively. For the frequency-domain, the pseudospectrum is used to account for the instability of the spectrum. For the time-domain, we add a small bump to the effective potential, and find that the new waveform does not differ significantly from the original one, where the comparison is characterized by the so-called mismatch functions. This means that quasinormal modes are stable in time-domain regardless of the shapes of the original effective potentials. In this way, our study reveals the non-equivalence of the stability of quasinormal modes in the frequency-domain and the time-domain. Besides, we also numerically investigate Price's law at both finite distances and infinity with the assistance of the hyperboloidal approach.
\end{abstract}

\maketitle

\section{Introduction}
Black holes in higher dimensions have gained more and more attention in high-energy physics, particularly with the advance of the string theory, the brane-world scenario, and the theoretical possibility of black hole production at the Large Hadron Collider (LHC)~\cite{Giddings:2001bu,Giddings:2008gr,Emparan:2008eg,Dimopoulos:2001hw,Kanti:2004nr,Antoniadis:1998ig,Randall:1999ee,Arkani-Hamed:1998sfv,Arkani-Hamed:1998jmv}. While general relativity (GR) provides a very successful framework for describing gravity, investigations into higher-dimensional black holes often need extensions of GR to address challenges and incorporate new physics~\cite{Charmousis:2008kc,Clifton:2011jh}. Einstein-Gauss-Bonnet (EGB) gravity theory, which is a natural generalization of GR suggested by Lovelock's theorem~\cite{Lovelock:1971yv,Lovelock:1972vz}, emerges as a candidate. EGB gravity theory incorporates a higher-order curvature term, i.e.,  the Gauss-Bonnet term, into the Einstein-Hilbert action. This term arises naturally in the low-energy effective action of string theory~\cite{Zwiebach:1985uq,Gross:1986iv,Boulware:1985wk}. This provides a well-motivated framework for exploring gravitational phenomena in extreme astrophysical and cosmological environments, such as black holes and the early universe. Among the solutions in EGB theory, the Boulware-Deser-Wheeler (BDW) black hole~\cite{Boulware:1985wk, Wheeler:1985nh} stands out as a static, spherically symmetric solution in higher-dimensional spacetimes and generalizes the Schwarzschild solution of GR to include Gauss-Bonnet corrections. This solution and its generalization incorporating cosmological constant~\cite{Cai:2001dz} have been extensively studied in the context of gravitational waves, black hole thermodynamics, and holographic property of gravity~\cite{Cai:2003gr, Brigante:2008gz, Brigante:2007nu, Cvetic:2001bk, Maeda:2008nz, Cao:2023zhy}.

A crucial aspect of a black hole is its stability under perturbations~\cite{chandrasekhar1998mathematical}. Linear perturbation theory provides a natural approach to investigate stability, leading to the concept of quasinormal modes (QNMs)~\cite{Nollert:1999ji, Kokkotas:1999bd,Berti:2009kk,Konoplya:2011qq}. QNMs are the solutions to the linear perturbation equations subject to the outgoing boundary conditions. The evolution of a perturbed black hole exhibits distinct stages~\cite{Leaver:1986gd}: an initial prompt response, followed by a damped oscillatory phase described by the superimpositions of QNMs, and finally, a late-time tail decay~\cite{Price:1971fb,Price:1972pw}. The real part of the QNM represents the oscillation frequency, while the imaginary part determines the damping rate, directly relating to stability. In this paper, with our notions, a mode with a negative imaginary part corresponds to an unstable mode. Moreover, QNMs serve as unique fingerprints for black hole identification and characterization in gravitational wave astronomy~\cite{Baibhav:2023clw, Giesler:2019uxc, Berti:2009kk, LIGOScientific:2016aoc, Berti:2018vdi, Maggiore}.

The linear stability and QNM spectra of the BDW black hole have gained a lot of attention~\cite{Dotti:2004sh, Dotti:2005sq, Gleiser:2005ra, Beroiz:2007gp, Konoplya:2008ix, Takahashi:2009dz, Takahashi:2009xh, Takahashi:2010ye, Takahashi:2010gz, Takahashi:2011cgy, Yoshida:2015vua}. The effective potentials of the BDW black hole for the test Klein-Gordon field, the gravitational perturbations of the scalar type, vector type, and tensor type have been presented in~\cite{Abdalla:2005hu,Dotti:2005sq, Gleiser:2005ra,Takahashi:2010ye}. The QNMs were calculated using the numerical characteristic integration method and the WKB method for the positive-definite, single-peak potentials~\cite{Konoplya:2008ix,Yoshida:2015vua}. However, for the BDW black hole, it is observed that the effective potential may possess a negative region which can lead to the instability. Previous work~\cite{Beroiz:2007gp} identified instabilities due to scalar perturbations in $5$ dimensions and tensor perturbations in $6$ dimensions, and this phenomenon is also corroborated by numerical simulations~\cite{Konoplya:2008ix}. Furthermore, the method for the stability of Lovelock black holes in arbitrary dimensions was developed~\cite{Takahashi:2009dz, Takahashi:2009xh, Takahashi:2010gz, Takahashi:2011cgy}. Applying this method to the EGB theory confirms no instability arises in other cases, and further gives out the explicit parameter regime leading to instability: when the black hole's mass parameter is sufficiently small compared to the coupling parameter. This specific range agrees with the numerical predictions in~\cite{Konoplya:2008ix}. It was commonly thought that gravitational instabilities arise at low multipole numbers $l$ since higher $l$ increases the centrifugal barrier thereby enhancing stability. However, higher $l$ also deepens the negative region. In fact, the instability in EGB theory is triggered by large $l$~\cite{Konoplya:2008ix, Takahashi:2010gz}.

Solving for QNMs involves dealing with the boundary conditions and the divergence of eigenfunctions at the boundaries. The hyperboloidal framework~\cite{Zenginoglu:2007jw,Ansorg:2016ztf,PanossoMacedo:2018hab,Zenginoglu:2011jz, PanossoMacedo:2023qzp, Zenginoglu:2024bzs, PanossoMacedo:2024nkw} offers a powerful approach to overcome these difficulties. By employing constant-$\tau$ slices that extend to null infinity and penetrate the event horizon, the hyperboloidal framework avoids the divergences of eigenfunctions. Furthermore, at boundaries, light-cones point outwards to the computational domain, the boundary conditions are naturally satisfied~\cite{Zenginoglu:2007jw,Ansorg:2016ztf,PanossoMacedo:2018hab,Zenginoglu:2011jz, PanossoMacedo:2023qzp, Zenginoglu:2024bzs, PanossoMacedo:2024nkw}. Compactification of spatial coordinates allows numerical computation to reach infinity, transforming the QNM problem into a simpler eigenvalue problem in the hyperboloidal framework. The hyperboloidal framework for the Kerr spacetime is constructed in~\cite{PanossoMacedo:2019npm}.

Recently, the spectrum instability of QNMs in black hole physics gains a lot of interest, especially considering the complicated astrophysical environments surrounding black holes~\cite{Barausse:2014tra}. The pioneering work of Nollert and Price showed that even small perturbations can significantly influence the higher overtone modes in the QNM spectra~\cite{Nollert:1996rf,Nollert:1998ys}. This phenomenon is known as the spectrum instability, which requires a thorough investigation of the robustness of QNMs against perturbations of the background spacetime. One approach to investigate QNM stability is to focus on specific, known modifications to the effective potential. Some of these works approximate the effective potential, which is equivalent to modifying the original effective potential~\cite{Nollert:1996rf,Daghigh:2020jyk,Qian:2020cnz}. Specifically, Ref.~\cite{Nollert:1996rf} uses a staircase approximation; Ref.~\cite{Daghigh:2020jyk} uses a piecewise linear approximation; and Ref.~\cite{Qian:2020cnz} introduces a cutoff of the potential. It is observed that for these approximations, a significant alteration in the corresponding QNM spectrum persists even after a moderate increase in the number of segments. Therefore, the spectrum instability of QNMs also presents a substantial challenge for their numerical computation, especially for the high overtones.

Another methodological different approach to investigate QNM spectrum instability utilizes the concept of pseudospectra, which does not require specifying the specific form of modification of the effective potential~\cite{Destounis:2023ruj, Jaramillo:2020tuu}. The dynamics of black hole systems are described by non-self-adjoint operators because of the dissipative boundary condition for QNMs. With the help of the hyperboloidal framework and the pseudo-spectral method, the QNMs problem is recast as an eigenvalue problem. Therefore, pseudospectra, which are originally developed to analyze non-self-adjoint operators and how their eigenvalues respond to small perturbations, offers a powerful framework for studying the stability of QNM spectra~\cite{trefethen2020spectra}. These perturbations include modifications to the effective potential. The $\epsilon$-pseudospectrum visually represents an upper bound on the displacement of QNM spectra under perturbations of norm $\epsilon$. Therefore, the open structure of contour lines of $\epsilon$-pseudospectrum in the complex plane indicates the instability of the QNM spectrum. The pseudospectrum has gained a lot of attention in asymptotically flat spacetime~\cite{Jaramillo:2020tuu, Destounis:2021lum,Cao:2024oud}, spacetime with a positive~\cite{Sarkar:2023rhp, Destounis:2023nmb,Luo:2024dxl,Warnick:2024usx} or negative~\cite{Arean:2024afl, Cownden:2023dam, Boyanov:2023qqf,Garcia-Farina:2024pdd,Arean:2023ejh} cosmological constant, and horizonless compact objects~\cite{Boyanov:2022ark}. Pseudospectrum can be calculated by evaluating the norm of the resolvent in hyperboloidal coordinates, which involves matrix approximations in numerical schemes~\cite{Jaramillo:2020tuu}. However, the convergence of the matrix-derived pseudospectrum to the true operator pseudospectrum in asymptotically flat spacetime as the grid resolution increases remains an open question. Choosing higher-order Sobolev norms may alleviate this problem~\cite{Boyanov:2023qqf}. Nevertheless, despite potential quantitative discrepancies, pseudospectra provide a valuable qualitative understanding of spectral migration and the sensitivity of QNMs to perturbations.

Note that the above instability for QNMs is encountered in the frequency-domain. Simultaneously, the question of time-domain stability of QNMs has been a focus of considerable research~\cite{Yang:2024vor, Ianniccari:2024ysv, Berti:2022xfj}. While frequency-domain analyses often suggest instability, the analysis in the time domain indicates stability. Some studies have shown that certain quantities resulting from the scattering properties are unaffected by spectrum instabilities~\cite{Kyutoku:2022gbr,Torres:2023nqg,Rosato:2024arw,Oshita:2024fzf,Wu:2024ldo}. This apparent contradiction needs further study to address these seemingly different results.

On the one hand, although the linear stability and QNM spectra of the BDW black hole have gained a lot of attention, the (in)stability of QNMs for the BDW black hole remains an open question. On the other hand, the effective potentials of the BDW black hole in EGB gravity have double peaks and the negative region, which are very different from those in GR. In the recent work~\cite{Spieksma:2024voy}, the authors have stated that despite the presence of spectrum instability, the QNMs from the viewpoints of the time domain are stable measured by mismatch function. However, whether the different shapes of the original effective potential have an impact on such stability is worth investigating. In this work, we investigate the dynamics of perturbations around the BDW black hole by analyzing its effective potentials and calculating QNMs in both the time domain and frequency domain. More importantly, we employ two complementary approaches to assess the (in)stability of QNMs: pseudospectrum in the frequency-domain and waveform mismatch in the time domain. In addition, we validate our numerical code by calculating the Price's law.

This paper is organized as follows. In Sec.\ref{sec: BDW}, the hyperboloidal framework of the BDW black hole is introduced. In Sec.\ref{QNMs}, we get the QNMs from the frequency domain and the time domain, respectively. In Sec.\ref{stability_analysis}, we study the stability of QNMs in frequency domain and time domain, respectively. Sec.\ref{sec: conclusions} contains the conclusions and discussion. Criterion for identifying physically relevant QNM spectra is given in Appendix \ref{sec: drift}. Appendix \ref{sec: evenprice} presents our computational investigation of Price's law for even dimensional spacetimes. Appendix \ref{sec: anothermismatch} shows additional computational results regarding time-domain stability.

\section{The hyperboloidal framework of the Boulware-Deser-Wheeler black hole}\label{sec: BDW}
We begin this section by reviewing the Boulware-Deser-Wheeler (BDW) black hole in the Einstein-Gauss-Bonnet gravity theory. The action of the EGB gravity theory is
\begin{eqnarray}\label{action}
    S=\frac{1}{16\pi G_{D}}\int\mathrm{d}^Dx\sqrt{-g}\Big(R+\tilde{\alpha} L_{\text{GB}}\Big)\, ,\quad D\ge5\, ,
\end{eqnarray}
where the Gauss-Bonnet term $L_{\text{GB}}$ is given by
\begin{eqnarray}\label{L_GB}
    L_{\text{GB}}=R_{abcd}R^{abcd}-4R_{ab}R^{ab}+R^2\, ,
\end{eqnarray}
$\tilde{\alpha}$ is a positive coupling constant and $G_{D}$ is the $D$-dimensional Newtonian constant of gravitation. The static vacuum solutions in EGB gravity were found long time ago by Boulware and Deser~\cite{Boulware:1985wk} and Wheeler~\cite{Wheeler:1985nh}. The metric of the solution can be expressed as
\begin{eqnarray}
    \mathrm{d}s^2=-f(r)\mathrm{d}t^2+f(r)^{-1}\mathrm{d}r^2+r^2\mathrm{d}\Omega^2_{D-2}\, ,
\end{eqnarray}
where the metric function is
\begin{eqnarray}
    f(r)=1+\frac{r^2}{2\tilde{\alpha}(D-3)(D-4)}\Bigg(1-\sqrt{1+\frac{8\tilde{\alpha}(D-3)(D-4)\tilde{\mu}}{(D-2)r^{D-1}}}\Bigg)\, ,
\end{eqnarray}
$\tilde{\mu}$ is the mass parameter of the black hole, and $\mathrm{d}\Omega_{D-2}^2$ is the line element of a $(D-2)$-dimensional unit sphere. This black hole has only one horizon, that is, the event horizon $r_{+}$. Therefore, by using $f(r_{+})=0$, the mass parameter $\tilde{\mu}$ can be expressed in terms of the black hole event horizon $r_+$ as follows
\begin{eqnarray}\label{relation_mu_and_alpha}
    \tilde{\mu}=\frac{(D-2)r_+^{D-3}}{2}\Big[1+\frac{\tilde{\alpha}(D-3)(D-4)}{r_+^2}\Big]\, .
\end{eqnarray}
For simplicity, we introduce two dimensionless quantities,
\begin{eqnarray}\label{dimensionless_quantities}
    \mu=\frac{\tilde{\mu}}{r_+^{D-3}}\, ,\quad\alpha=\frac{\tilde{\alpha}}{r_+^2}\, .
\end{eqnarray}
From the definitions (\ref{dimensionless_quantities}) and Eq. (\ref{relation_mu_and_alpha}), two dimensionless quantities have the following relation,
\begin{eqnarray}
    \mu=\frac{(D-2)}{2}\Big[1+\alpha(D-3)(D-4)\Big]\,.
\end{eqnarray}
The above equation indicates that there is only one independent dimensionless parameter in the BDW black hole. For convenience, $\alpha$ is chosen as the free dimensionless parameter. One thing that needs to be emphasized is that even though the actual $\tilde{\alpha}$ may be small, $\alpha$ can also take a large value in the context of a small black hole, i.e., $r_{+}$ is small.

For the BDW black hole we are currently considering, we consider not only a test scalar field, but also the gravitational perturbations. The test scalar field obeys the Klein-Gordon equation, and linear perturbations in EGB gravity can  be classified into scalar, vector, and tensor types generally. The equations of gravitational perturbations and the Klein-Gordon equation for the BDW black hole can all be decoupled into radial and angular parts. The radial part reduces to a wave-like equation,
\begin{eqnarray}\label{perturbation eq}
    \Big(\frac{\partial^2}{\partial t^2}-\frac{\partial^2}{\partial r_{\star}^2}+\tilde{V}_{X}(r)\Big)\Psi(t,r)=0\, ,\quad \mathrm{d} r_{\star}=\frac{\mathrm{d}r}{f(r)}\, ,
\end{eqnarray}
in which $X$ refers to $K$, $S$, $V$, $T$ and they represent the effective potentials of KG equation, scalar perturbation, vector perturbation and tensor perturbation, respectively. The explicit expressions of the effective potentials for these four sectors are very cumbersome and can be found in~\cite{Cao:2021sty, Abdalla:2005hu,Konoplya:2008ix,Konoplya:2010vz,Dotti:2004sh,Yoshida:2015vua,Dotti:2005sq,Gleiser:2005ra}. One should be aware that the potential $\tilde{V}_{X}(r)$ depends on the multipole number $l$. For the KG equation, the integer $l$ starts from $0$, but for gravitational perturbations, the integer $l$ starts from $2$~\cite{Konoplya:2008ix}.

Following~\cite{PanossoMacedo:2023qzp}, for the BDW black hole, we study the compact hyperboloidal coordinates $(\tau, \sigma, \theta, \phi)$ related with original coordinate $(t, r, \theta, \phi)$ as follows
\begin{eqnarray}\label{compact_hyperboloidal_coordinates}
    t&=&r_+\big[\tau-h(\sigma)\big]\, ,\nonumber\\
    r&=&\frac{r_+}{\sigma}\, ,
\end{eqnarray}
where $h(\sigma)$ is called the height function. The domain of $\sigma$ is restricted to $[0, 1]$, where $\sigma=0$ and $\sigma=1$ corresponds to the null infinity $\mathscr{I}^+$ and the event horizon $\mathscr{H}$, respectively.

Since perturbations propagate at the speed of light, they inevitably reach the future null infinity $\mathscr{I}^{+}$ or the future event horizon $\mathscr{H}^+$. However, in traditional coordinates $(t, r, \theta, \phi)$, the QNM eigenfunctions grow exponentially near the bifurcation surface and spatial infinity, as constant-$t$ slices squeeze together in these regions. Hyperboloidal coordinates overcome this limitation by introduciting a height function that characterizes the deformation of constant-$\tau$ surfaces relative to constant-$t$ surfaces. With an appropriate choice of height function, one can construct a foliation by constant-$\tau$ surfaces that smoothly extends to $\mathscr{I}^{+}$ and penetrates the future event horizon $\mathscr{H}^{+}$.

\begin{figure}[htbp]
    \centering
    \includegraphics{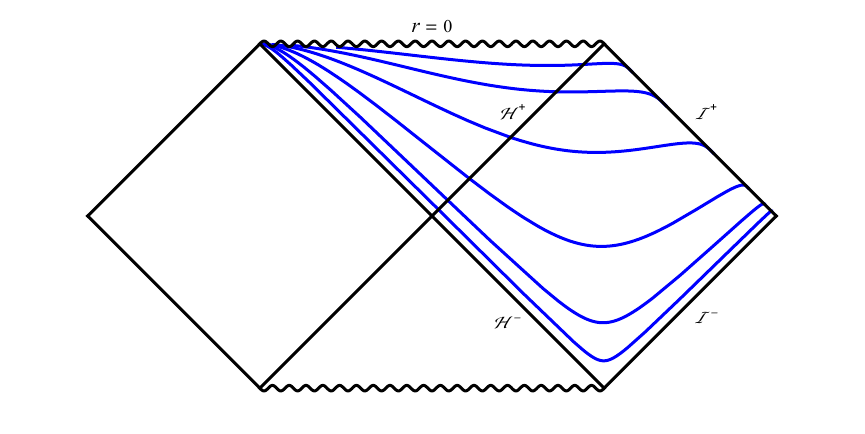}
    \caption{The Penrose diagram for $D=5$ and $\alpha=0.1$ of the BDW black hole. The blue lines represent the time foliation by constant-$\tau$ surfaces, obtained using out-in strategy.}
    \label{penrose}
\end{figure}

The method provided by~\cite{PanossoMacedo:2023qzp} states that there are two common and alternative approaches to derive the explicit expressions of $h(\sigma)$: in-out and out-in strategies, which gives out different height functions
\begin{eqnarray}\label{height_functions}
    h_{\text{in-out}}^{\prime}(\sigma) &=& \frac{2}{\sigma^2}-\frac{1}{\sigma^2F(\sigma)}\, ,\\
    h_{\text{out-in}}^{\prime}(\sigma) &=& -\frac{1+2(D-3)(D-4)\alpha}{(D-3)[1+(D-4)(D-5)\alpha]}\frac{2}{1-\sigma}+\frac{1}{\sigma^2F(\sigma)}\, ,
\end{eqnarray}
where
\begin{eqnarray}
    F(\sigma) \equiv f\big(r(\sigma)\big)=1+\frac{1}{2(D-3)(D-4)\alpha\sigma^2}\Big(1-\sqrt{1+4(D-4)(D-3)[1+(D-4)(D-3)\alpha]\alpha\sigma^{D-1}}\Big)\, ,
\end{eqnarray}
and prime denotes differentiation with respect to $\sigma$. The in-out strategy starts with ingoing null coordinate $v=t+r_{\star}$, which ensures that the time foliation penetrates the event horizon. It then reverses the sign of the singular contribution within $r_{\star}$ near $\sigma=0$. Consequently, the time foliation, which would otherwise reach past null infinity, is instead guided toward future null infinity. Finally, the constructed coordinate is rendered dimensionless and denoted as $\tau$. Conversely, the out-in strategy begins with the  outgoing null coordinate $u=t-r_{\star}$, and reverses the sign of the singular contribution within $r_{\star}$ around $\sigma=1$. The resulting coordinate is also made dimensionless and denoted by $\tau$. The time foliation by constant-$\tau$ surfaces is shown in the Penrose diagram, Fig. \ref{penrose}, for $D=5$ and $\alpha=0.1$, using the out-in strategy.

Using the compact hyperboloidal coordinates (\ref{compact_hyperboloidal_coordinates}), we can recast Eq. (\ref{perturbation eq}) into a hyperbolic equation,
\begin{eqnarray}\label{maineq}
    \partial_\tau u=Lu\, ,\quad
    L=
    \begin{pmatrix}
        0   & 1   \\
        L_1 & L_2
    \end{pmatrix}\, ,\quad
    u=\begin{pmatrix}
        \Psi \\
        \Pi
    \end{pmatrix}\, ,
\end{eqnarray}
where $\Pi=\partial_\tau \Psi$ is introduced to reduce the time derivative order in the equation. The differential operators $L_1$ and $L_2$ in Eq. (\ref{maineq}) are written in Sturm-Liouville form,
\begin{eqnarray}
    L_{1}&=&\frac{1}{w(\sigma)}\Big[\partial_{\sigma}(p(\sigma)\partial_{\sigma})-q_X(\sigma)\Big]\, ,\label{operator_L1}\\
    L_{2}&=&\frac{1}{w(\sigma)}\Big[2\gamma(\sigma)\partial_{\sigma}+\partial_{\sigma}\gamma(\sigma)\Big]\, ,\label{operator_L2}
\end{eqnarray}
where the explicit forms of the above functions are given as follows
\begin{eqnarray}\label{function_p_gamma_w_qx}
    p(\sigma)=\sigma^2F(\sigma)\, ,\quad \gamma(\sigma)=h^{\prime}(\sigma)p(\sigma)\, ,\quad w(\sigma)=\frac{1-\gamma(\sigma)^2}{p(\sigma)}\, ,\quad q_X(\sigma)=\frac{1}{p(\sigma)}V_X(\sigma)\, ,
\end{eqnarray}
and $V_X(\sigma)=r_+^2\tilde{V}_X(\sigma)$ is the dimensionless effective potential with $\tilde{V}_X(\sigma)=\tilde{V}_X(r(\sigma))$. Moreover, it can be found that $w(0)=0$ for $D>5$ within the in-out strategy, indicating that the hypersurfaces $\tau=\text{constant}$ will become asymptotically null. Therefore, we adopt solely the out-in strategy and omit the subscript. So far, we have established the hyperboloidal framework of the BDW black hole. Here, such a hyperboloidal framework refers to the PDE system [Eqs. (\ref{maineq})-(\ref{function_p_gamma_w_qx})], in which $L$ is the infinitesimal generator of time evolution.

\begin{figure}[htbp]
    \centering
    \subfigure[]{\includegraphics[width=0.45\linewidth]{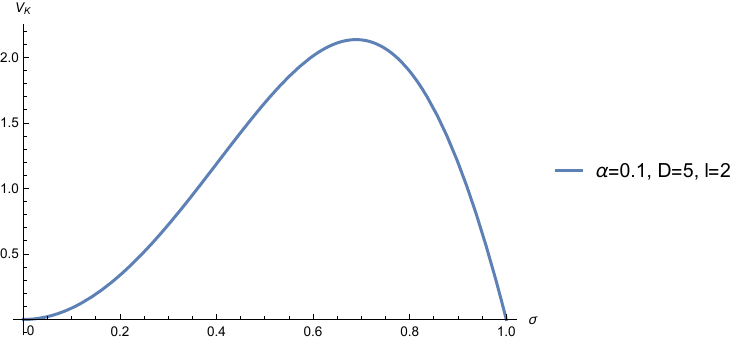}}
    \hfill
    \subfigure[]{\includegraphics[width=0.45\linewidth]{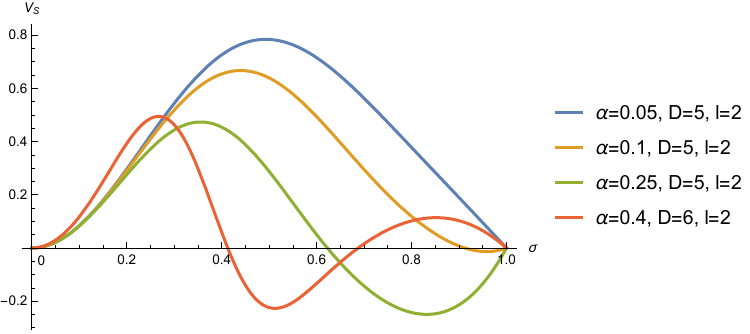}}\\
    \subfigure[]{\includegraphics[width=0.45\linewidth]{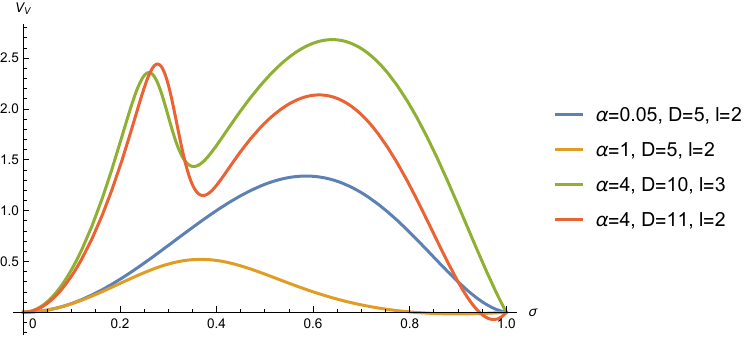}}
    \hfill
    \subfigure[]{\includegraphics[width=0.45\linewidth]{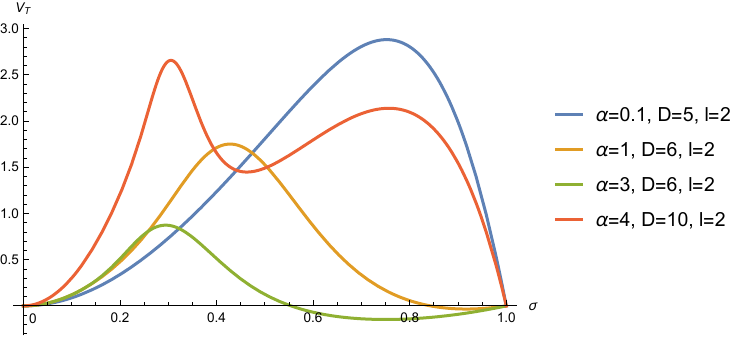}}
    \caption{Some typical behaviors of effective potentials $V_X(\sigma)$ in terms of the coordinate $\sigma$. With the exception of the KG case, four typical effective potentials are shown for all three other cases. The top row shows the potentials for the KG equation and scalar perturbation, while the bottom row depicts the potentials for vector and tensor perturbations. $\sigma=0$ and $\sigma=1$ corresponds to null infinity $\mathscr{I}^+$ and event horizon $\mathscr{H}$ respectively.}
    \label{fig:potentialfig}
\end{figure}

Since the shape of the effective potential can significantly influence the time-domain behavior and QNM spectra, in Fig. \ref{fig:potentialfig}, we show several typical effective potentials, which will be served as benchmarks for subsequent studies. Four panels of Fig. \ref{fig:potentialfig} represent the effective potentials $V_X(\sigma)$, with the coordinate $\sigma$ as the variable, for KG type, scalar type, vector type, tensor type perturbations in turn. From the first panel, the KG case presents a straightforward single-peaked positive potential barrier. In contrast, the scalar, vector, and tensor perturbation cases display more intricate behaviors. These potentials can feature double-peaked structures, with or without negative regions that may arise either adjacent to or separate from the event horizon. It is worth mentioning that although the second and third parameters of both the scalar and tensor perturbations share similar characteristics (see the yellow and green lines of the second panels and the fourth panels of Fig. \ref{fig:potentialfig}), exhibiting a single peak and a negative region adjacent to the horizon, their QNM spectra and time-domain behaviors differ unexpectedly. The second parameters (yellow lines) are stable, showing no unstable QNMs and remaining well-behaved over time, but instability arises in the third parameters (green lines) when the potential’s negative region exceeds a critical depth. Therefore, we treat them as two distinct cases. These structures emerge naturally from EGB theory under varying parameters rather than artificially imposed ones. These cases are quite different from the situation under the framework of general relativity where the effective potential exhibits only a single peak.  The presence of double-peaked structures and negative regions renders some traditional methods like WKB approximation invalid to get the QNM frequencies, while at the same time the irrational nature of the metric function $f(r)$ makes the continued fraction method inapplicable~\cite{Berti:2009kk}. A significant advantage of the pseudo-spectral method is its ability to circumvent these difficulties and its high versatility.

We include a remark regarding the scope of the parameters of our analysis to end this section. Our analysis is performed for a fixed $l$ and perturbation type. To ensure the legitimacy of our analysis, it is necessary to verify that the analysis of the stability of QNMs is calculated exclusively for stable black hole solutions, which must be stable under any perturbation type and for any value of $l$. Being consistent with the previous study of linear perturbation theory for BDW black holes~\cite{Dotti:2004sh,Dotti:2005sq,Gleiser:2005ra,Beroiz:2007gp,Konoplya:2008ix,Takahashi:2009dz,Takahashi:2009xh,Takahashi:2010gz,Takahashi:2011cgy}, we use the hyperboloidal framework and identify only two specific cases where dynamically unstable QNMs may exist: scalar type perturbations in $5$-dimensions and tensor type perturbations in $6$-dimensions. In all other cases, no dynamically unstable QNMs ($\text{Im} \omega<0$) were found for various multipole numbers $l$. Furthermore, in the cases where unstable QNMs may exist, higher multipole numbers $l$ are more likely to lead to instability~\cite{Konoplya:2008ix,Takahashi:2010gz}. The explicit parameter regime leading to instability can be further determined through applying the analytical method developed by~\cite{Takahashi:2009dz, Takahashi:2009xh, Takahashi:2010gz, Takahashi:2011cgy} within the context of EGB theory. Note that we adopt the dimensionless coupling parameter $\alpha$ and dimensionless mass parameter $\mu$, where the event horizon radius $r_+$ is the unit (see Eq. (\ref{dimensionless_quantities}), similar to ~\cite{Konoplya:2008ix,Konoplya:2017ymp}). In this convention, the ``small black hole'' in ~\cite{Takahashi:2010gz} corresponds to a black hole with a large $\alpha$, which is also the case in ~\cite{Konoplya:2008ix,Konoplya:2017ymp}. From the results of~\cite{Takahashi:2010gz}, the BDW black hole will be unstable when
\begin{eqnarray}
    \alpha>\frac{\sqrt{2}-1}{4}\approx0.104
\end{eqnarray}
for $5$-dimensions under scalar perturbation and
\begin{eqnarray}
    \alpha>\frac{\sqrt{25+10\sqrt{6}}-1}{12}\approx0.503
\end{eqnarray} for $6$-dimensions under tensor perturbation. These two thresholds closely match the numerical predication of~\cite{Konoplya:2008ix}, where one should be aware that the definition of $\alpha$ in~\cite{Konoplya:2008ix,Konoplya:2017ymp} is twice as ours. Except for three parameters ($\{\alpha=0.25, D=5\}, \{\alpha=3, D=6\}, \{\alpha=1, D=6\}$), all other parameters we selected avoid unstable parameter regimes. Although the black hole with the parameter $\{\alpha=1, D=6\}$ is stable under the multipole number $l=2$, it is unstable under perturbations with $l=4$. We will present the results of the dynamic (in)stability in terms of QNMs for these parameters in the next section.

\section{The quasinormal modes}\label{QNMs}
In this section, we start from Eq. (\ref{maineq})-Eq. (\ref{function_p_gamma_w_qx}) to analyze the QNMs of the BDW black hole by computing spectra in the frequency-domain and simulating the time-domain behavior. As a basis, we first introduce the pseudo-spectral method used. We discretize the spatial domain on a grid $\{\sigma_i\}_{i=0}^N$, converting the original PDE system to a coupled ODE system. The state vector $u(\tau,\sigma)$ is evaluated in such a spatial grid so that $\Psi(\tau,\sigma)$ is converted to $\Psi_i(\tau)\equiv\Psi(\tau,\sigma_i)$, and $\Pi(\tau,\sigma)$ is converted into $\Pi_i(\tau)\equiv\Pi(\tau,\sigma_i)$. Thus, the problem reduces to the solution of ODEs involving the variables $\Psi_i(\tau)$ and $\Pi_i(\tau)$ for $i=0,\cdots,N$. Finally, Eq. (\ref{maineq}) amounts to a system of $(2N+2)$ ODEs associated with the time variable $\tau$ via
\begin{eqnarray}\label{odes}
    \frac{\mathrm{d}}{\mathrm{d}\tau}\begin{bmatrix}
        \Psi_i \\
        \Pi_i
    \end{bmatrix}
    =\sum_{j=0}^{N}
    \begin{bmatrix}
        0                                                        & \delta_{ij}                       \\
        C_i(\mathbf{D}^2)_{ij}+E_i\mathbf{D}_{ij}+W_i\delta_{ij} & A_i\mathbf{D}_{ij}+B_i\delta_{ij}
    \end{bmatrix}
    \begin{bmatrix}
        \Psi_j \\
        \Pi_j
    \end{bmatrix}\, ,\quad i=0,1,\cdots,N-1,N\, ,
\end{eqnarray}
where $A_i=A(\sigma_i)$, $B_i=B(\sigma_i)$, $C_i=C(\sigma_i)$, $E_i=E(\sigma_i)$, $W_i=W(\sigma_i)$ and $\delta_{ij}$ is the Kronecker delta. These five functions $C(\sigma)$, $E(\sigma)$, $W(\sigma)$, $A(\sigma)$ and $B(\sigma)$ are directly derived from Eq. (\ref{operator_L1}) and Eq. (\ref{operator_L2}). There is no summation over $i$ in the above equation (\ref{odes}). The specific form of the differential matrix $\mathbf{D}$\footnote{We use bold symbols to represent matrices and vectors.} is determined by the chosen Chebyshev-Lobatto grid, given by
\begin{eqnarray}
    \sigma_j=\frac{1}{2}\left(1+\cos\Big(\frac{j\pi}{N}\Big)\right)\, ,\quad j=0,1,\cdots,N-1,N\, .
\end{eqnarray}
The expression of the differential matrix $\mathbf{D}$ associated with the Chebyshev-Lobatto grid can be found in~\cite{boyd2001chebyshev,Jaramillo:2020tuu,Cao:2024oud}.

In frequency-domain, consider the Fourier transform of $u(\tau,\sigma)$ with respect to time $\tau$, we have
\begin{eqnarray}
    u(\tau, \sigma)=e^{i\tilde{\omega}t}\tilde{u}(r_\star)=e^{i\omega\tau}u(\sigma)\, ,
\end{eqnarray}
where $\omega=r_+\tilde{\omega}$ is the nondimensionalized frequency, and $u(\sigma)=e^{-i\omega h(\sigma)}\tilde{u}(r_\star)$. For Eq. (\ref{perturbation eq}) within the Schwarzschild-like coordinates, since all the constant-$t$ surfaces intersect the spatial infinity $i^0$ and the bifurcation sphere $\mathcal{B}$ simultaneously, the QNM eigenfunction $\tilde{u}(r_\star)$ diverges at both boundaries. Therefore, QNM boundary conditions must be carefully imposed, and it is required that a prior separation of the divergent part of $\tilde{u}(r_\star)$. In contrast, the hyperboloidal formalism, where the constant-$\tau$ surface ends at null infinity $\mathscr{I}^+$ and penetrates the future event horizon $\mathscr{H}^+$, naturally extracts the divergent part, $e^{i\omega h(\sigma)}$, through a coordinate transformation. Moreover, the light-cones point outward at the boundary of the computation domain, simplifying boundary conditions to merely requiring a regular solution $u(\sigma)$, which is trivially satisfied in numerical calculations. This is evident from the structure of the ingoing ($\bar{l}^a$) and outgoing ($\bar{k}^a$) conformal null vectors in hyperboloidal coordinates,
\begin{eqnarray}
    \bar{l}^a=\nu\left(\delta^a_\tau-\frac{1+\gamma}{w}\delta^a_\sigma\right)\, ,\quad \bar{k}^a=\frac{w}{2\nu}\left(\delta^a_\tau+\frac{1-\gamma}{w}\delta^a_\sigma\right)\, ,
\end{eqnarray}
where $\nu(\sigma)$ is a freely specifiable boost parameter under Lorentz transformation~\cite{PanossoMacedo:2023qzp}. Crucially, the time evolution vector $(\partial_\tau)^a$ aligns with the outgoing null vector $\bar{k}^a$ at future null infinity $\mathscr{I}^+$ and with the ingoing null vector $\bar{l}^a$ at event horizon $\mathscr{H}^+$. Consequently, at these boundaries, the local light cones are oriented outwards from the computational domain. This causal structure ensures that no boundary information propagates into the domain, thereby eliminating the need for explicit boundary data and requiring only numerical regularity of the solution. Accordingly, the hyperboloidal formalism provides a significant advantage for computing QNMs~\cite{PanossoMacedo:2024nkw}.

In the numerical calculations, the operator $L$ is approximated by a matrix operator $\mathbf{L}$ by using the Chebyshev-Lobatto grid. Therefore, the problem of solving QNM spectra now turns into a finite-dimenisonal algebraic eigenvalue problem
\begin{eqnarray}\label{eigenvalue_porblem}
    \mathbf{L}\mathbf{u}=i\omega \mathbf{u}\, ,
\end{eqnarray}
where the matrix $\mathbf{L}$ above is exactly the coefficient matrix of the right-hand of Eq. (\ref{odes}). Although numerical approaches can yield modes that closely approximate true physical modes, they are always accompanied by undesired numerical spurious modes. This requires us to find a way to distinguish between the two. A reliable method based on defining the drifts among the QNM spectra can be found in~\cite{boyd2001chebyshev,Chen:2024mon,BOYD199611,Cownden:2023dam}. One can refer to Appendix \ref{sec: drift} to get more technical details about such method. The spherical symmetry of spacetime results in the matrix $\mathbf{L}$ being pure real, which implies that both $\omega$ and $-\bar{\omega}$ are eigenvalues of Eq. (\ref{eigenvalue_porblem}), where bar represents the complex conjugate. Consequently, QNM frequencies are symmetric about the imaginary axis in the complex frequency plane. Therefore, we only present modes with a non-negative real part. The branch cut in our asymptotically flat spacetime leads to a continuous spectrum along the positive imaginary axis. So we exclude this region from our definitions of the nearest drift (see Appendix \ref{sec: drift}).

The above statements concern the calculation of QNM spectra in the frequency domain. Now we turn to the time-domain behavior of QNMs. For the time-domain case, we have ODEs (\ref{odes}) from Eq. (\ref{maineq}). To solve the ODEs numerically, we adopt a discrete time evolution scheme via a $6$th-order Hermite integration method~\cite{markakis2019timesymmetry,Markakis:2023pfh, DaSilva:2024yea,OBoyle:2022lek,OBoyle:2022yhp}, which is actually an implicit scheme. Selecting a constant time step $\Delta\tau$, we arrive at
\begin{eqnarray}
    \mathbf{u}\left((i+1)\Delta\tau\right) = \mathbf{U}\cdot\mathbf{u}(i\Delta\tau),\quad i=0,1,\cdots\, ,
\end{eqnarray}
where $\mathbf{U}$ is called evolution matrix whose explicit form can be expressed as
\begin{eqnarray}
    \mathbf{U}=\mathbf{I}+(\Delta \tau \mathbf{L})\cdot\left(\mathbf{I}+\frac{1}{60}(\Delta \tau \mathbf{L})\cdot(\Delta \tau \mathbf{L})\right)\cdot\left(\mathbf{I}-\frac{\Delta \tau}2\mathbf{L}\cdot\left(\mathbf{I}-\frac{\Delta \tau}5\cdot\left(\mathbf{I}-\frac{\Delta \tau}{12}\mathbf{L}\right)\right)\right)^{-1},
\end{eqnarray}
and $\mathbf{I}$ is the identity matrix whose dimension is consistent with that of $\mathbf{L}$. Because such scheme is unconditionally stable, there is no Courant limit on the time step $\Delta\tau$. Meanwhile, there is no requirement for constant time steps, i.e., $\Delta\tau$ can depend on $i$. For convenience, we simply take even time steps in our practice.

For initial data, we choose a Gaussian pulse on the coordinate $\sigma$ as the initial perturbation, i.e.,
\begin{eqnarray}
    \Psi(\tau=0,\sigma)&=&a_0\exp\bigg[-\frac{(\sigma-c_0)^2}{2b_0^2}\bigg]\, ,\nonumber\\
    \Pi(\tau=0,\sigma)&=&0\, ,
    \label{initial condition}
\end{eqnarray}
where $a_0$ is the amplitude, $b_0$ is the width and $c_0$ is the position of the Gaussian pulse. We employ the hyperboloidal foliation, which terminates at future null infinity rather than spacelike infinity. The perturbations we consider propagate at the speed of light and thus can reach future null infinity. Therefore, there is no need to resrict initial data to having compact support, which would require rapid decay near the boundary. From a mathematical perspective, the hyperboloidal framework requires only regular initial data. In addition, within this framework, many studies also adopt the initial condition as a Gaussian wave packet~\cite{Besson:2024adi,Cardoso:2024jme}.

\begin{figure}[htbp]
    \subfigure[]{\includegraphics[width=0.24\textwidth]{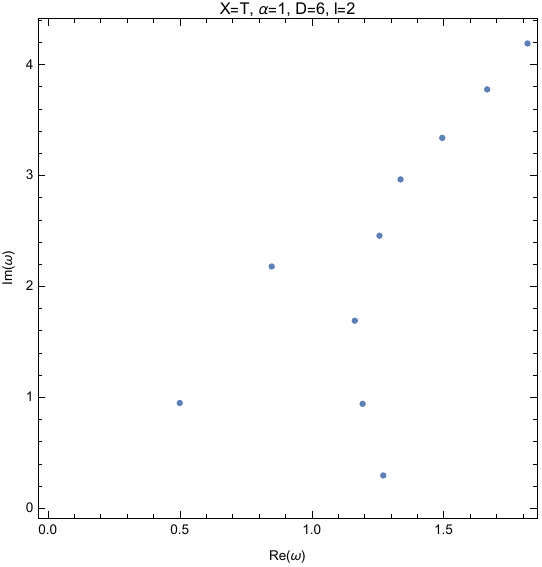}\label{qnmT20}}~
    \subfigure[]{\includegraphics[width=0.246\textwidth]{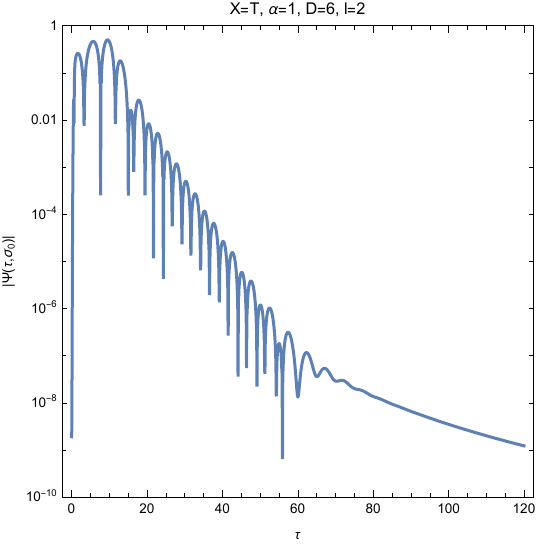}\label{timeT20}}
    \hfill
    \subfigure[]{\includegraphics[width=0.24\textwidth]{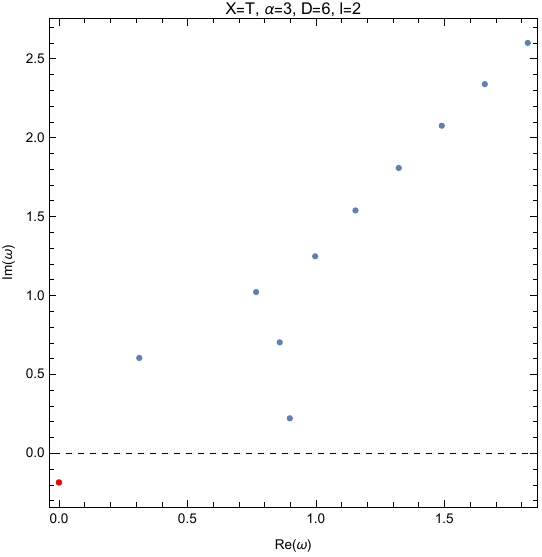}\label{qnmT2}}~
    \subfigure[]{\includegraphics[width=0.246\textwidth]{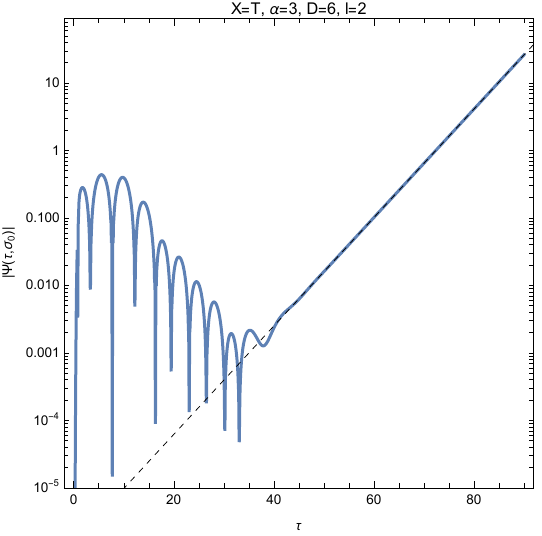}\label{timeT2}}
    \\
    \subfigure[]{\includegraphics[width=0.24\textwidth]{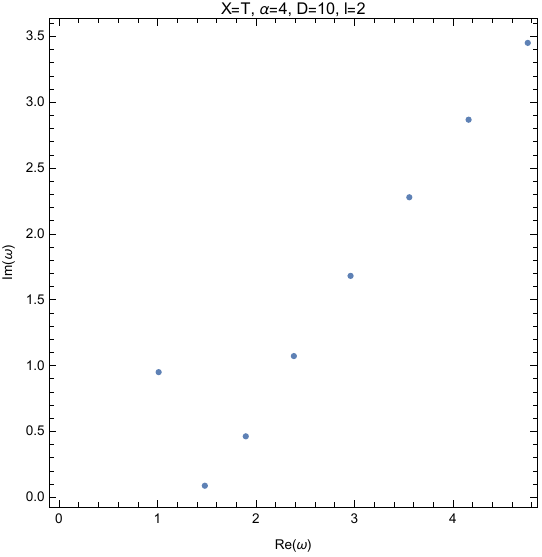}\label{qnmT3}}~
    \subfigure[]{\includegraphics[width=0.246\textwidth]{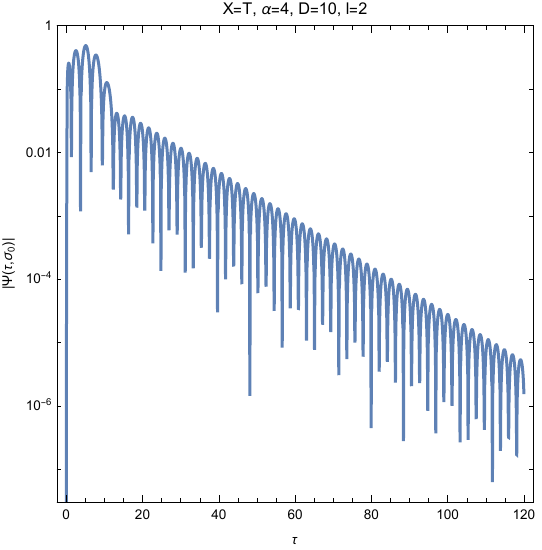}\label{timeT3}}
    \caption{The QNM spectra and time-domain evolution for the parameter choices in Fig. \ref{fig:potentialfig}. The initial condition in time domain calculation is chosen as $a_0=1$, $b_0=1/(10\sqrt{10})$ and $c_0=1/5$. Both the calculation of QNM spectra and the time-domain evolution is performed on the resolution with $N=160$. The time evolution employs a time step $\Delta \tau=0.075$, with the observer positioned at infinity (i.e., $\sigma_0=0$). Instable modes are labeled in red color.}
    \label{fig:qnmfig}
\end{figure}

In the hyperboloidal framework, the compactified coordinates ensure that the black hole horizon and null infinity can be reached, while in the usual Schwarzschild-like coordinates, we cannot strictly discuss infinity without using extra methods. We show the main results of this section in Fig. \ref{fig:qnmfig}, where the parameters used are from Fig. \ref{fig:potentialfig}, and the time evolution is seen by an observer located at infinity ($\sigma_0=0$). Both the QNM spectra and time-domain evolution are computed at the resolution $N=160$. The time integration proceeds with time step $\Delta\tau=0.075$. The initial condition for time domain is chosen as Eqs. (\ref{initial condition}) with $a_0=1$, $b_0=1/(10\sqrt{10})$, and $c_0=1/5$. Note that Fig. \ref{fig:qnmfig} has a total of $26$ subfigures (a)-(z). The odd sequences [like (a), (c), $\cdots$, (y)] show the QNM spectra and the even sequences [like (b), (d), $\cdots$, (z)] show the corresponding time-domain waveforms. These time-domain figures show the logarithm of the absolute value of the waveform. The phase dominated by a single QNM in the time-domian figure exhibits a series of interconnected bell-like shapes. Such shape reflects the characteristics of the QNM. A wider bell opening corresponds to a smaller real part of the QNM spectrum, and a steeper slope of the envelope at the bell's peak indicates a larger imaginary part of the QNM spectrum. The results of these $13$ sets of parameters present different characteristics of QNMs, which we summarize as follows.

\begin{figure}[htbp]
    \centering
    \subfigure[]{\includegraphics[width=0.24\textwidth]{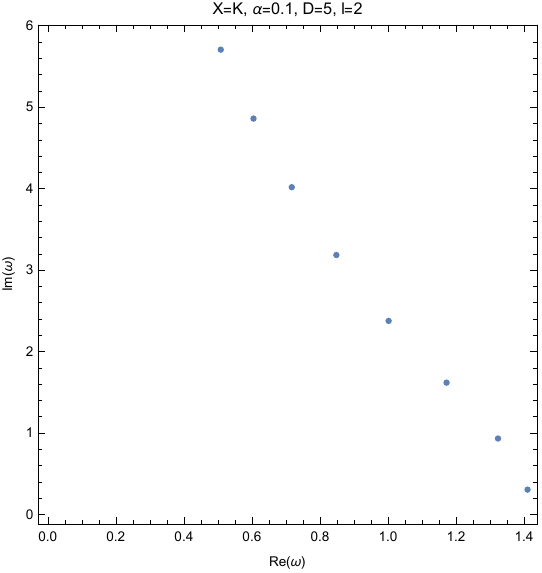}\label{qnmK}}~
    \subfigure[]{\includegraphics[width=0.246\textwidth]{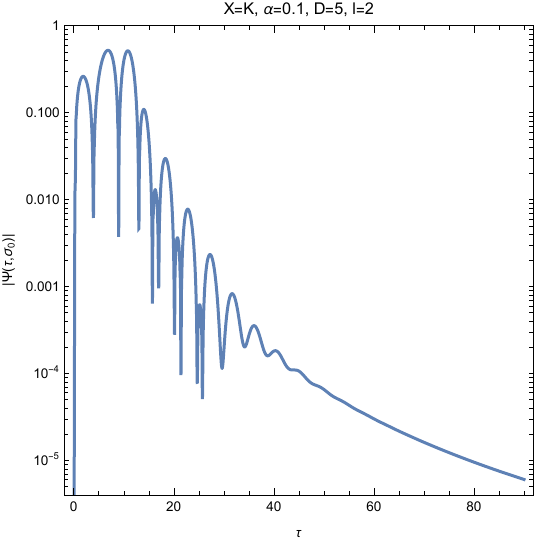}\label{timeK}}
    \hfill
    \subfigure[]{\includegraphics[width=0.24\textwidth]{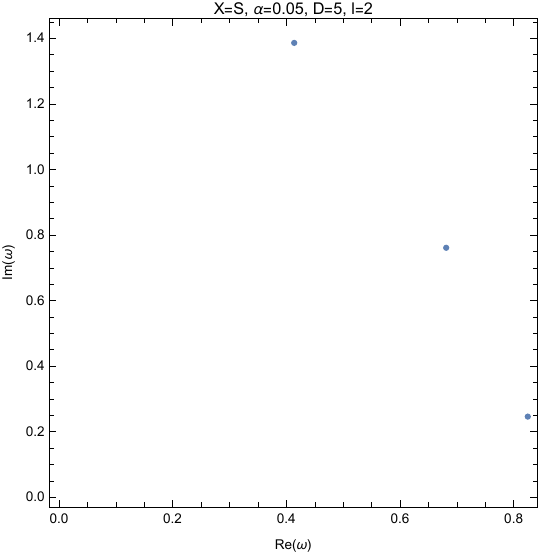}\label{qnmS1}}~
    \subfigure[]{\includegraphics[width=0.246\textwidth]{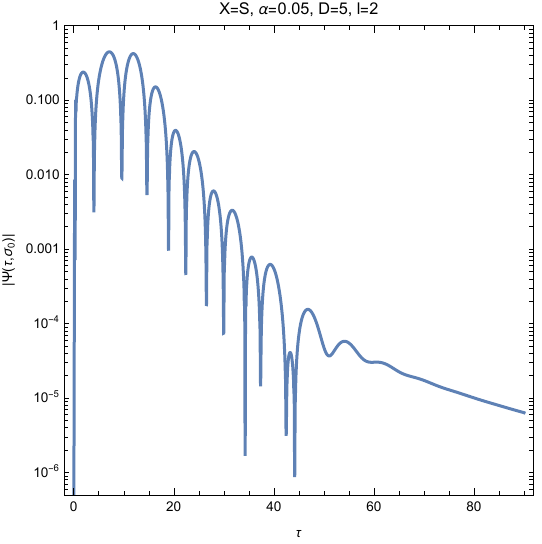}\label{timeS1}}
    \\
    \subfigure[]{\includegraphics[width=0.24\textwidth]{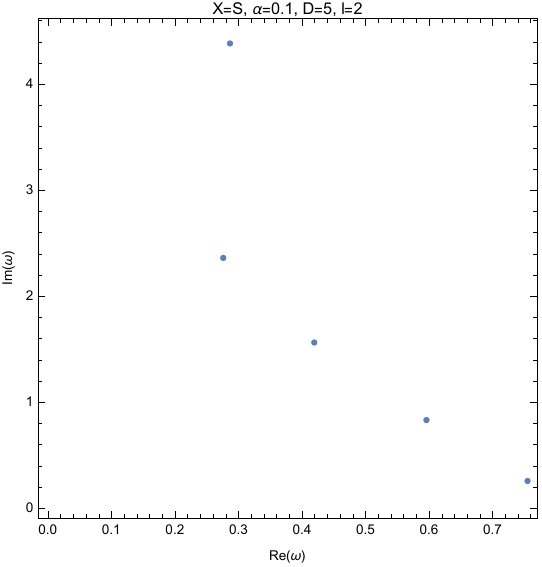}\label{qnmS20}}~
    \subfigure[]{\includegraphics[width=0.246\textwidth]{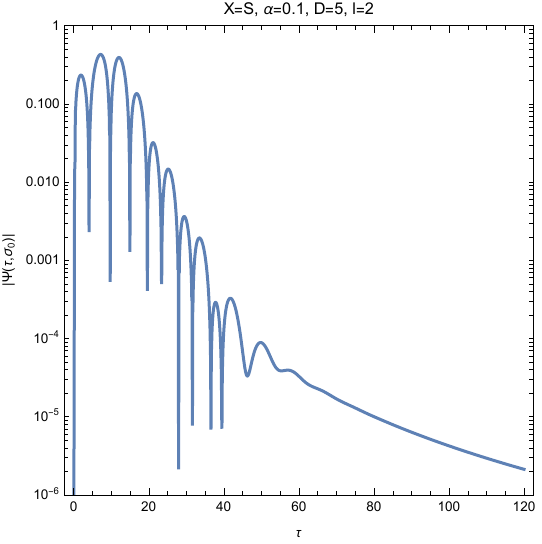}\label{timeS20}}
    \hfill
    \subfigure[]{\includegraphics[width=0.24\textwidth]{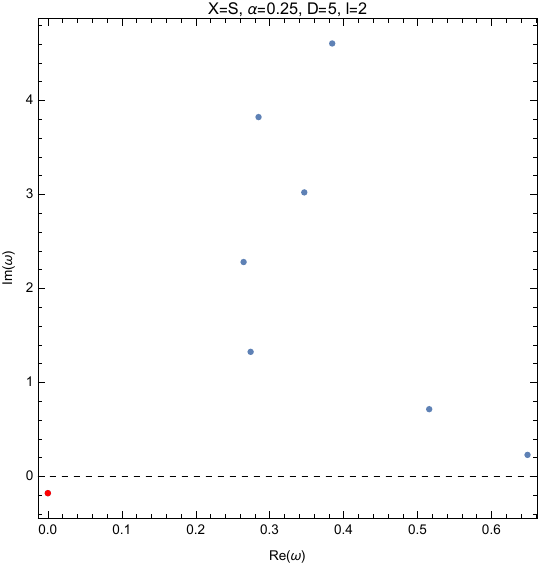}\label{qnmS2}}~
    \subfigure[]{\includegraphics[width=0.246\textwidth]{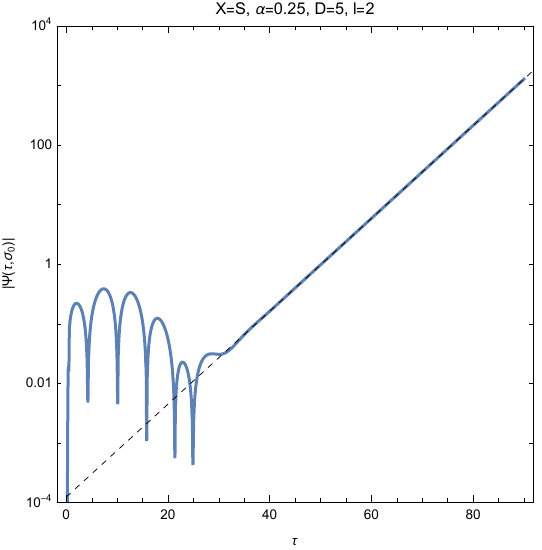}\label{timeS2}}
    \\
    \subfigure[]{\includegraphics[width=0.24\textwidth]{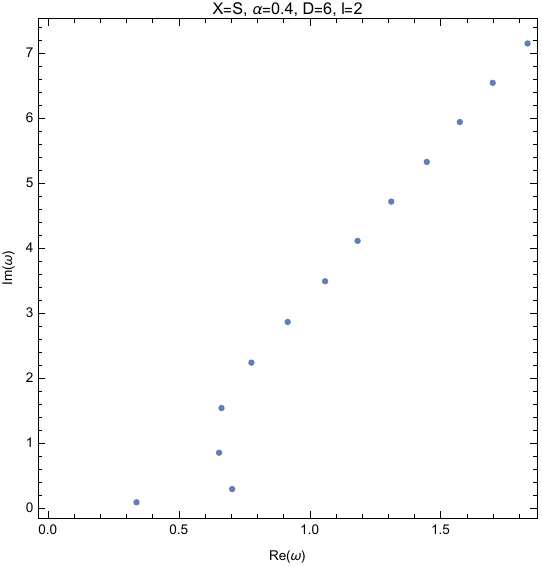}\label{qnmS3}}~
    \subfigure[]{\includegraphics[width=0.246\textwidth]{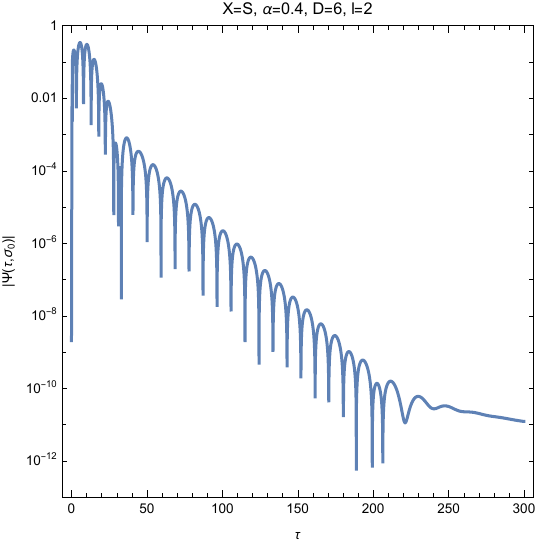}\label{timeS3}}
    \hfill
    \subfigure[]{\includegraphics[width=0.24\textwidth]{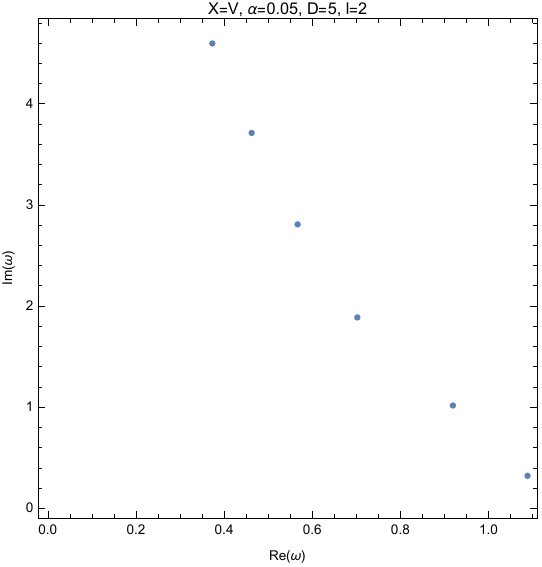}\label{qnmV1}}~
    \subfigure[]{\includegraphics[width=0.246\textwidth]{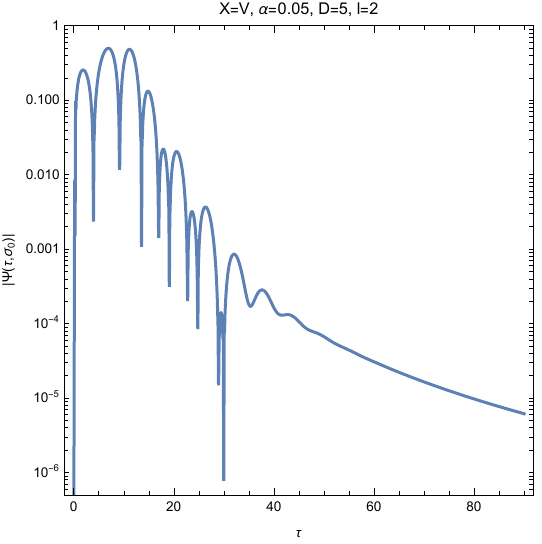}\label{timeV1}}
    \\
    \subfigure[]{\includegraphics[width=0.24\textwidth]{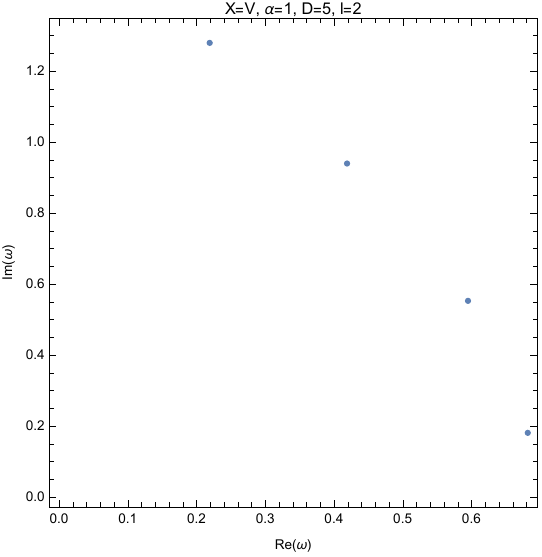}\label{qnmV2}}~
    \subfigure[]{\includegraphics[width=0.246\textwidth]{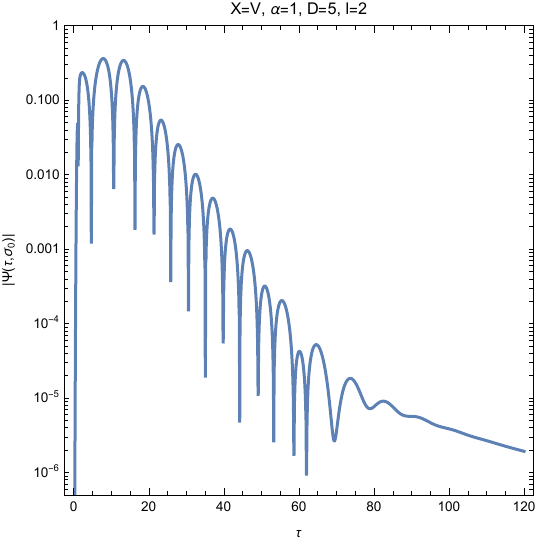}\label{timeV2}}
    \hfill
    \subfigure[]{\includegraphics[width=0.24\textwidth]{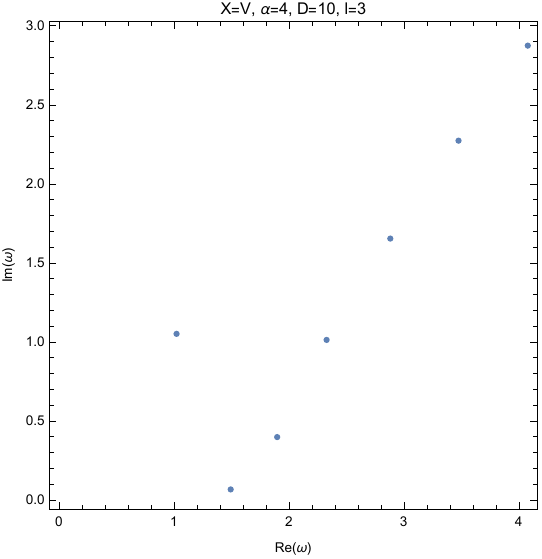}\label{qnmV3}}~
    \subfigure[]{\includegraphics[width=0.246\textwidth]{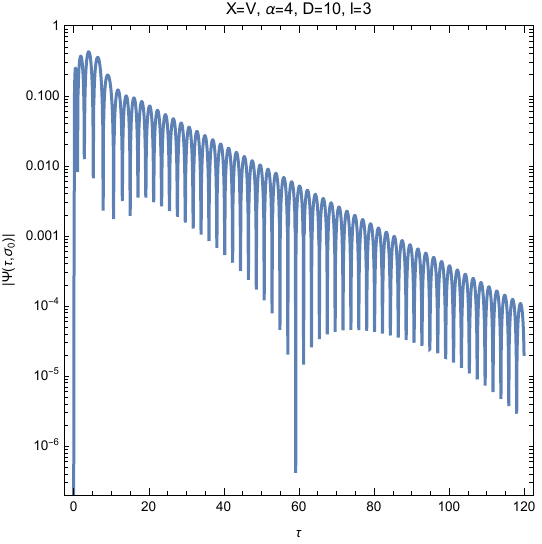}\label{timeV3}}
    \\
    \subfigure[]{\includegraphics[width=0.24\textwidth]{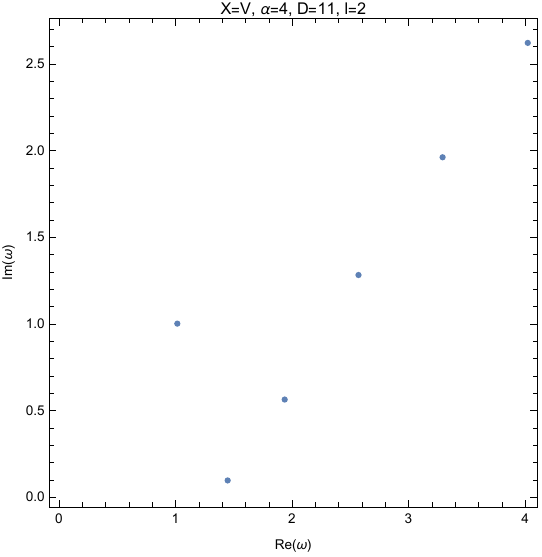}\label{qnmV4}}~
    \subfigure[]{\includegraphics[width=0.246\textwidth]{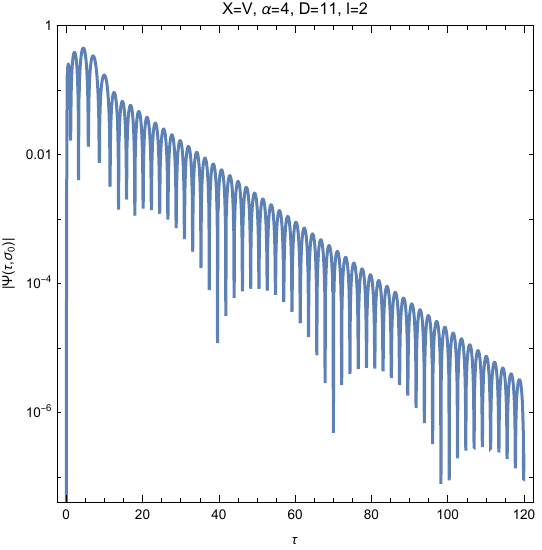}\label{timeV4}}
    \hfill
    \subfigure[]{\includegraphics[width=0.24\textwidth]{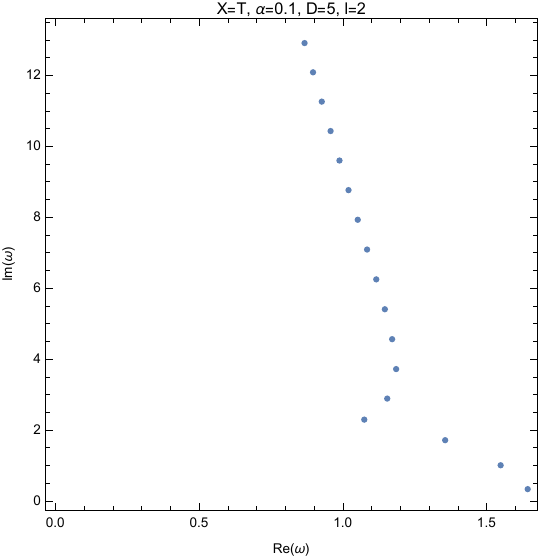}\label{qnmT1}}~
    \subfigure[]{\includegraphics[width=0.246\textwidth]{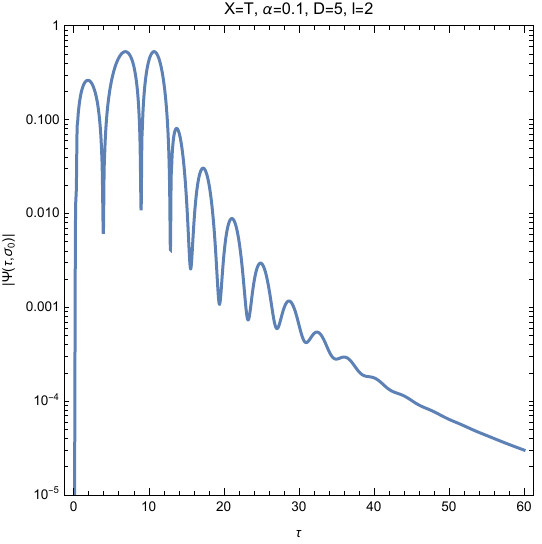}\label{timeT1}}
\end{figure}

\begin{figure}[htbp]
    \centering
    \subfigure[]{\includegraphics[width=0.4\textwidth]{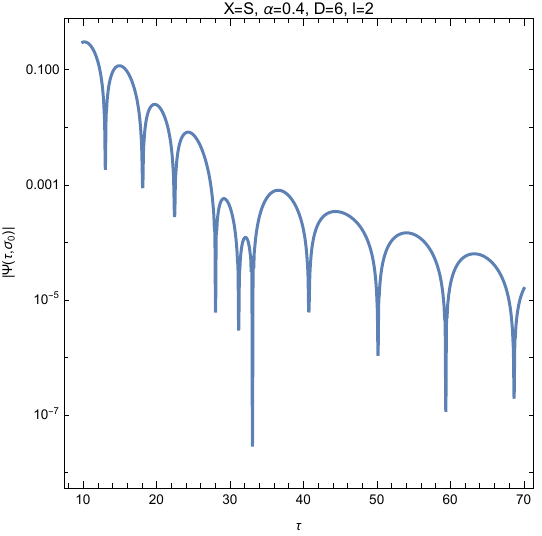}\label{timeS3zoomin}}\hspace{0.5cm}
    \subfigure[]{\includegraphics[width=0.41\textwidth]{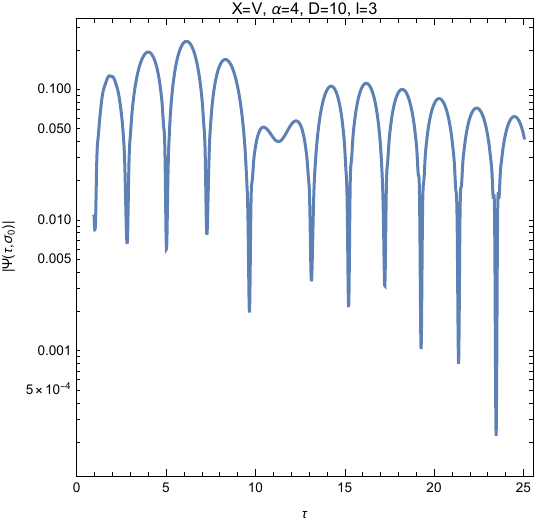}\label{timeV3zoomin}}
    \hfill
    \subfigure[]{\includegraphics[width=0.41\textwidth]{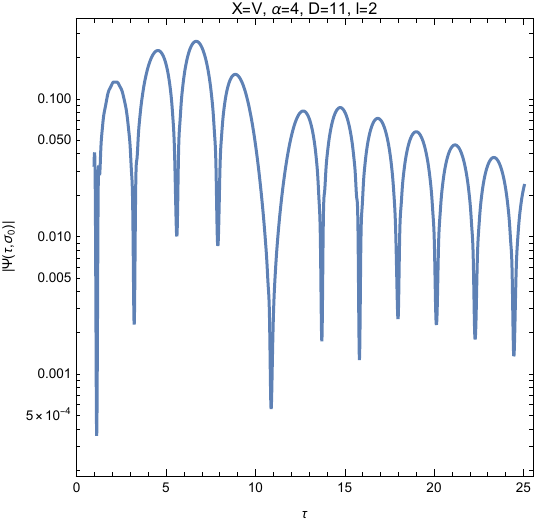}\label{timeV4zoomin}}\hspace{0.5cm}
    \subfigure[]{\includegraphics[width=0.4\textwidth]{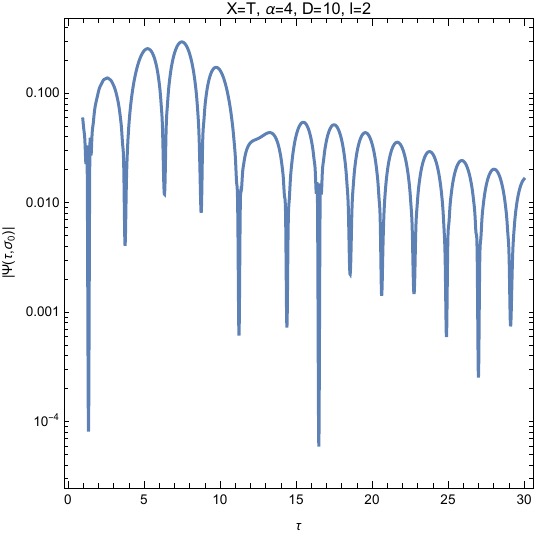}\label{timeT3zoomin}}
    \caption{The zoomed-in view of Figs. \ref{timeS3}, \ref{timeV3}, \ref{timeV4}, and \ref{timeT3}, where the rise phase followed after the first decay phase is illustrated. The parameters in Figs. \ref{timeV3zoomin}, \ref{timeV4zoomin}, and \ref{timeT3zoomin} are the same as those in Figs. \ref{timeV3}, \ref{timeV4}, and \ref{timeT3} except the initial condition $b_0=1/(10\sqrt{10})$ is changed as $b_0=1/100$.}
    \label{fig:timezoomin}
\end{figure}

First, it can be seen that for the parameters leading to double-peaked effective potentials, there exist long-lived modes depicted in Figs. \ref{qnmS3}, \ref{qnmV3}, \ref{qnmV4} and \ref{qnmT3}, and at the same time the echoes will occur in Figs. \ref{timeS3}, \ref{timeV3}, \ref{timeV4}, \ref{timeT3}. Long-lived modes indicate the echo pattern, which decays slowly and persists for a long period, covering up the power-law tail for a long time. Furthermore, the echo pattern also displays a rise phase after the first decay phase, as is shown in Fig. \ref{fig:timezoomin}. Fig. \ref{fig:timezoomin} presents four panels: the first zooms in on Fig. \ref{timeS3} (with identical parameters), while the remaining three zoom in on Figs.  \ref{timeV3}, \ref{timeV4}, and \ref{timeT3} (using their parameters but with initial conditions $b_0 = 1/100$ to highlight the echo pattern). In these four plots, it is evident that the dominant QNM during the first decay phase differs from the one dominating the subsequent rise phase. This transition marks the emergence of the echo mode, which begins to dominate thereafter.

\begin{figure}[htbp]
    \centering
    \subfigure[]{\includegraphics[width=0.24\textwidth]{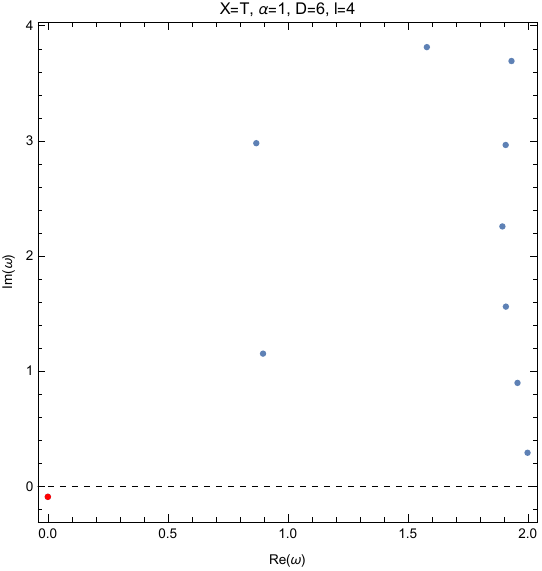}\label{qnmT20unstable}}~
    \subfigure[]{\includegraphics[width=0.253\textwidth]{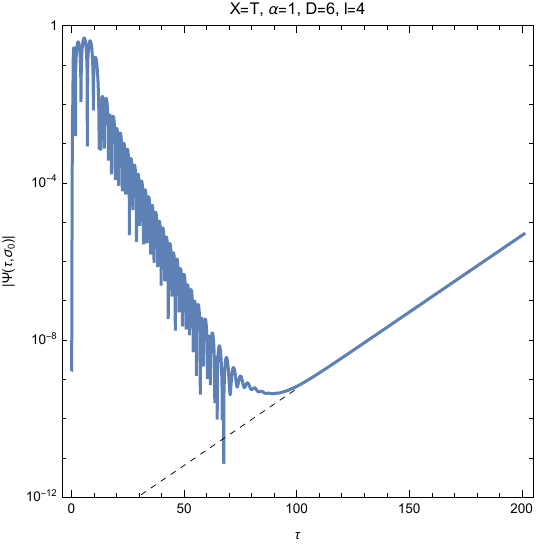}\label{timeT20unstable}}
    \caption{The QNM spectra and time-domain evolution for the parameter $\{X=T, \alpha=1, D=6, l=4\}$. The initial condition in time-domain calculation is chosen as $a_0=1$, $b_0=1/(10\sqrt{10})$ and $c_0=1/5$. Both the calculation of QNM spectra and the time-domain evolution is performed on the resolution with $N=160$. The time evolution employs a time step $\Delta \tau=0.075$, with the observer positioned at infinity (i.e., $\sigma_0=0$). Instable modes are labeled in red color.}
    \label{fig:qnmfigT20unstable}
\end{figure}

Second, the dynamics instability is previously reported by~\cite{Dotti:2004sh, Dotti:2005sq, Beroiz:2007gp, Konoplya:2008ix, Takahashi:2009dz, Takahashi:2009xh, Takahashi:2010gz, Takahashi:2011cgy}, which is triggered by large $\alpha$ for scalar type in $5$-dimensional spacetime and for tensor type in $6$-dimensional spacetime. Similarly, it is also observed that there exists the dynamics instability, i.e., an exponential growth in the time-domain simulations in Figs. \ref{timeS2}, \ref{timeT2}, and \ref{timeT20unstable}. The QNM spectra corresponding to these subfigures are shown in Figs. \ref{qnmS2}, \ref{qnmT2} and \ref{qnmT20unstable}, and all of them have the mode with a negative imaginary part. The increase in $l$ leads to a transition from dynamically stable [Figs. \ref{qnmT20}, \ref{timeT20}] to dynamically unstable [Figs. \ref{qnmT20unstable}, \ref{timeT20unstable}], which is consistent with~\cite{Konoplya:2008ix, Takahashi:2010gz}. Additionally, we find that these unstable modes always appear on the negative half of the imaginary axis. These modes do not migrate from stable modes, but rather, they appear abruptly on the imaginary axis. The slope of the fitting lines precisely matches the imaginary part of the corresponding unstable modes.

\begin{figure}[htbp]
    \centering
    \subfigure[]{\includegraphics[width=0.4\linewidth]{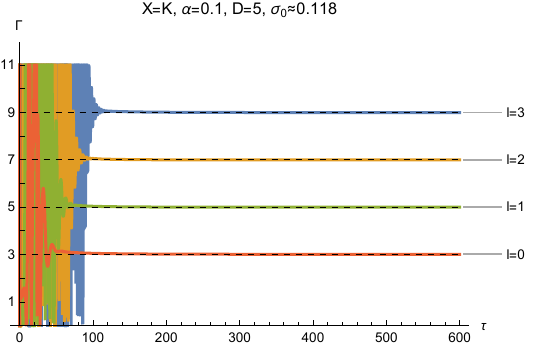}}
    \hspace{0.3cm}
    \subfigure[]{\includegraphics[width=0.4\linewidth]{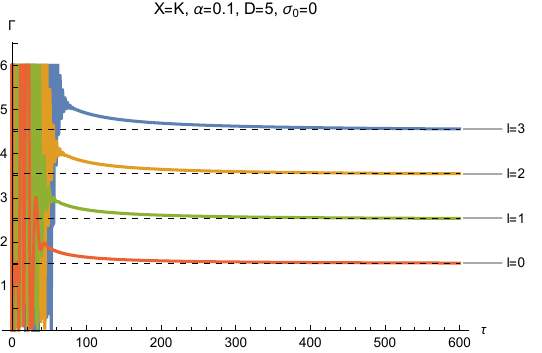}}
    \\
    \subfigure[]{\includegraphics[width=0.4\linewidth]{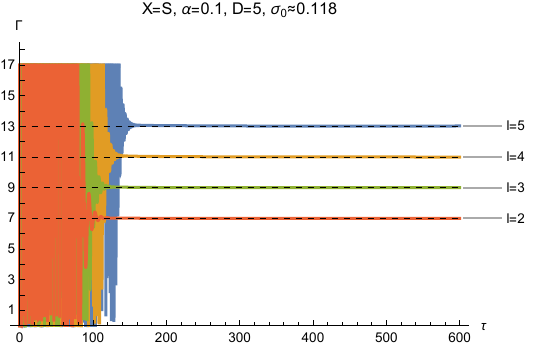}}
    \hspace{0.3cm}
    \subfigure[]{\includegraphics[width=0.4\linewidth]{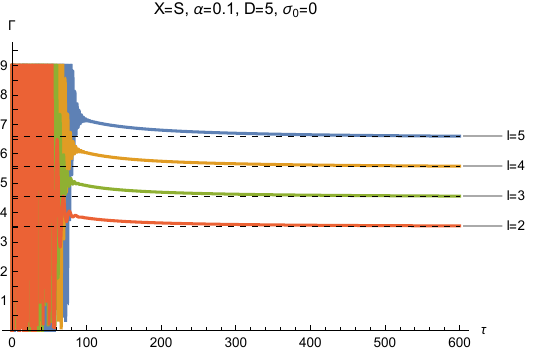}}
    \\
    \subfigure[]{\includegraphics[width=0.4\linewidth]{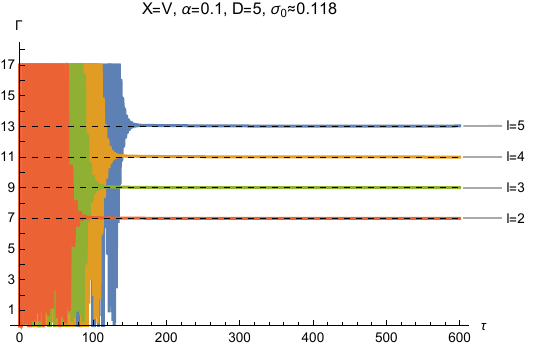}}
    \hspace{0.3cm}
    \subfigure[]{\includegraphics[width=0.4\linewidth]{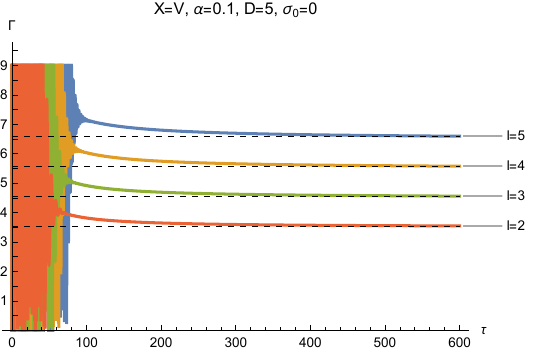}}
    \\
    \subfigure[]{\includegraphics[width=0.4\linewidth]{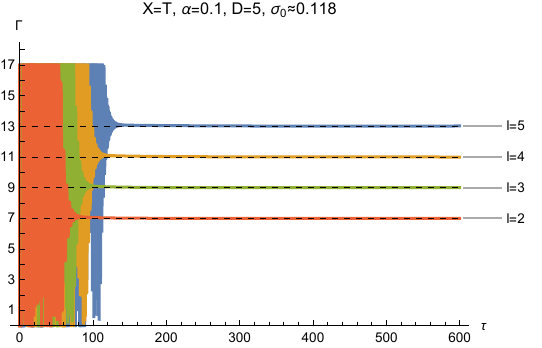}}
    \hspace{0.3cm}
    \subfigure[]{\includegraphics[width=0.4\linewidth]{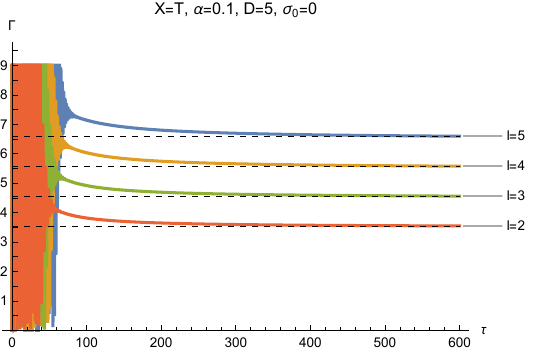}}
    \caption{The effective power-law index $\Gamma$ is computed using the initial conditions specified in Eqs. (\ref{initial condition}) with $a_0=1, b_0=1/(10\sqrt{10})$ and $c_0=1/5$. The observer is located at $\sigma_0\approx0.118$ in the first column and infinity, i.e., $\sigma_0=0$, in the second column, with the computations performed on the resolution $N=300$ using time step $\Delta \tau=0.075$.}
    \label{fig:pricelaw}
\end{figure}

We end this section with an analysis of the Price's law~\cite{Price:1971fb,Price:1972pw}. The analysis of the Price's law can further verify the reliability of our code. The analytical method of~\cite{Ching:1994bd,Ching:1995tj} shows that the Price's law obtained by a finite-distant observer is determined by the asymptotic behavior of the effective potential near infinity. Although the forms of the effective potentials for different perturbations vary, these effective potentials exhibit similar asymptotic behavior at infinity, i.e.,
\begin{eqnarray}
    \frac{V_X}{r_+^2}=\frac{\nu(\nu+1)}{r_{\star}^{2}}+\frac{A_X}{r_{\star}^{D-1}}+o\left(\frac{1}{r_{\star}^{D-1}}\right)\, ,
\end{eqnarray}
where
\begin{eqnarray}\label{nu}
    \nu=\frac{1}{2}(D-4+2l)\, ,
\end{eqnarray}
$\mathrm{d}r_{\star}=\mathrm{d}r/f(r)$ is the tortoise coordinate, $A_X$ depends on $\alpha$, $D$, $l$ and does not vanish except when $X=K$ and $l=0$. In this case, the next leading term is given by $r_{\star}^{-(2D-4)}$. Therefore, from~\cite{Ching:1994bd,Ching:1995tj}, the Price's law index should be $D-2+2l$ for observers at a finite position in odd dimensions of spacetime. Following~\cite{OBoyle:2022yhp,DaSilva:2023xif}, we introduce an effective power-law index $\Gamma$ for an observer located at $\sigma_0$:
\begin{eqnarray}\label{effective_power_law_index}
    \Gamma(\tau,\sigma_0)=\left\lvert\frac{\tau\Pi(\tau,\sigma_0)}{\Psi(\tau,\sigma_0)}\right\rvert\, ,
\end{eqnarray}
which should tend to the true Price's law index at late times, i.e., $\tau\to\infty$. Fig. \ref{fig:pricelaw} presents the computed effective power-law index $\Gamma$. The initial condition is chosen to be Eqs. (\ref{initial condition}) with $a_0=1$, $b_0=1/(10\sqrt{10})$, and $c_0=1/5$. The observation position $\sigma_0$ is approximately $0.118$ for the left column and is $0$ for the right column. The computation is performed on a grid with resolution $N=300$ and the time step $\Delta \tau=0.075$. After an initial period of strong oscillation, the index $\Gamma$ stabilizes to a constant, indicating the presence of a power-law tail and marking the transition from the damping stage to the power-law tail stage. For a finite-distant observer, $\Gamma$ converges to $1$, $3$, $5$, and $7$ for $l=0$, $1$, $2$, and $3$, respectively, in the case of the KG case. Similarly, it converges to $5$, $7$, $9$, and $11$ for $l=2$, $3$, $4$, and $5$, respectively, in the case of gravitational perturbations. For observers located at infinity, the index $\Gamma$ stabilizes at approximately half of the corresponding values observed by the finite observer. Rigorous proofs using analytical methods need using analytical method need to be considered further consideration, and we leave it for the future. These numerical results for finite-distance observers show excellent convergence and agree with the analytical prediction $D-2+2l$ in odd dimensions, which does not depend on $\alpha$ and is the same as in GR. This behavior is a feature of odd-dimensional spacetimes and does not depend on the presence of a black hole~\cite{Abdalla:2005hu,cardoso:2003jf}. This demonstrates the long-time accuracy of our numerical calculations. For even dimensions of spacetime, we leave the results in Appendix \ref{sec: evenprice}.

\section{The stability analysis}\label{stability_analysis}
\begin{figure}[htbp]
    \subfigure[]{\includegraphics[width=0.45\textwidth]{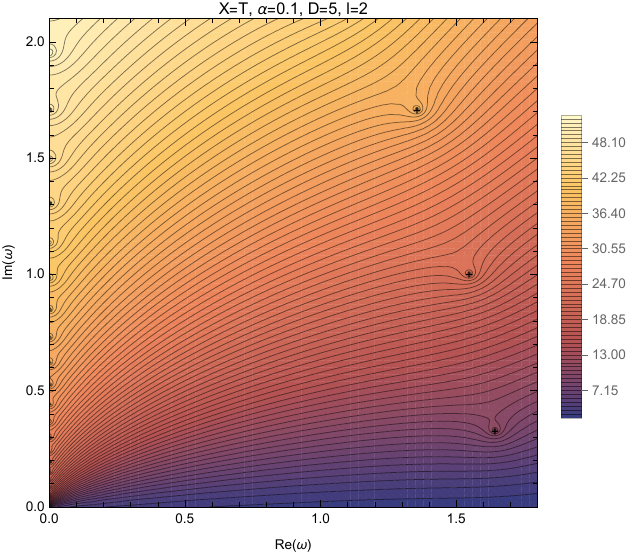}\label{fig:pseudoT1}}
    \subfigure[]{\includegraphics[width=0.45\textwidth]{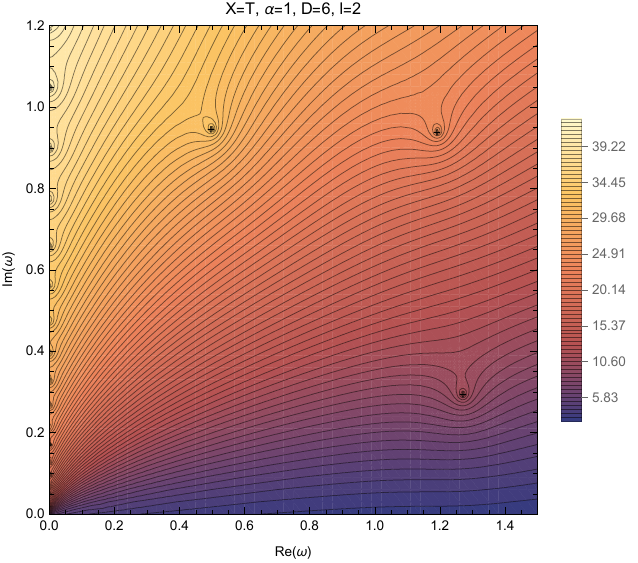}\label{fig:pseudoT20}}
    \\
    \subfigure[]{\includegraphics[width=0.45\textwidth]{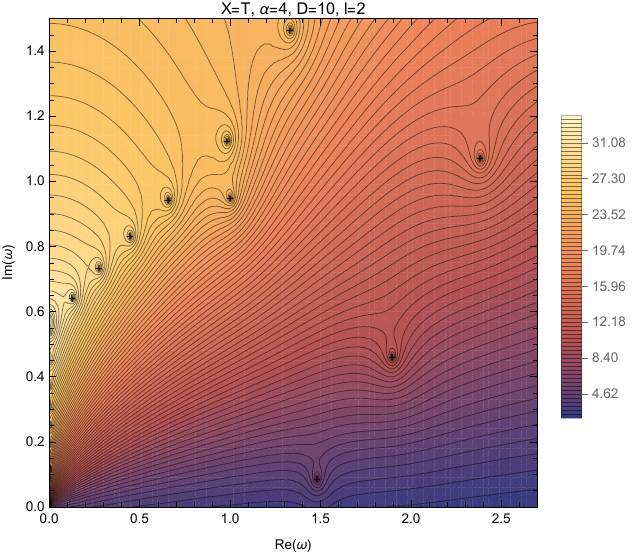}\label{fig:pseudoT3}}
    \caption{The contour plots of $-\ln(1/\lVert(\mathbf{L}/i-\omega\mathbf{I})^{-1}\rVert_{\text{E}})$ for the stable spacetime parameters in Fig. \ref{fig:qnmfig}. QNM spectra calculated using the same grid points and parameters are shown in each panel by a black plus sign. The calculation is performed on the resolution with $N=100$.}
    \label{fig:pseudo}
\end{figure}

In this section, we will perform the stability analysis from the perspective of both frequency and time domains. Due to their non-conservative nature, black hole systems are described by non-Hermitian operators. Unlike Hermitian operators, whose spectra fully characterize their properties, the spectra of non-Hermitian operators provide only partial information and are susceptible to potential instabilities~\cite{trefethen2020spectra,Ashida:2020dkc}. Non-Hermitian operators are characterized by pseudospectrum, which also captures the instability of their eigenvalues. Given $\epsilon>0$, and a norm $\lVert\cdot\rVert$, the $\epsilon$-pseudospectrum for an operator $A$ is defined as~\cite{trefethen2020spectra}
\begin{eqnarray}\label{def of pseudo 1}
    \sigma_\epsilon(A)=\{\omega\in\mathbb{C}:\lVert(A-\omega I)^{-1}\rVert>\epsilon^{-1}\}\,,
\end{eqnarray}
where $I$ is the identity operator, $R_{A}(\omega)=(A-\omega I)^{-1}$ is called the resolvent operator. This definition is the most suitable for computing the pseudospectrum. Another equivalent and more intuitive definition of the pseudospectrum is~\cite{trefethen2020spectra}
\begin{eqnarray}\label{def of pseudo 2}
    \sigma_{\epsilon}(A) =\{\omega\in\mathbb{C}: \exists\delta A \;\text{with}\; \lVert\delta A\rVert<\epsilon \;\text{such that}\; \omega\in\sigma(A+\delta A)\}\, .
\end{eqnarray}
In the limit $\epsilon\to0$, the set $\sigma_\epsilon(A)$ reduces to the spectrum set $\sigma(A)$, whose elements are the spectrum $\omega_n$. The quantity $\epsilon$ serves as a measure of the ``proximity" between points in $\sigma_\epsilon(A)$ and the spectrum $\omega_n$, offering a clear interpretation of perturbations to the underlying operator. The boundary of $\sigma_\epsilon(A)$ is just the constant-$\epsilon$ contour. The choice of norm strongly affects the shape and structure of the pseudospectrum. Here, we choose the above norm to be the so-called energy norm~\cite{Jaramillo:2020tuu, Gasperin:2021kfv, Besson:2024adi}. For a function $u(\sigma)$, it is defined as
\begin{eqnarray}\label{energy_norm}
    \lVert u\rVert_{\text{E}}=\sqrt{\langle u, u\rangle_{\text{E}}}\, ,
\end{eqnarray}
which is induced by the energy inner product
\begin{eqnarray}\label{EnergyScalarProduct}
    \langle u_1, u_2\rangle_{\text{E}} &=& \left\langle\begin{pmatrix}
        \Psi_1 \\
        \Pi_1
    \end{pmatrix}, \begin{pmatrix}
        \Psi_2 \\
        \Pi_2
    \end{pmatrix}\right\rangle_{\text{E}}=	\frac{1}{2} \int_{0}^{1} \Big(w(\sigma)\bar{\Pi}_1 \Pi_2 + p(\sigma)  \partial_\sigma \bar{\Psi}_1\partial_\sigma\Psi_2 + q_X(\sigma)\bar{\Psi}_1 \Psi_2 \Big)\mathrm{d}\sigma\, ,
\end{eqnarray}
where a bar represents the complex conjugate, and three functions $w(\sigma)$, $p(\sigma)$, $q_X(\sigma)$ are given by Eqs. (\ref{function_p_gamma_w_qx}). The choice of this energy norm is motivated by~\cite{Jaramillo:2020tuu}, where the wave equation has the same form as ours. This form of the wave equation (Eq. (\ref{perturbation eq})) suggests that the wave functions $\Psi$ can be viewed as propagating in an effective $(1+1)$-dimensional Minkowski spacetime with an effective potential $V_X$. Therefore, from this perspective, using the energy-momentum tensor for a complex scalar field in $(1+1)$-dimensional Minkowski spacetime and the timelike Killing field $(\partial_\tau)^a = r_+(\partial_t)^a$, one can obtain the total energy on a spacelike hypersurface thus defining the energy norm. In numerical computations, integrals Eq. (\ref{EnergyScalarProduct}) can be efficiently approximated using the Chebyshev-Lobatto grid and the Clenshaw-Curtis quadrature. This approach involves computing a weighted sum of the function values at the grid points, leading to an inner product that incorporates a Gram matrix $\mathbf{G}$. This implies that Eq. (\ref{EnergyScalarProduct}) is approximated by the following discrete version
\begin{eqnarray}
    \left\langle\begin{pmatrix}
        \Psi_1 \\
        \Pi_1
    \end{pmatrix},
    \begin{pmatrix}
        \Psi_2 \\
        \Pi_2
    \end{pmatrix}\right\rangle_{\text{E}}
    =\begin{pmatrix}
        \mathbf{\Psi}_1^\star & \mathbf{\Pi}_1^\star
    \end{pmatrix} \cdot \mathbf{G} \cdot
    \begin{pmatrix}
        \mathbf{\Psi}_2 \\ \mathbf{\Pi}_2
    \end{pmatrix}\, ,
\end{eqnarray}
where the asterisk denotes the Hermitian conjugate. With the help of the Gram matrix, the energy norm of any operator $A$ is given by
\begin{eqnarray}\label{matrixnorm}
    \lVert A\rVert_{\text{E}}=\lVert \mathbf{W} \cdot\mathbf{A}\cdot\mathbf{W}^{-1}\rVert_{2}\, ,
\end{eqnarray}
where $\mathbf{A}$ is the matrix approximation of the operator $A$, $\mathbf{W}$ is the Cholesky decomposition of $\mathbf{G}$, i.e., $\mathbf{G}=\mathbf{W}^{\star}\cdot \mathbf{W}$ and $\lVert\cdot\rVert_2$ represents the matrix $2$-norm. For any point $\omega$ in the complex plane (excluding the exact QNM spectra), the norm of the corresponding resolvent $\lVert R_{L/i}(\omega)\rVert_{\text{E}}$ can be calculated using Eq. (\ref{matrixnorm}) with $A=R_{L/i}(\omega)$. Given a positive constant $\epsilon$, the $\epsilon$-pseudospectrum of the operator $L/i$, i.e., $\sigma_\epsilon(L/i)$ is defined as the set of all $\omega$ such that $\lVert R_{L/i}(\omega)\rVert_{\text{E}}> \epsilon^{-1}$ [see (\ref{def of pseudo 1})]. In practice, we generate the contour plots using $-\ln(\lVert1/R_{L/i}(\omega)\rVert_{\text{E}})$. Further computational details can be found in~\cite{Jaramillo:2020tuu,Cao:2024oud,Chen:2024mon}. Unlike in general relativity, spacetimes in EGB gravity can possess (linear) unstable modes, resulting in a non-positive definite energy inner product that cannot define a norm. In addition to the above computational challenges, spacetimes exhibiting dynamically unstable modes are considered non-physical, rendering the analysis of QNM spectrum stability meaningless. Fig. \ref{fig:pseudo} presents the pseudospectra corresponding to the parameters chosen in Fig. \ref{fig:potentialfig} except for those parameters where dynamically unstable modes exist. Although the parameter $\{\alpha=1, D=6\}$ corresponds to an unstable black hole, there is no dynamically unstable mode presented for $l=2$. So the pseudospectra for $\{X=T, \alpha=1, D=6, l=2\}$ can still be computed. Therefore, there are only $11$ panels in Fig. \ref{fig:pseudo}. The calculation is performed at resolution $N=100$. The QNM spectra are computed with the same resolution and they are represented by a black plus sign in each panel.

It can be read from Fig. \ref{fig:pseudo}, when $\omega$ is not in close proximity to a QNM spectrum, $\lVert R_{L/i}(\omega)\rVert_{\text{E}}$ increases with the increasing of the imaginary part of $\omega$. However, for those $\omega$ in the vicinity of a QNM spectrum, $\lVert R_{L/i}(\omega)\rVert_{\text{E}}$ increases rapidly as $\omega$ approaches the QNM spectrum. A perturbation with norm $\epsilon$ displaces the QNM spectra from their original positions, with the $\epsilon$-pseudospectrum providing a bound on their displacement [see (\ref{def of pseudo 2})]. For small $\epsilon$, the constant-$\epsilon$ contour line encloses the QNM spectrum, forming a closed curve. However, for slightly larger $\epsilon$, the constant-$\epsilon$ contour line extends well, forming an open structure that can even reach infinity. It suggests that perturbations of this magnitude may cause significant migration of the QNM spectrum. For each QNM spectrum, there exists a critical value $\epsilon_c$ that separates closed contour lines from the open contour lines in the pseudospectrum. For perturbations with $\epsilon<\epsilon_c$, the QNM spectrum remains stable, while for $\epsilon>\epsilon_c$, the QNM spectrum can become highly unstable. A QNM spectrum's stability can be characterized by its critical value $\epsilon_c$, with larger values indicating greater stability. It can be seen that $\epsilon_c$ decreases as the overtone number $n$ increases, which demonstrates that higher overtones exhibit more spectrum instability.

\begin{figure}[htbp]
    \subfigure[]{\includegraphics[width=0.45\textwidth]{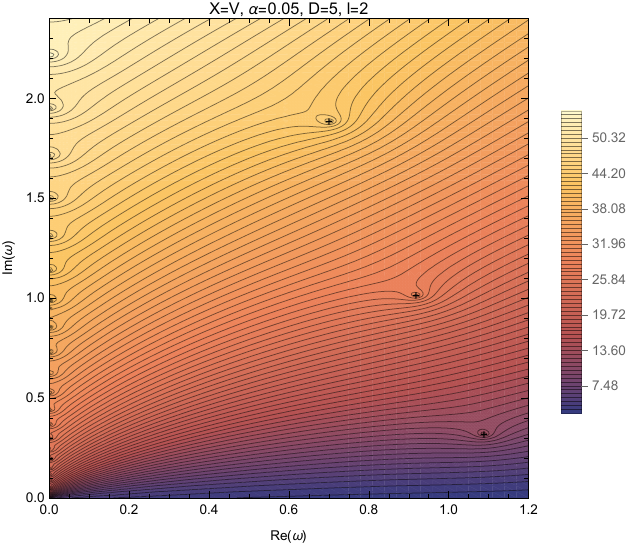}\label{fig:pseudoV1}}
    \subfigure[]{\includegraphics[width=0.45\textwidth]{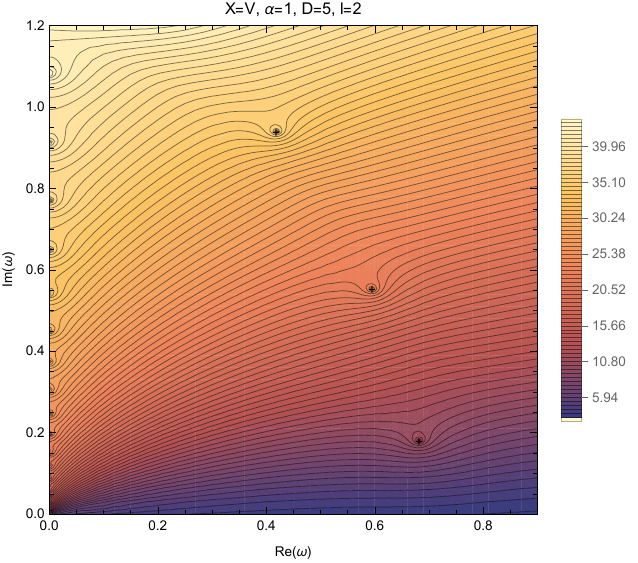}\label{fig:pseudoV2}}
    \\
    \subfigure[]{\includegraphics[width=0.45\textwidth]{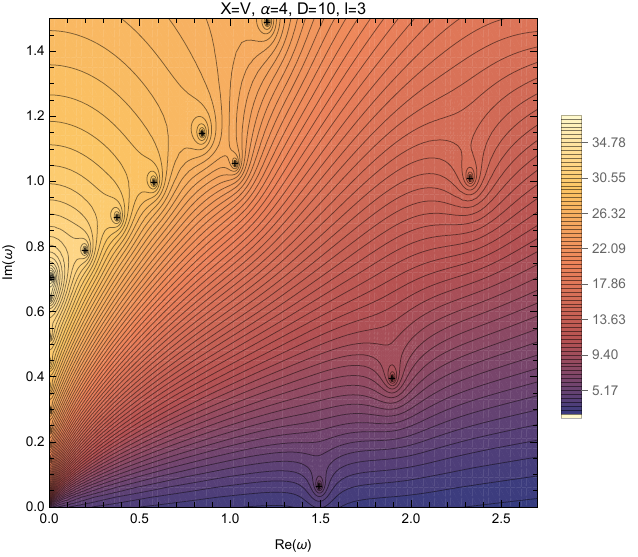}\label{fig:pseudoV3}}
    \subfigure[]{\includegraphics[width=0.45\textwidth]{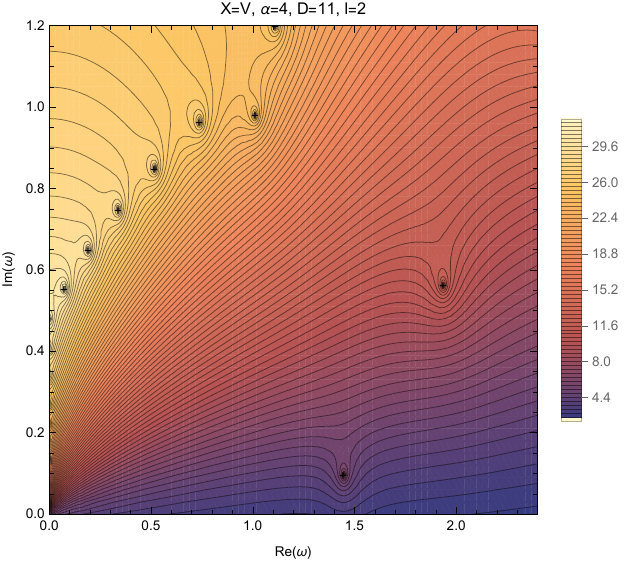}\label{fig:pseudoV4}}
\end{figure}

In addition, in Figs. \ref{fig:pseudoV3}, \ref{fig:pseudoV4}, and \ref{fig:pseudoT3}, we find each panel contains a branch of discrete spurious QNM spectra located near the imaginary axis. These QNM spectra, identified as spurious using the method described in Appendix \ref{sec: drift}, originate from the branch cut of the Green's function along the positive imaginary semi-axis, which causes the discrete matrix to capture a set of QNM spectra along this semi-axis. At low resolution, these QNM spectra detach from the axis, forming the observed spurious branches. As resolution increases, these spurious QNM spectra gradually return to the imaginary axis. This behavior is a characteristic feature of asymptotically flat spacetimes, and our model is no exception. In fact, the process of discreteness can also be regarded as perturbations to the original analytical system. Increasing the resolution essentially reduces such perturbation. Therefore, this reduction in the perturbation causes spurious modes to progressively return to the imaginary axis, while simultaneously enhancing the convergence of the computed QNM spectra toward the true physical QNM spectra. More importantly, it still remains an open question whether the norm of the resolvent of the matrix operator $\mathbf{L}$ obtained from the hyperboloidal framework in asymptotically flat spacetime converges to the true norm of the resolvent of the original operator $L$ as the resolution increases. Nonetheless, while the quantitative calculations of the resolvent may be under suspicion, the qualitative insights derived from these calculations remain valid~\cite{Boyanov:2023qqf}.

\begin{figure}[htbp]
    \centering
    \subfigure[]{\includegraphics[width=0.45\textwidth]{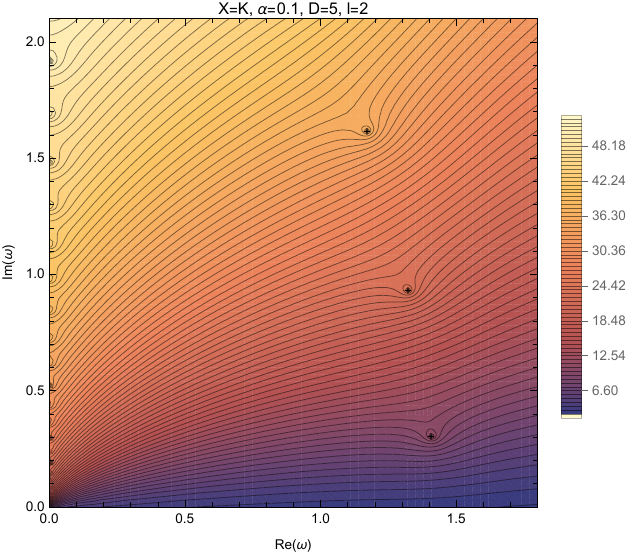}\label{fig:pseudoK}}
    \subfigure[]{\includegraphics[width=0.45\textwidth]{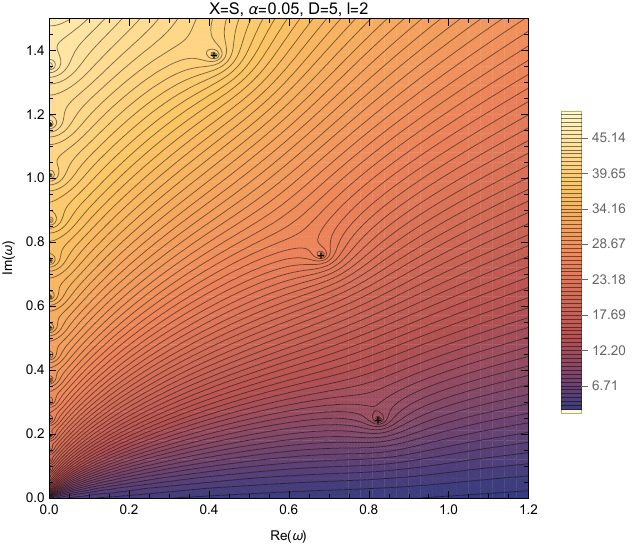}\label{fig:pseudoS1}}
    \\
    \subfigure[]{\includegraphics[width=0.45\textwidth]{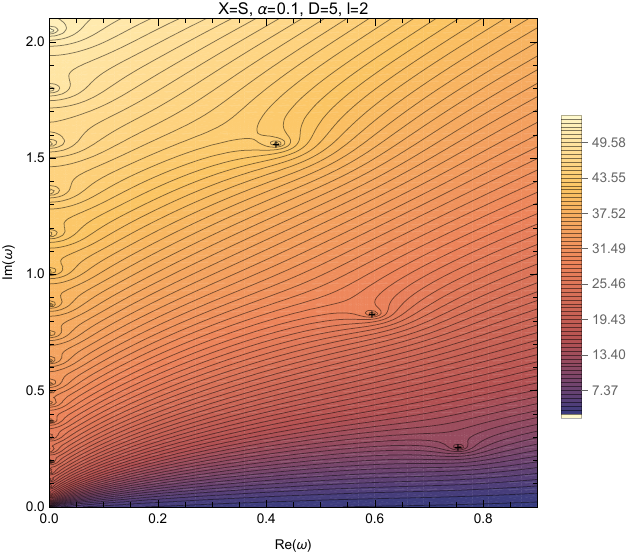}\label{fig:pseudoS20}}
    \subfigure[]{\includegraphics[width=0.45\textwidth]{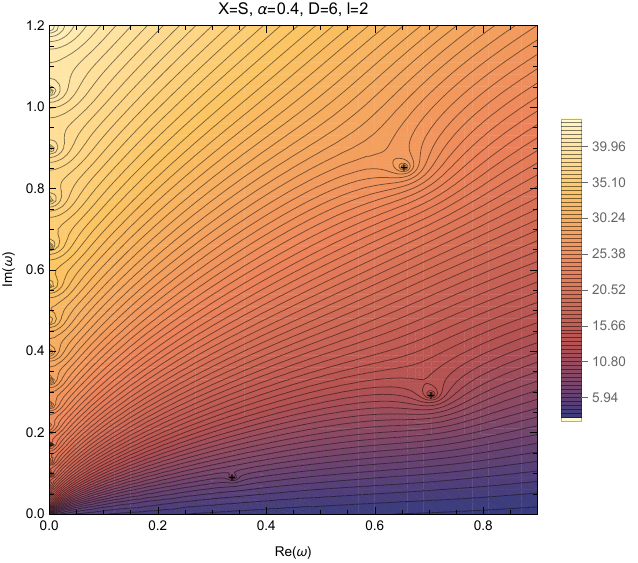}\label{fig:pseudoS3}}
\end{figure}

As shown by the second definition of the pseudospectrum (\ref{def of pseudo 2}), considering the pseudospectra of $\mathbf{L}/i$ is equivalent to considering any perturbation on the matrix, including those who could unphysically disturb the differential part of the operator $L$. True physical perturbations should be confined to alterations in the effective potential, potentially induced by surrounding matter in the environment. The preceding discussion focuses on the spectrum stability in the frequency-domain. We now shift our attention to stability analysis of QNMs in the time domain. In the following, we investigate the effects of specific perturbations on the effective potential in the time domain. Considering the potential impact of the surrounding environment on the black hole's effective potential, we introduce a specific bump in the potential to model this effect
\begin{equation} \label{bump}
    \delta V(r(\sigma)) = a \left(1 - \frac{r_{+}}{r}\right)^2 \exp\left[-\frac{\left(r/r_{+} - 1/c\right)^2}{2b^2}\right] = a (1 - \sigma)^2 \exp\left[-\frac{\left(1/\sigma - 1/c\right)^2}{2b^2}\right],
\end{equation}
where $a$ is the amplitude of $\delta V$, $c$ approximately represents the center of the bump in the $\sigma$ coordinate, as $r_{+}/c$ approximately gives the center of the bump in the $r$ coordinate. The width of the bump in the $r$ coordinate is characterized by $b r_{+}$. Utilizing the relationship $\sigma = r_{+}/r$, we can estimate the width of the bump in the $\sigma$ coordinate $\Delta \tau$ as
\begin{equation}\label{width}
    \Delta \sigma = \frac{\Delta \sigma}{\Delta r}\Delta r =  \left\lvert\left.\frac{\mathrm{d}\sigma}{\mathrm{d}r}\right|_{r = \xi}\right\rvert \Delta r = \frac{r_{+}}{\xi^2} b r_{+} \approx c^2 b\, ,
\end{equation}
where in the second step, the Lagrange's mean value theorem guarantees the existence of $\xi$, and in the final step, we approximate $\xi \approx r_{+}/c$. Note that $a$, $b$, $c$ should not be confused with the parameters $a_0$, $b_0$, $c_0$ that characterize the initial waveform in Eqs. (\ref{initial condition}). The factor $\left(1 - r_{+}/r\right)^2$ is included to prevent the disturbances near the horizon, while simultaneously maintaining the asymptotic behavior of the effective potential near the horizon. To quantify the time-domain impact of this perturbation, we calculate the mismatch between the time-domain signals generated by unperturbed and perturbed potentials observed at infinity ($\sigma_0=0$)~\cite{Spieksma:2024voy}, which can be easily achieved for the hyperboloidal framework. The mismatch between two time-domain signals, $h_1(\tau)$ and $h_2(\tau)$, is defined as~\cite{Spieksma:2024voy}
\begin{eqnarray}\label{mismatch_function}
    \mathcal{M}[h_1(\tau),h_2(\tau)]=1-\max_{\delta \tau}\frac{\langle h_1(\tau), h_2(\tau+\delta\tau)\rangle}{\sqrt{\langle h_1(\tau), h_1(\tau)\rangle\langle h_2(\tau+\delta\tau), h_2(\tau+\delta\tau)\rangle}}\, ,
\end{eqnarray}
where we allow the waveform $h_2$ to go through a time shift $\delta \tau$, and then maximize the inner product between $h_1$ and shifted $h_2$ over $\delta \tau$, the inner product is defined as
\begin{eqnarray}\label{inner_product_h1_h2}
    \langle h_{1}, h_{2}\rangle=4 \operatorname{Re} \int_{0}^{\infty} \mathcal{F}\{h_{1}\}(\omega) \overline{\mathcal{F}\{h_{2}\}(\omega)}\mathrm{d} \omega\, ,
\end{eqnarray}
$\mathcal{F}\{\cdot\}$ indicates Fourier transformation and overbar denotes taking the complex conjugation. The shift $\delta \tau$ in time is achieved by utilizing the time shift property of the Fourier transformation,
\begin{eqnarray}\label{Fourier_transformation}
    \mathcal{F}\{h_2(\tau + \delta\tau)\}(\omega) = e^{i\omega\delta\tau} \mathcal{F}\{h_2(\tau)\}(\omega)\, .
\end{eqnarray}
$h_1$ and $h_2$ are selected as the time-domain waveforms corresponding to the effective potentials before and after adding the bump seen by observers located at infinity. In numerical calculations, we work with values defined only at discrete time grid points. Consequently, we use the discrete Fourier transform (DFT) as an approximation of the true Fourier transform, and the vector inner product as an approximation of the integral. The same initial conditions are adopted to generate the signals. Waveforms are aligned in time such that their peaks coincide. After this alignment, it is observed that $\delta \tau$ remains consistently small.

\begin{figure}[htbp]
    \centering
    \subfigure[]{\includegraphics[width=0.48\linewidth]{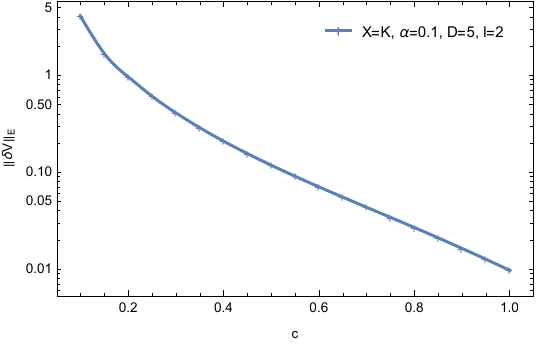}\label{normc0}}  \hfill
    \subfigure[]{\includegraphics[width=0.48\linewidth]{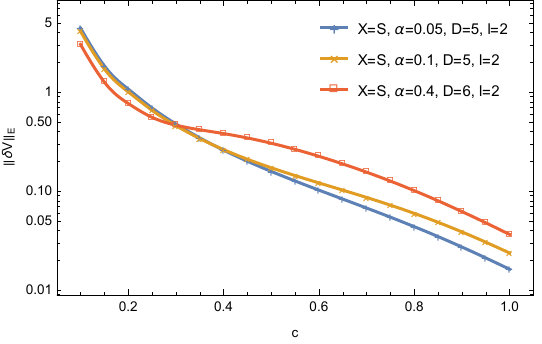}\label{normc1}}  \\
    \subfigure[]{\includegraphics[width=0.48\linewidth]{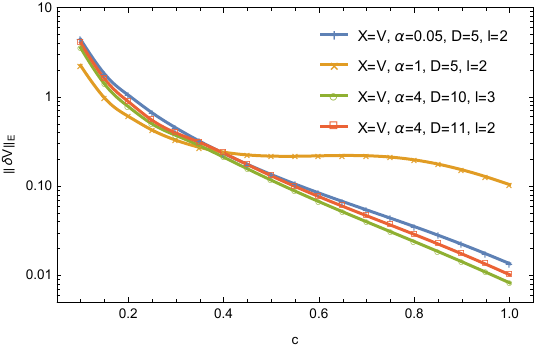}\label{normc2}}  \hfill
    \subfigure[]{\includegraphics[width=0.48\linewidth]{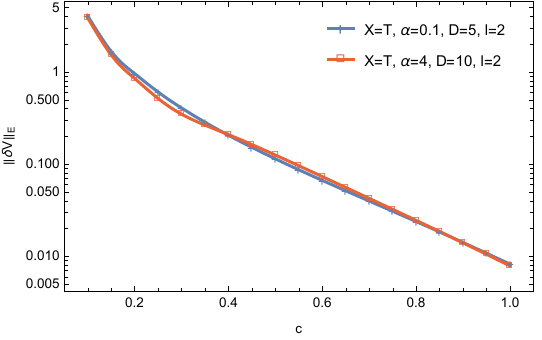}\label{normc3}}
    \caption{The norm for bump $\delta V$ in Eq. (\ref{bump}) with different $c$, $b$ is fixed as $1/5$ and $a$ is fixed as $1$. The computation is performed on resolution $N=200$. Lines of different colors represent different types of perturbations with their own parameters chosen from Fig. \ref{fig:potentialfig}, where those where parameters corresponding to dynamically unstable conditions are excluded.}
    \label{fig:norm and amp}
\end{figure}

The norm rather than simply the amplitude $a$ should be used to quantify the magnitude of the bump properly~\cite{Cheung:2021bol, Boyanov:2024fgc}. The bump's norm is defined using the matrix norm from Eq. \eqref{matrixnorm}, specifically as the norm of the difference between the matrices $L$ from Eq. \eqref{maineq} constructed with and without the bump, i.e.,
\begin{eqnarray}
    \lVert\delta V\rVert_{\text{E}}:=\left\lVert
    \frac{1}{p(\sigma)w(\sigma)}\begin{pmatrix}
        0                & 0 \\
        \delta V(\sigma) & 0
    \end{pmatrix}\right\rVert_{\text{E}}\, ,
\end{eqnarray}
where $\delta V(\sigma)$ is given by Eq. (\ref{bump}) and its matrix representation is a diagonal matrix. Fig. \ref{fig:norm and amp} illustrates that the bump's norm can vary dramatically with respect to $c$ despite a fixed amplitude $a=1$, in which $b$ is fixed as $1/5$ and the parameters are chosen from Fig. \ref{fig:potentialfig}. The norm satisfies $\lVert a\delta V\rVert_{\text{E}}=\lvert a\rvert\lVert\delta V\rVert_{\text{E}}$, so we simply fix $a=1$ in the Fig. \ref{fig:norm and amp}.
Unlike the amplitude, the relationship between the position $c$ of the bump and the norm is not simply linear, when other parameters are held constant. It is worth mentioning that for the case with $\{X=V, \alpha=1, D=5, l=2\}$, the norm for bump $\delta V$ does not change monotonically with $c$. As $c$ approaches zero, the bump's position shifts further away from the black hole, resulting in its norm diverges which is in agreement with~\cite{Boyanov:2024fgc}. Simultaneously, the bump becomes increasingly narrow in the $\sigma$ coordinate as shown in (\ref{width}), making it more challenging for numerical discrete grids to capture its features accurately. The norms of $\delta V$ are carried out up to $c=0.1$. A grid resolution of $N = 200$ is found to be sufficient, as increasing $N$ does not notably affect the obtained norm for bumps with $c>0.1$. This indicates that the current grid resolution adequately captures the characteristics of the bumps, and the subsequent calculations will maintain this resolution. While it is possible to extend the computation of $\lVert\delta V\rVert_{\text{E}}$ to $c<0.1$, doing so would require a higher resolution than we currently used.

\begin{figure}[htbp]
    \subfigure[]{\includegraphics[width=0.48\linewidth]{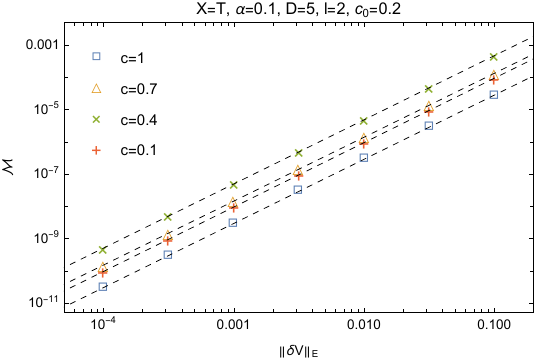}}  \hfill
    \subfigure[]{\includegraphics[width=0.48\linewidth]{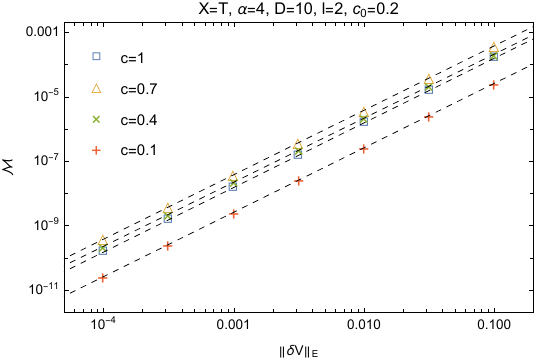}}
    \caption{
        The relationship between mismatch $\mathcal{M}$ and $\lVert\delta V\rVert_{\text{E}}$ for varying values of $c$ of the bumps. The widths $b$ of the bumps are fixed as $1/5$. The initial wave is chosen to be Eqs. (\ref{initial condition}) with $a_0=1, b_0=1/(10\sqrt{10})$ and $c_0=1/5$. The dashed lines represent the fitted lines and all of them exhibit a slope of approximately $2$. The computations are performed on the resolution $N=200$ using time step $\Delta \tau=0.075$.}
    \label{fig:Mnorm}
\end{figure}

Fig. \ref{fig:Mnorm} shows the relationship between the mismatch $\mathcal{M}$ and the norm of the perturbation, $\lVert\delta V\rVert_{\text{E}}$, for $c=1/10$, $4/10$, $7/10$, $1$. The width of the bump, which is determined by the parameter $b$, is fixed as $1/5$. The initial wave is Eqs. (\ref{initial condition}) with $a_0=1, b_0=1/(10\sqrt{10})$ and $c_0=1/5$. For fixed $b$ and $c$, the norm $\lVert\delta V\rVert_{\text{E}}$ is proportional to the amplitude $a$. To determine the amplitude $a$ corresponding to a desired norm, we first compute $\lVert\delta V\rVert_{\text{E}}$ for $a=1$. Finally, using the proportionality between the norm and ampulitude, we determine the value of $a$ for the given desired norm $\lVert\delta V\rVert_{\text{E}}$ by $a=\lVert\delta V\rVert_{\text{E}}/\lVert\delta V(a=1)\rVert_{\text{E}}$. In the case where $a$, $b$, and $c$ are known, one can use Eq. (\ref{mismatch_function}) to determine the mismatch $\mathcal{M}$.  For fixed $c$, the linear fit (dotted line) with a slope of approximately $2$ indicates a quadratic relationship, $\mathcal{M}\propto\lVert\delta V\rVert_{\text{E}}^2$, where log-log plots are used. Since the mismatch cannot exceed $1$, this quadratic relationship only holds when the norm of the perturbation is small enough. Since the norm $\lVert\delta V\rVert_{\text{E}}$ is proportional to amplitude $a$ for fixed $b$ and $c$, our results are consistent with~\cite{Spieksma:2024voy}, where they find that the mismatch scales quadratically with the amplitude (and consequently the norm) of the deformation on the effective potential. Despite our effective potentials and the shape of deformation differs significantly with theirs, the quadratic relationship remains remarkably intact. This raises the questions of the underlying mechanism for the quadratic dependence and the generality of this relationship for other deformation profiles. Nevertheless, this demonstrates the stability of the time-domain waveform: small perturbations in the effective potential lead to proportionally smaller mismatches, implying a smooth, rather than abrupt, change in the waveforms.
\begin{figure}[htbp]
    \centering
    \subfigure[]{\includegraphics[width=0.48\linewidth]{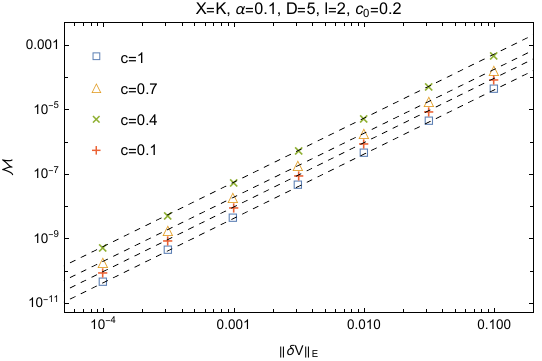}}  \hfill
    \subfigure[]{\includegraphics[width=0.48\linewidth]{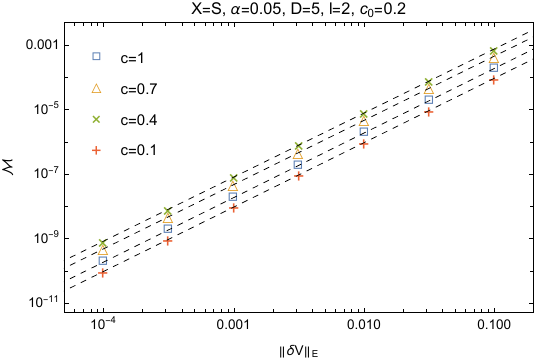}}  \\
    \subfigure[]{\includegraphics[width=0.48\linewidth]{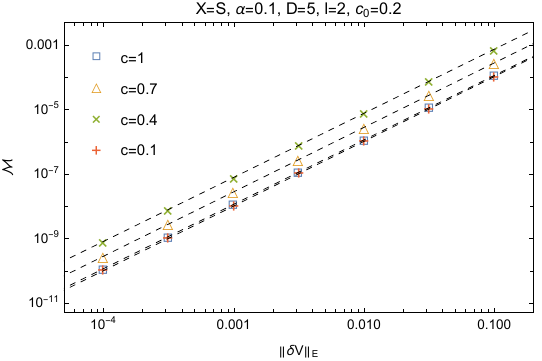}}  \hfill
    \subfigure[]{\includegraphics[width=0.48\linewidth]{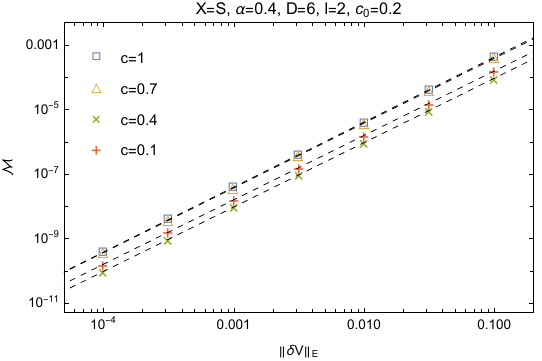}}  \\
    \subfigure[]{\includegraphics[width=0.48\linewidth]{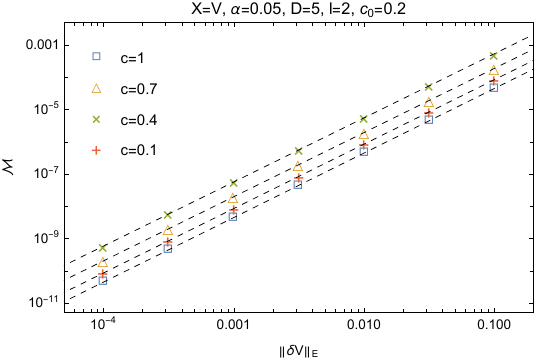}}  \hfill
    \subfigure[]{\includegraphics[width=0.48\linewidth]{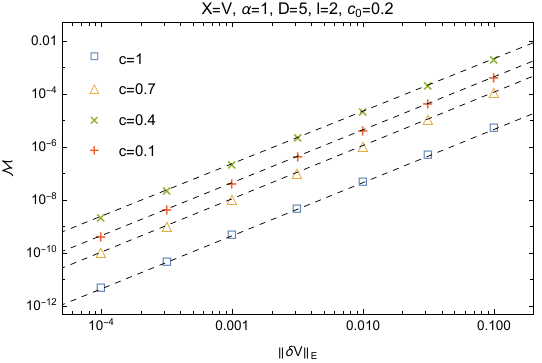}}  \\
    \subfigure[]{\includegraphics[width=0.48\linewidth]{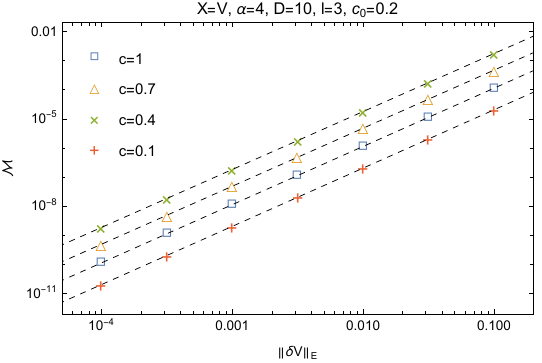}}  \hfill
    \subfigure[]{\includegraphics[width=0.48\linewidth]{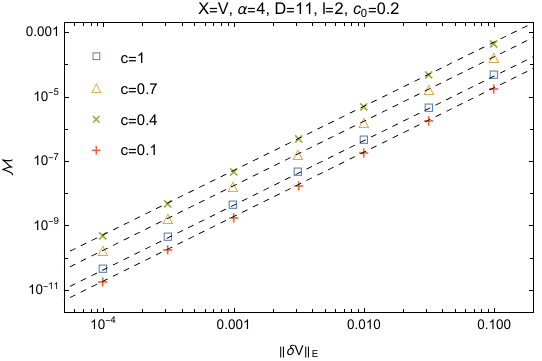}}
\end{figure}

\begin{figure}[htbp]
    \centering
    \subfigure[]{\includegraphics[width=0.48\linewidth]{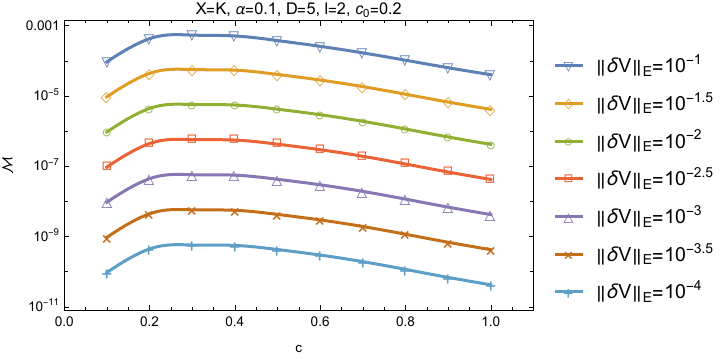}} \hfill
    \subfigure[]{\includegraphics[width=0.48\linewidth]{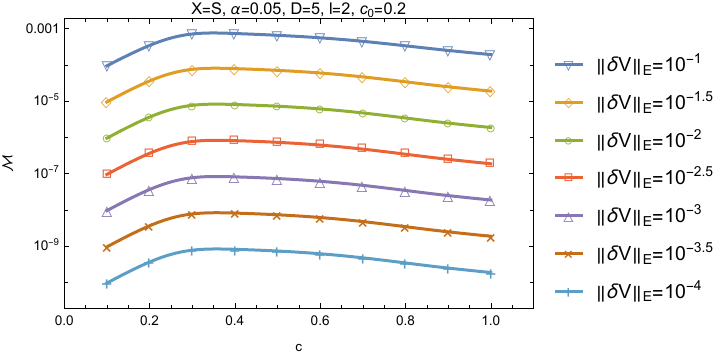}} \\
    \subfigure[]{\includegraphics[width=0.48\linewidth]{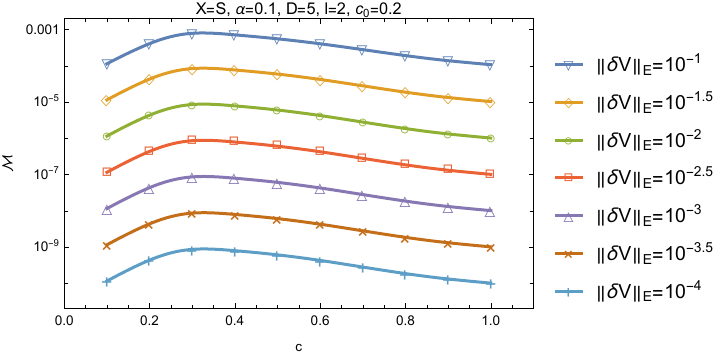}} \hfill
    \subfigure[]{\includegraphics[width=0.48\linewidth]{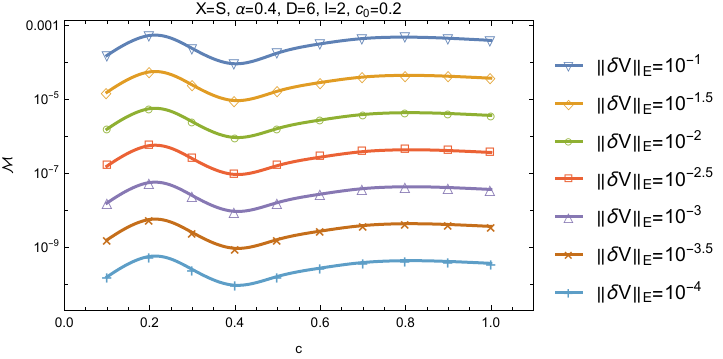}} \\
    \subfigure[]{\includegraphics[width=0.48\linewidth]{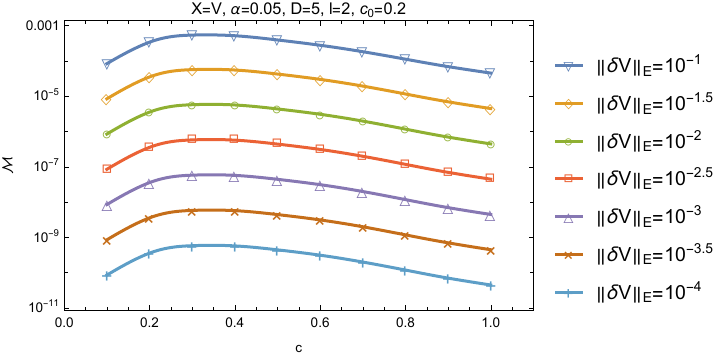}} \hfill
    \subfigure[]{\includegraphics[width=0.48\linewidth]{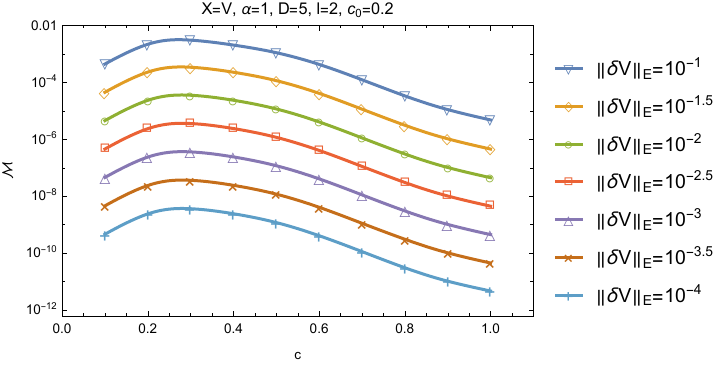}} \\
    \subfigure[]{\includegraphics[width=0.48\linewidth]{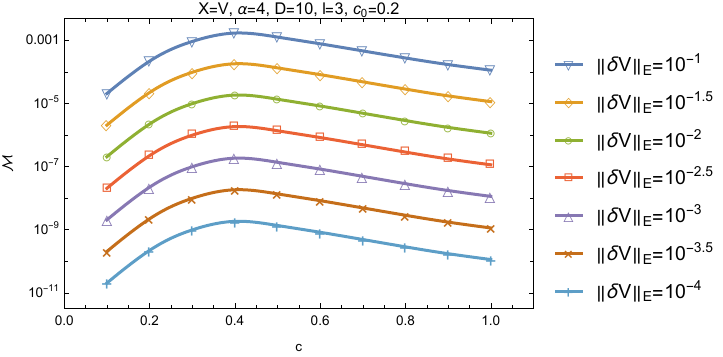}} \hfill
    \subfigure[]{\includegraphics[width=0.48\linewidth]{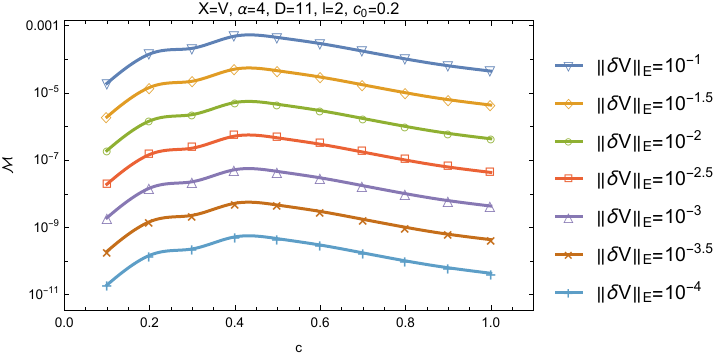}} \\
    \subfigure[]{\includegraphics[width=0.48\linewidth]{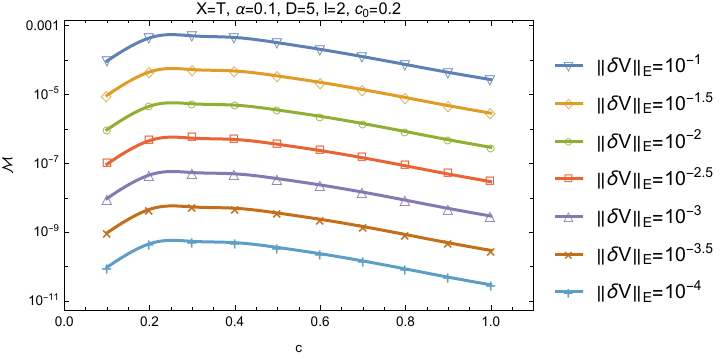}} \hfill
    \subfigure[]{\includegraphics[width=0.48\linewidth]{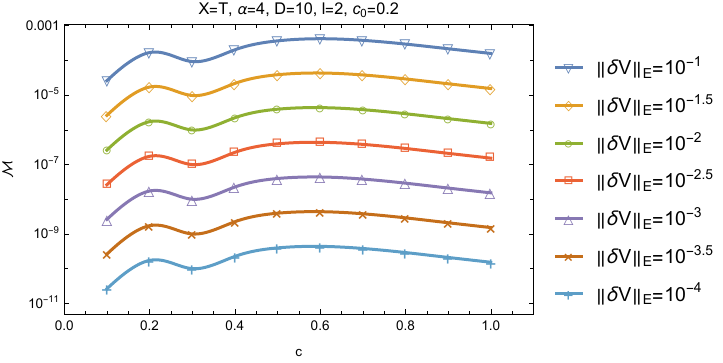}}
    \caption{The relationship between mismatch $\mathcal{M}$ and the location of the bump, which is described by the parameter $c$, for varying values of the norm $\lVert\delta V\rVert_{\text{E}}$. The width of the bump $b$ is fixed as $1/5$. The initial wave is specified as Eqs. (\ref{initial condition}) with $a_0=1, b_0=1/(10\sqrt{10})$ and $c_0=1/5$. The computations are performed on the resolution $N=200$ using time step $\Delta \tau=0.075$.}
    \label{fig:Mc}
\end{figure}

Having established the link between $\mathcal{M}$ and $\lVert\delta V\rVert_{\text{E}}$, we proceed to decode the relationship between $\mathcal{M}$ and $c$. Fig. \ref{fig:Mc} shows the relationship between the mismatch $\mathcal{M}$ and the location of the perturbation, which is described by the parameter $c$, for given $\lVert\delta V\rVert_{\text{E}}=10^{-4}, 10^{-3.5}, 10^{-3}, 10^{-2.5}, 10^{-2}, 10^{-1.5}, 10^{-1}$. The width of the bump is fixed as $b=1/5$. The initial wave is chosen to be Eqs. (\ref{initial condition}) with $a_0=1, b_0=1/(10\sqrt{10})$ and $c_0=1/5$. The lines connecting points with the same norm are determined through fitting. Fig. \ref{fig:Mc} presents a comprehensive view of the data, which is partially shown in Fig. \ref{fig:Mnorm}. Due to the small magnitude of differences between different sets of the data, plotting all data points in Fig. \ref{fig:Mnorm} results in significant overlap. For a fixed $\lVert\delta V\rVert_{\text{E}}$, it can be seen that the relationship between $c$ and the mismatch $\mathcal{M}$ is not monotonic as shown in Fig. \ref{fig:Mc}. There are one or two maximum points among the mismatch $\mathcal{M}$ for different perturbations. The variation range of the mismatch function caused by changes in $c$ does not exceed approximately $3$ orders of magnitude. For the sake of eliminating the impact of initial conditions on the conclusion, we present Fig. \ref{fig:Mnorm1} and Fig. \ref{fig:Mc1} with a different initial condition in Appendix \ref{sec: anothermismatch}, where $c_0 = 1/5$ has been replaced by $c_0 = 4/5$. Therein, the linear fits in Fig. \ref{fig:Mnorm1} still exhibit slopes of approximately $2$, consistent with Fig. \ref{fig:Mnorm}, demonstrating that the time-domain stability is independent of the initial conditions.

\section{Conclusions and discussion}\label{sec: conclusions}
In this paper, we explore the QNMs and their (in)stability of the BDW black holes within the framework of EGB gravity theory in both frequency domain and time domain using the hyperboloidal formalism. The effective potential of gravitational perturbations in BDW black holes exhibits a variety of shapes depending on the parameters. These potentials can display single or double peaks and may also feature negative regions, either adjacent to or distant from the event horizon. Such variations in the effective potential are inherent to the theoretical framework and are not the result of artificial perturbations or external conditions. Additional perturbations could lead to even more complex configurations. We calculated the QNMs for these typical effective potentials in both the frequency and time domains, identifying characteristic QNMs such as echoes and instabilities. Spurious modes, arising from numerical artifacts, are eliminated through joint computation at multiple resolutions. We further validated our code by computing the Price's law. Our results align with the expected behavior in odd spacetime dimensions, but we found limited evidence supporting the Price's law in even dimensions, similar to the findings of Abdalla \textit{et al.}~\cite{Abdalla:2005hu}.

The (in)stability of QNMs in frequency-domain is assessed by calculating the pseudospectrum. The $\epsilon$-pseudospectrum provides an upper bound on the migration of QNM spectra induced by a perturbation to the effective potential with a norm of $\epsilon$. However, certain parameter configurations of spacetimes reveal dynamically unstable QNMs, indicated by a negative imaginary part. This instability leads to energy inner products that are not positive definite. As a result, one cannot define an energy norm or calculate the pseudospectrum. From a physical perspective, the stability of QNMs and pseudospectrum in unstable spacetimes is also meaningless. The pseudospectrum accounts for arbitrary perturbations on the time-evolution operator, including those that may be unphysical, e.g., changing the structure of the derivative part. However, even within its limitations, ~\cite{Jaramillo:2020tuu} suggests that high-frequency perturbations on the effective potential can possibly saturate this upper bound given by pseudospectrum. In numerical computations of asymptotically flat spacetimes using hyperboloidal formalism, finite-rank approximations of the resolvent operator $R_{L/i}(\omega)$ may exhibit quantitative inaccuracies but do capture the qualitative features of the pseudospectrum.

The (in)stability of QNMs in the time domain is studied by comparing the waveforms evolved before and after adding a bump on the effective potential. We use the mismatch to quantify the difference between the two waveforms. Results show that for a bump with a small norm on the effective potential, the mismatch remains small and scales quadratically with the bump's norm, which is consistent with~\cite{Spieksma:2024voy}. Although our effective potentials and deformation shapes differ significantly from those in~\cite{Spieksma:2024voy}, such quadratic relationship, i.e., $\mathcal{M}\propto\lVert\delta V\rVert_{\text{E}}^2$ persists. In other words, the time-domain stability is unrelated to the shape of the original effective potential. The underlying mechanism responsible for this quadratic scaling requires further investigation. Nevertheless, it underscores the stability of the time-domain waveform, as minor disturbances in the effective potential result in correspondingly smaller mismatches, suggesting a gradual rather than sudden alteration in the waveform. This time-domain stability is of particular importance for gravitational-wave astronomy, where detectors observe the waveform directly in the time domain. The observed robustness of the black hole ringdown signal under small environmental perturbations enhances confidence in its use for probing fundamental properties of black holes, such as in tests of the no-hair theorem.

The apparent contradiction between the frequency-domain instabilities and time-domain stability highlights limitations in our current understanding of black hole spectral instabilities. Since the time-domain waveform is not a simple linear superposition of quasinormal modes, the resolution of this contradiction may lie in the analytical structure of the Green's function, which remains partially understood. Our hypothesis is that QNMs migrate in a specific pattern, and the waveform changes caused by different QNMs tend to cancel each other out, resulting in little net change in the final waveform. Resolving this discrepancy between frequency-domain instability and time-domain stability requires further analytical and numerical investigation.


\section*{Acknowledgement}
This work is supported in part by the National Key R\&D Program of China Grant No.2022YFC2204603, by the National Natural Science Foundation of China with grants No.12475063, No.12075232 and No.12247103. This work is also supported by the National Natural Science Foundation of China with grants No.12235019 and No.11821505.

\appendix
\section{Criteria for identifying physically relevant QNMs}
\label{sec: drift}
\begin{figure}[htbp]
    \subfigure[]{\includegraphics[width=0.44\textwidth]{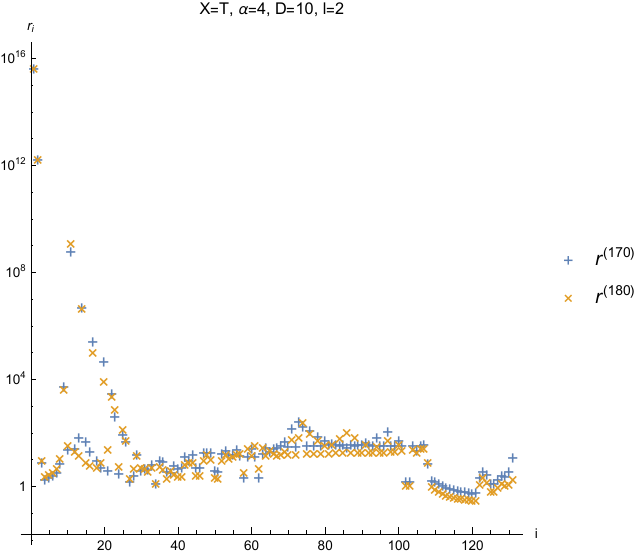}}\hfill
    \subfigure[]{\includegraphics[width=0.44\textwidth]{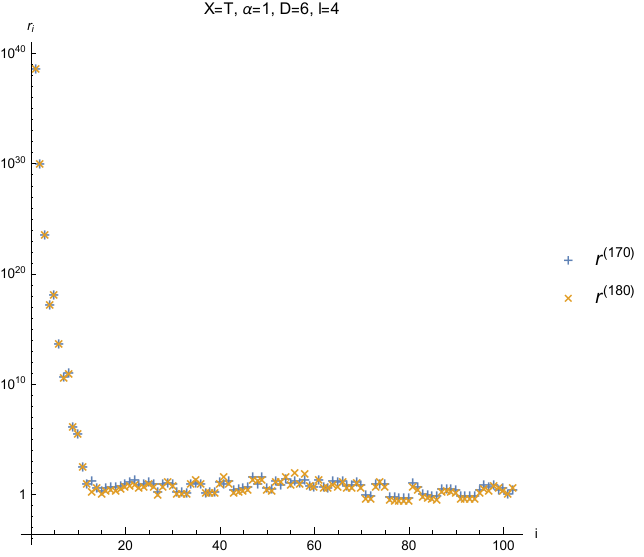}}
    \caption{The drift ratios $r_i$ corresponding to QNM spectra discussed previously, where $i$ is the mode index. The calculations for all panels are performed on the resolutions with $N=160$, $N_1=170$, $N_2=180$. The last one stands for the tensor perturbation with $\alpha=1$, $D=6$, $l=4$, where the corresponding effective potential does not appear in Fig. \ref{fig:potentialfig}.}
    \label{fig:drift}
\end{figure}

In this appendix, we give a criterion for identifying physically relevant QNMs~\cite{boyd2001chebyshev,Chen:2024mon,BOYD199611,Cownden:2023dam}. Identifying true physical QNMs from contaminated ones requires careful inspection. A common strategy is to examine whether the calculated modes converge to a fixed point as the grid resolution increases, or alternatively, if they align with previously computed modes. For a given grid resolution $N$ associated with eigenvalues $\omega_i, i=1,2,\cdots, n$, sorted in ascending order by their imaginary parts, we define a drift ratio relative to another higher resolution $N_1$,
\begin{eqnarray}
    r_i^{(N_1)}=\frac{\min\{\lvert\omega_i\rvert,\sigma_i\}}{\delta_i^{(N_1)}}\, ,
\end{eqnarray}
where the intermodal separation $\sigma_{i}$ is
\begin{eqnarray}
    \sigma_{i}&=&
    \begin{cases}
        \lvert\omega_{1}-\omega_{2}\rvert,                                                            & i=1\, ,   \\
        \frac{1}{2}\big(\lvert\omega_{i}-\omega_{i-1}\rvert+\lvert\omega_{i+1}-\omega_{i}\rvert\big), & 1<i<n\, , \\
        \lvert\omega_{n-1}-\omega_n\rvert,                                                            & i=n\, ,
    \end{cases}
\end{eqnarray}
the nearest neighbor drift is
\begin{eqnarray}
    \delta_i^{(N_1)}&=&\min_j\lvert\omega_i-\omega_j^{(N_1)}\rvert\, ,
\end{eqnarray}
and $\omega_{i}^{(N_1)}$ are eigenvalues of resolution $N_1$. Therefore, the eigenvalues corresponding to a physical mode will exhibit a significant drift ratio, whereas the numerical spurious eigenvalues rely heavily on resolution and show smaller drift ratios. To avoid potential coincidences between spurious eigenvalues at the resolutions $N$ and $N_1$, we compute $r_i^{(N_1)}$ and $r_i^{(N_2)}$ with $N_2 > N_1 > N$. Only modes $\omega_i$ that satisfy $\min\{r_i^{(N_1)}, r_i^{(N_2)}\}>10^3$ are considered good approximations of true physical QNMs. In asymptotically flat spacetime, the branch cut along the positive imaginary axis leads to a continuous spectrum in this region. Consequently, there are a series of discrete modes located in this region for numerical computation, therefore we excluded modes in this region from the calculation of the drift ratio. The drift ratios associated with the discussed QNM spectra are presented in Fig. \ref{fig:drift}.

\begin{figure}[htbp]
    \centering
    \subfigure[]{\includegraphics[width=0.44\textwidth]{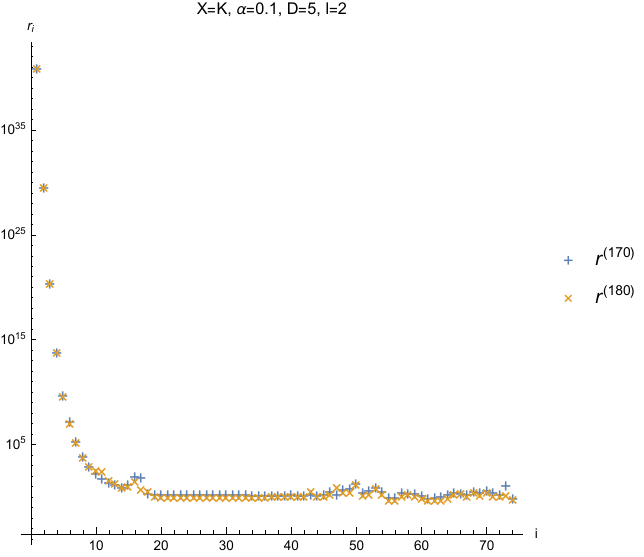}}\hfill
    \subfigure[]{\includegraphics[width=0.44\textwidth]{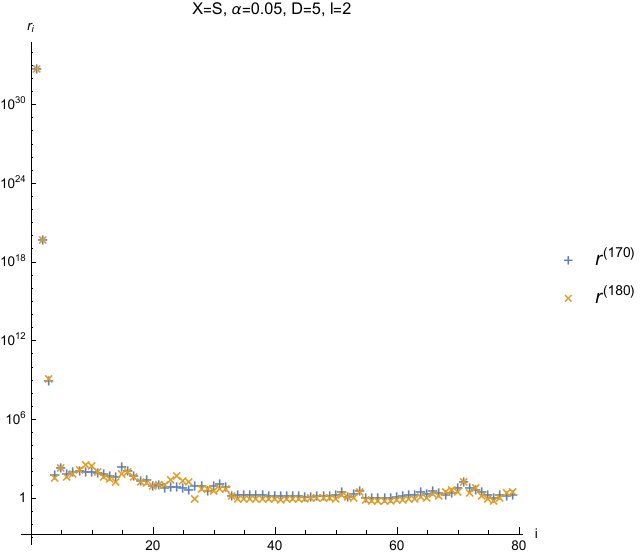}}
    \\
    \subfigure[]{\includegraphics[width=0.44\textwidth]{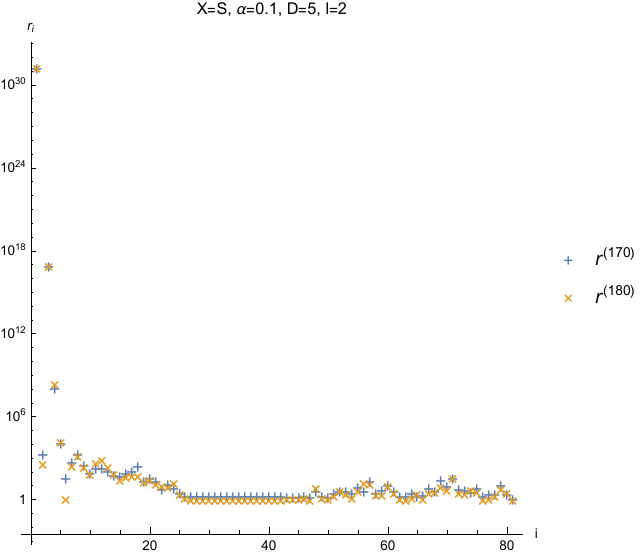}}\hfill
    \subfigure[]{\includegraphics[width=0.44\textwidth]{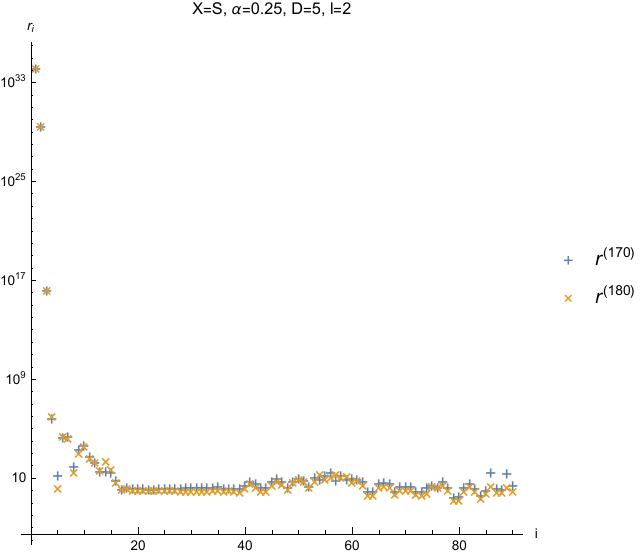}}
    \\
    \subfigure[]{\includegraphics[width=0.44\textwidth]{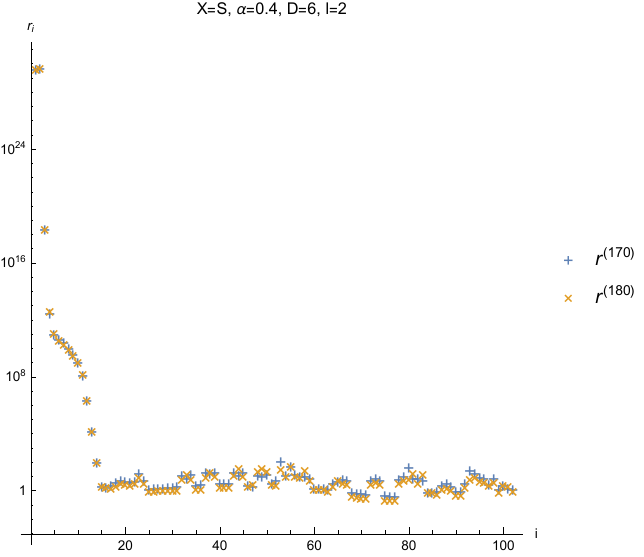}}\hfill
    \subfigure[]{\includegraphics[width=0.44\textwidth]{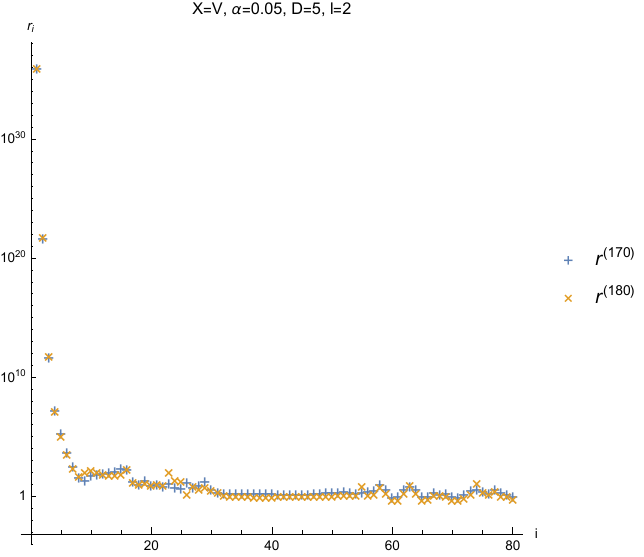}}
\end{figure}
\begin{figure}[htbp]
    \subfigure[]{\includegraphics[width=0.44\textwidth]{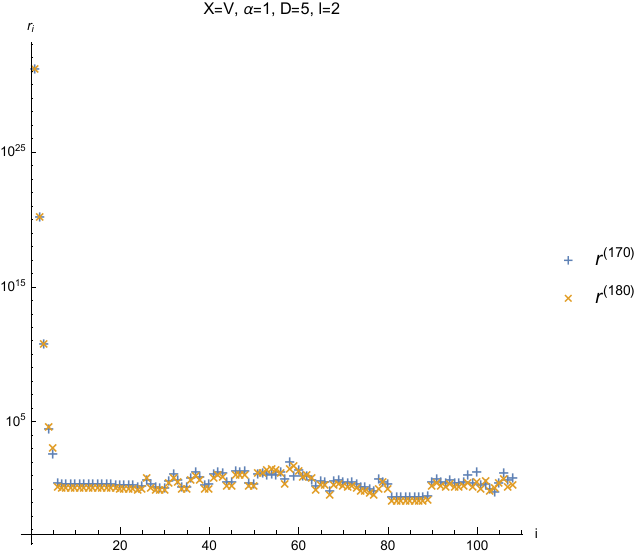}}\hfill
    \subfigure[]{\includegraphics[width=0.44\textwidth]{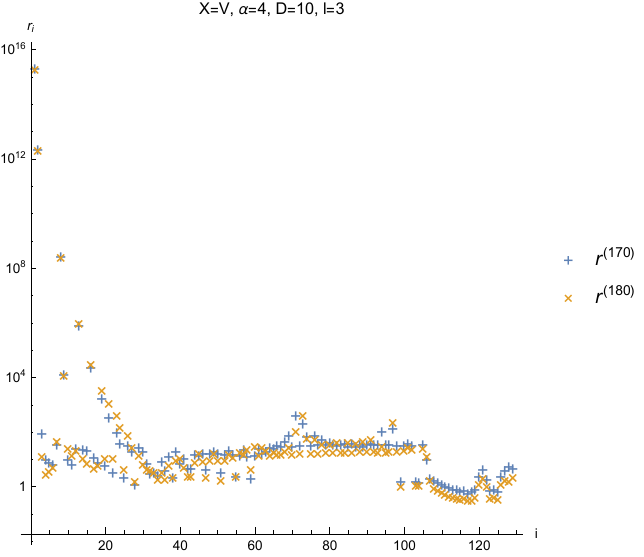}}
    \\
    \subfigure[]{\includegraphics[width=0.44\textwidth]{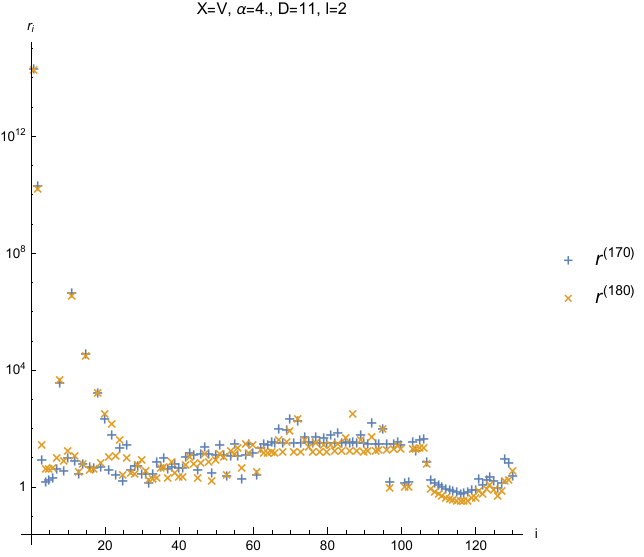}}\hfill
    \subfigure[]{\includegraphics[width=0.44\textwidth]{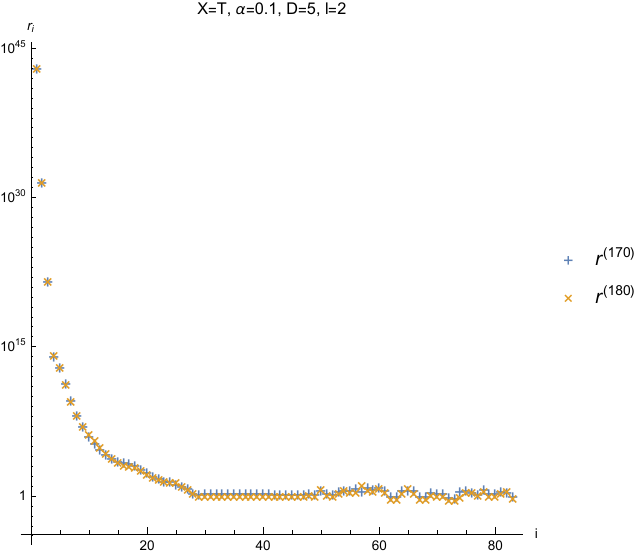}}
    \\
    \subfigure[]{\includegraphics[width=0.44\textwidth]{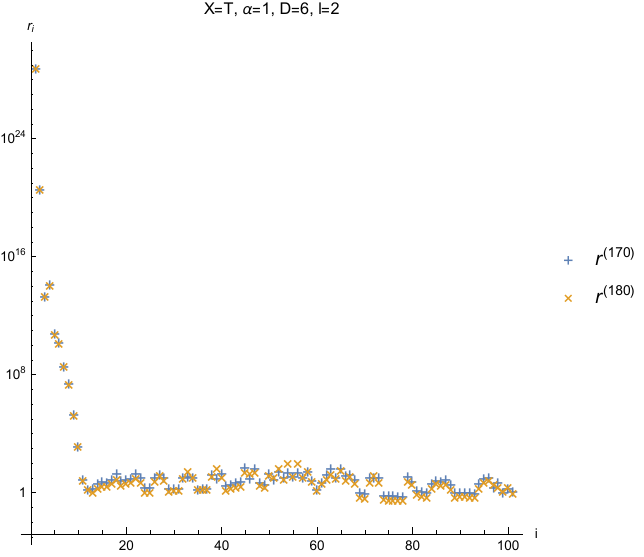}}\hfill
    \subfigure[]{\includegraphics[width=0.44\textwidth]{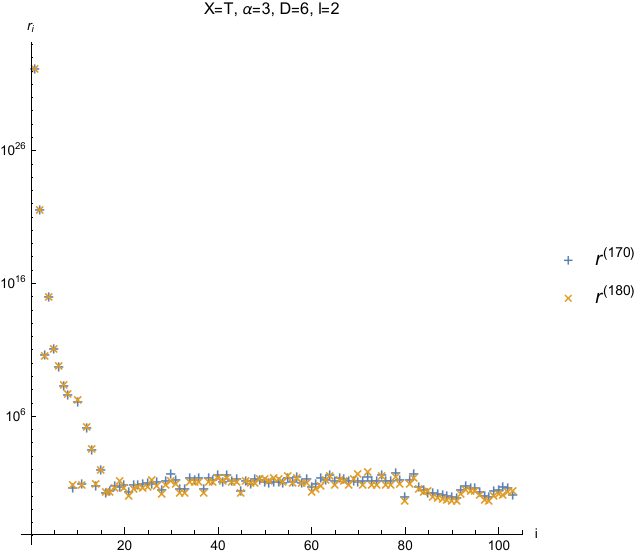}}
\end{figure}

\section{Price's law in even dimensions}
\label{sec: evenprice}
\begin{figure}[htbp]
    \centering
    \subfigure[]{\includegraphics[width=0.4\linewidth]{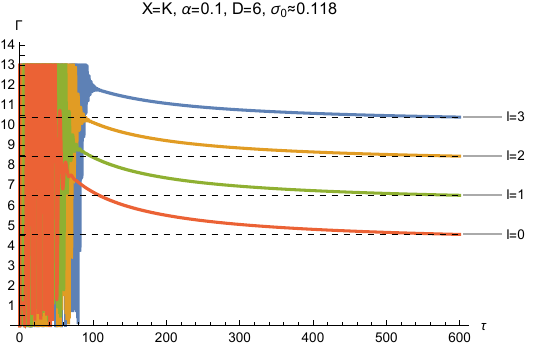}}
    \hspace{0.3cm}
    \subfigure[]{\includegraphics[width=0.4\linewidth]{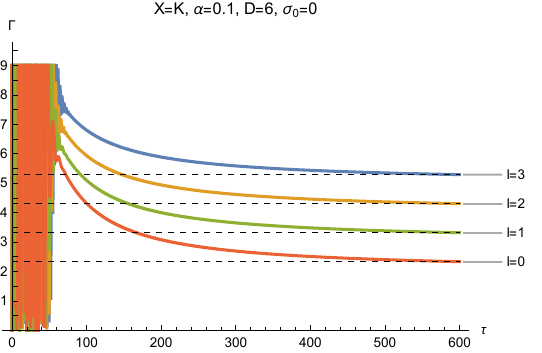}}
    \\
    \subfigure[]{\includegraphics[width=0.4\linewidth]{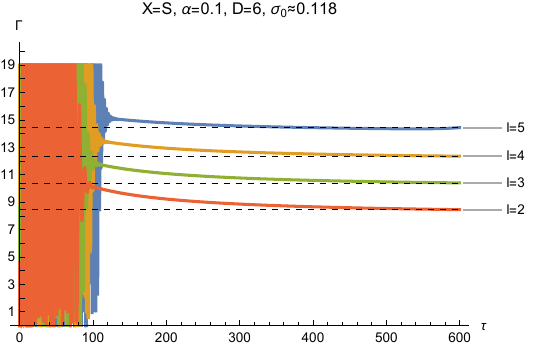}}
    \hspace{0.3cm}
    \subfigure[]{\includegraphics[width=0.4\linewidth]{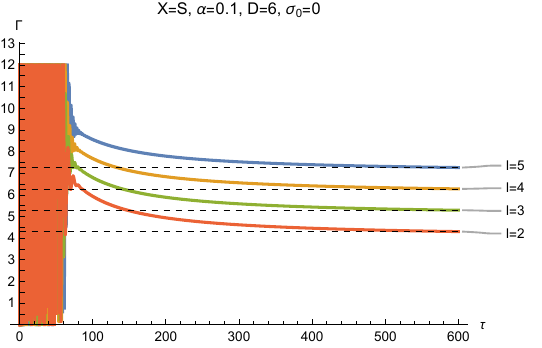}}
    \\
    \subfigure[]{\includegraphics[width=0.4\linewidth]{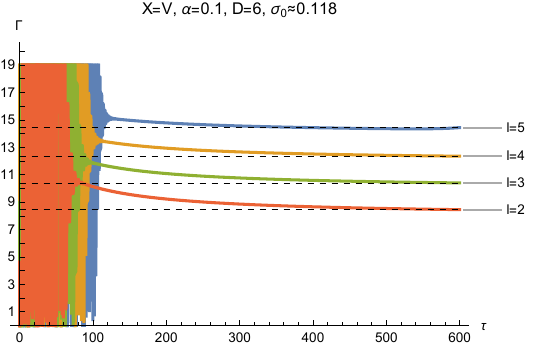}}
    \hspace{0.3cm}
    \subfigure[]{\includegraphics[width=0.4\linewidth]{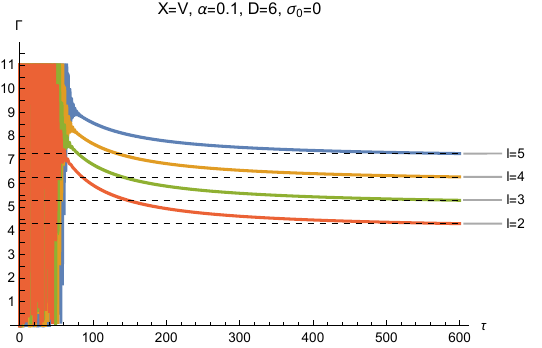}}
    \\
    \subfigure[]{\includegraphics[width=0.4\linewidth]{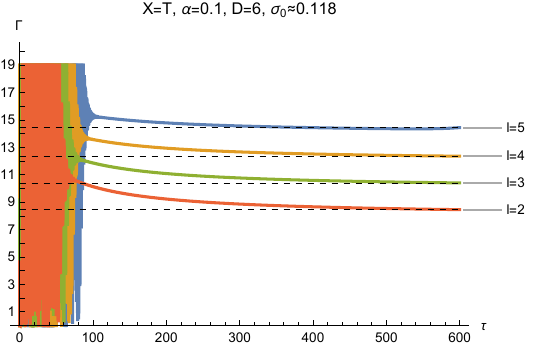}}
    \hspace{0.3cm}
    \subfigure[]{\includegraphics[width=0.4\linewidth]{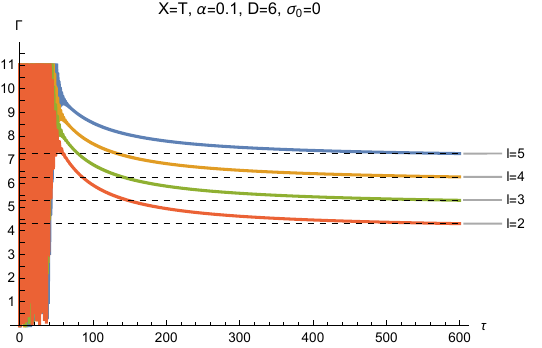}}
    \caption{The effective power-law index $\Gamma$ is computed using the initial conditions specified in Eqs. (\ref{initial condition}) with $a_0=1, b_0=1/(10\sqrt{10})$ and $c_0=1/5$. The observer is located at $\sigma_0\approx0.118$ in the first column and infinity, i.e., $\sigma_0=0$, in the second column, with the computations performed on the resolution $N=300$ using time step $\Delta \tau=0.075$.}
    \label{fig:pricelaweven}
\end{figure}

In this appendix, we show the Price's law in even dimensions as a supplement to the text. For observers at finite locations in even dimensions spacetimes, ~\cite{Ching:1994bd,Ching:1995tj} indicate that there exists a power-law tail with index $3D-8$ (when $X=K$ and $l=0$), or provide a lower bound on the index of $2D-5+2l$. However, in~\cite{Abdalla:2005hu}, they did not observe a power-law tail for even dimensions. Fig. \ref{fig:pricelaweven} shows our computational results for observers located both in finite distance and infinity in even dimensions, where power-law tails emerge, but their indices differ from the analytical predictions.

\section{Mismatch with different initial condition} \label{sec: anothermismatch}
\begin{figure}[htbp]
    \subfigure[]{\includegraphics[width=0.48\linewidth]{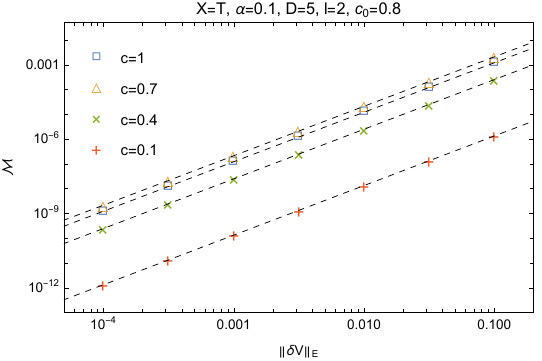}}  \hfill
    \subfigure[]{\includegraphics[width=0.48\linewidth]{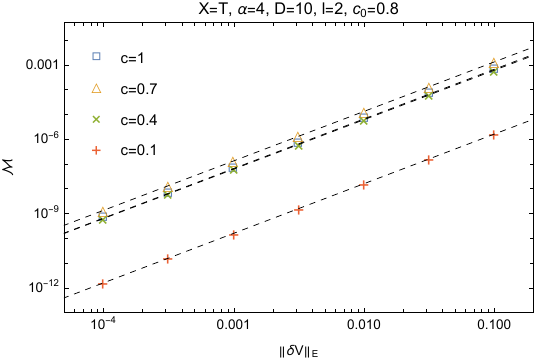}}
    \caption{
        The relationship between mismatch $\mathcal{M}$ and $\lVert\delta V\rVert_{\text{E}}$ for different $c$ values of the bump. The width $b$ of the bump is fixed as $1/5$. The initial wave is chosen to be Eqs. (\ref{initial condition}) with $a_0=1, b_0=1/(10\sqrt{10})$ and $c_0=4/5$. The dashed lines represent the fitted lines, which have slopes of approximately $2$. The computations are performed on the resolution $N=200$ using time step $\Delta \tau=0.075$.}
    \label{fig:Mnorm1}
\end{figure}

\begin{figure}[htbp]
    \centering
    \subfigure[]{\includegraphics[width=0.48\linewidth]{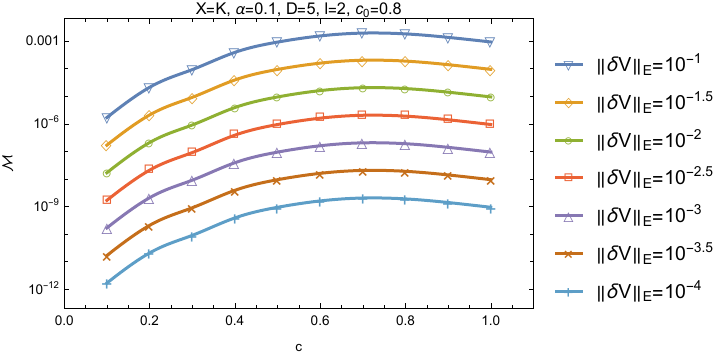}} \hfill
    \subfigure[]{\includegraphics[width=0.48\linewidth]{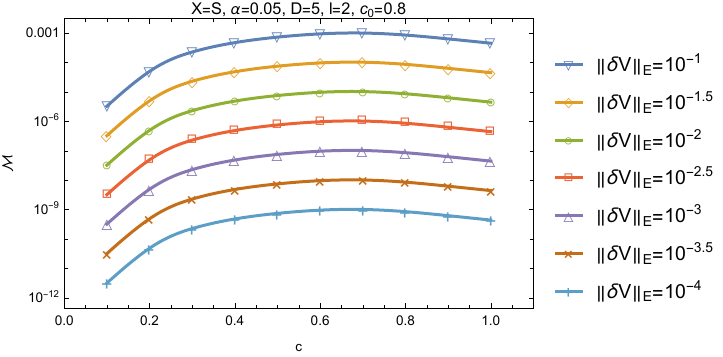}} \\
    \subfigure[]{\includegraphics[width=0.48\linewidth]{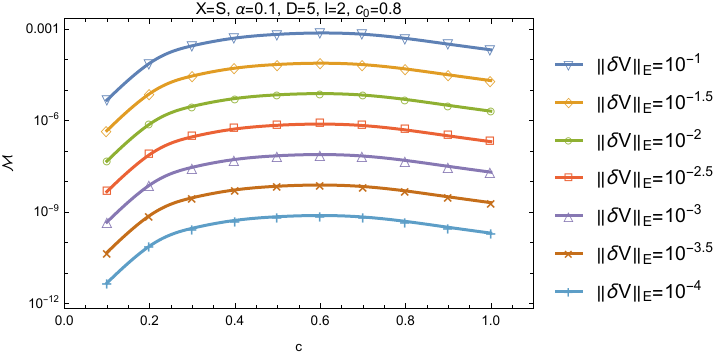}} \hfill
    \subfigure[]{\includegraphics[width=0.48\linewidth]{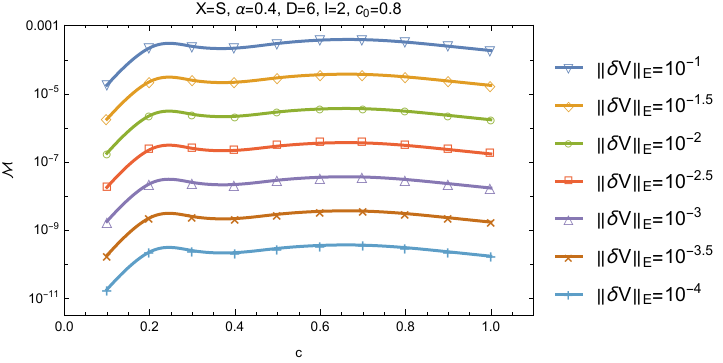}} \\
    \subfigure[]{\includegraphics[width=0.48\linewidth]{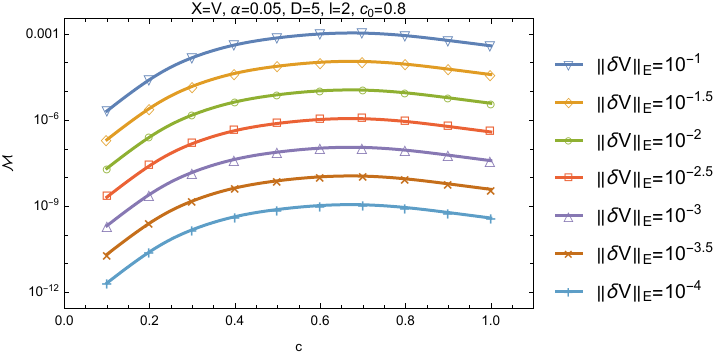}} \hfill
    \subfigure[]{\includegraphics[width=0.48\linewidth]{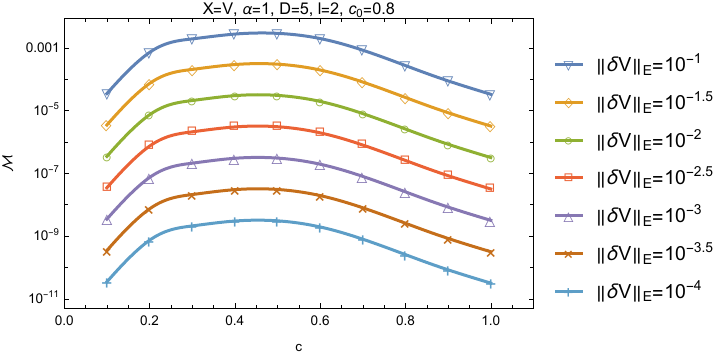}} \\
    \subfigure[]{\includegraphics[width=0.48\linewidth]{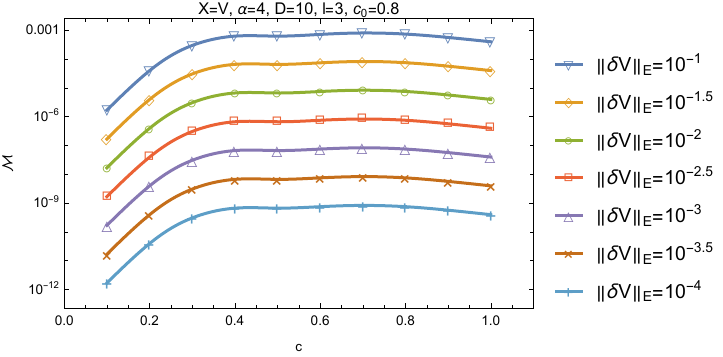}} \hfill
    \subfigure[]{\includegraphics[width=0.48\linewidth]{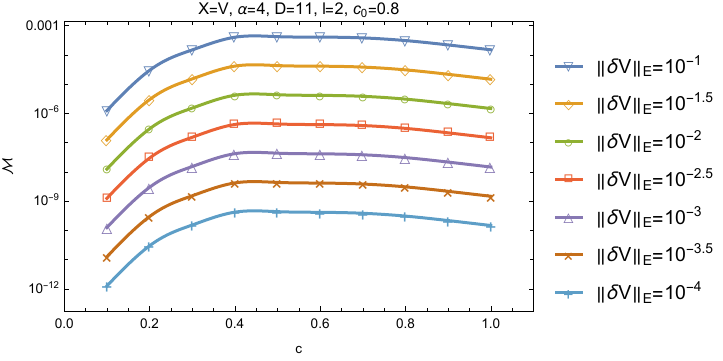}} \\
    \subfigure[]{\includegraphics[width=0.48\linewidth]{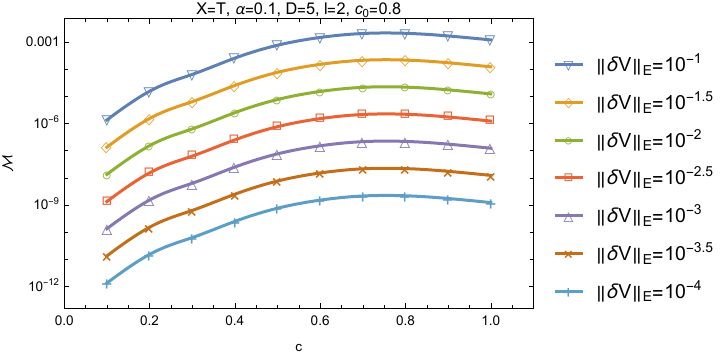}} \hfill
    \subfigure[]{\includegraphics[width=0.48\linewidth]{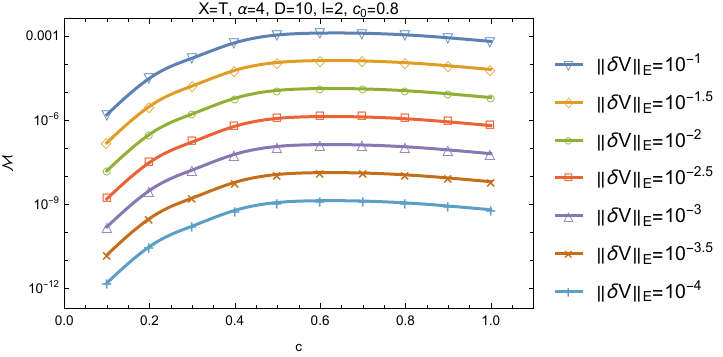}}
    \caption{The relationship between mismatch $\mathcal{M}$ and the location of the bump, which is described by the parameter $c$, for varies norm $\lVert\delta V\rVert_{\text{E}}$. The width of the bump $b$ is fixed as $1/5$. The initial wave is specified as Eqs. (\ref{initial condition}) with $a_0=1, b_0=1/(10\sqrt{10})$ and $c_0=4/5$. The computations are performed on the resolution $N=200$ using time step $\Delta \tau=0.075$.}
    \label{fig:Mc1}
\end{figure}

In this appendix, we show the results of the relationship between mismatch and the width and the location of the bump with a different initial condition in Figs. \ref{fig:Mnorm1} and \ref{fig:Mc1}. The linear fits in Fig. \ref{fig:Mnorm1} and Fig. \ref{fig:Mnorm} have slopes of approximately $2$, indicating that time-domain stability is independent of the initial data.
\begin{figure}[htbp]
    \centering
    \subfigure[]{\includegraphics[width=0.48\linewidth]{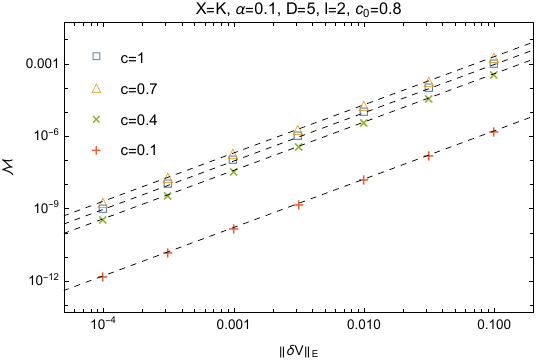}}  \hfill
    \subfigure[]{\includegraphics[width=0.48\linewidth]{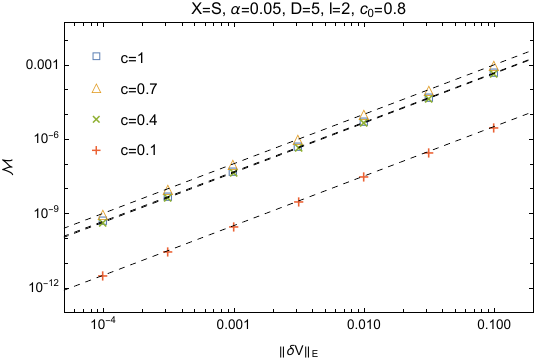}}  \\
    \subfigure[]{\includegraphics[width=0.48\linewidth]{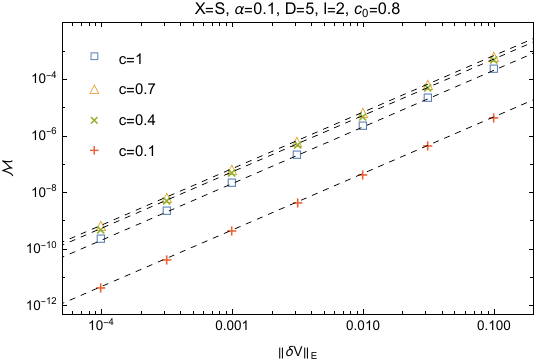}}  \hfill
    \subfigure[]{\includegraphics[width=0.48\linewidth]{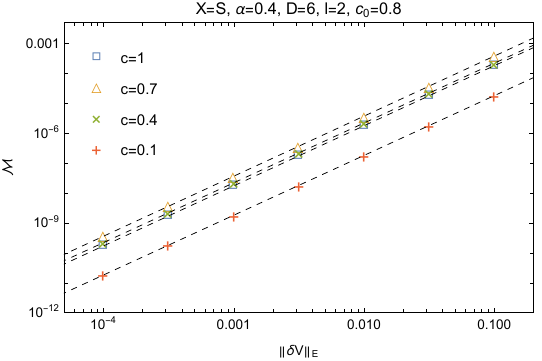}}  \\
    \subfigure[]{\includegraphics[width=0.48\linewidth]{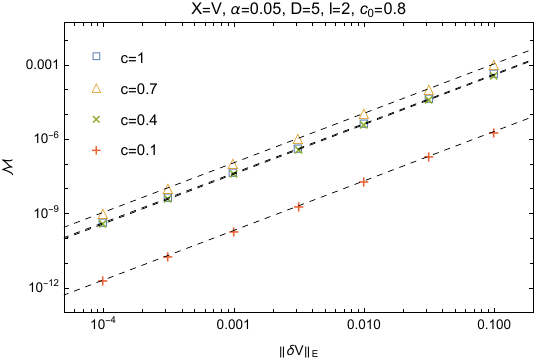}}  \hfill
    \subfigure[]{\includegraphics[width=0.48\linewidth]{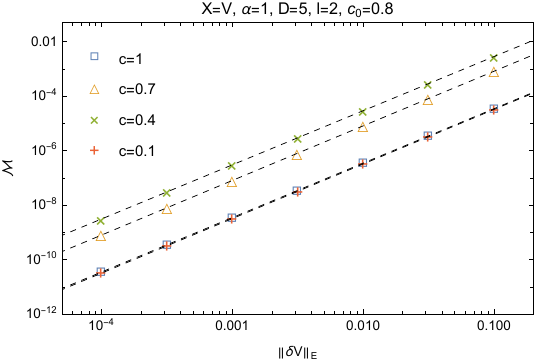}}  \\
    \subfigure[]{\includegraphics[width=0.48\linewidth]{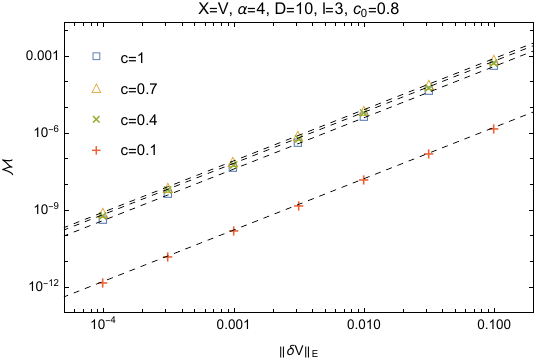}}  \hfill
    \subfigure[]{\includegraphics[width=0.48\linewidth]{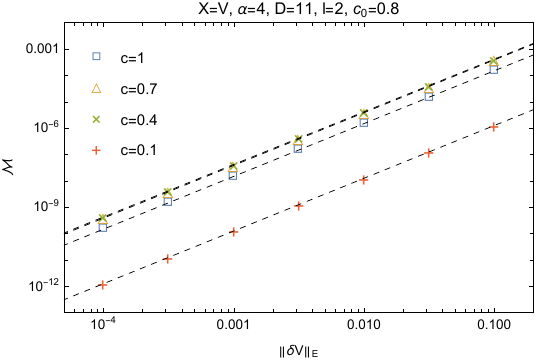}}
\end{figure}

\bibliography{mainreference}

\begin{thebibliography}{107}%
\makeatletter
\providecommand \@ifxundefined [1]{%
 \@ifx{#1\undefined}
}%
\providecommand \@ifnum [1]{%
 \ifnum #1\expandafter \@firstoftwo
 \else \expandafter \@secondoftwo
 \fi
}%
\providecommand \@ifx [1]{%
 \ifx #1\expandafter \@firstoftwo
 \else \expandafter \@secondoftwo
 \fi
}%
\providecommand \natexlab [1]{#1}%
\providecommand \enquote  [1]{``#1''}%
\providecommand \bibnamefont  [1]{#1}%
\providecommand \bibfnamefont [1]{#1}%
\providecommand \citenamefont [1]{#1}%
\providecommand \href@noop [0]{\@secondoftwo}%
\providecommand \href [0]{\begingroup \@sanitize@url \@href}%
\providecommand \@href[1]{\@@startlink{#1}\@@href}%
\providecommand \@@href[1]{\endgroup#1\@@endlink}%
\providecommand \@sanitize@url [0]{\catcode `\\12\catcode `\$12\catcode `\&12\catcode `\#12\catcode `\^12\catcode `\_12\catcode `\%12\relax}%
\providecommand \@@startlink[1]{}%
\providecommand \@@endlink[0]{}%
\providecommand \url  [0]{\begingroup\@sanitize@url \@url }%
\providecommand \@url [1]{\endgroup\@href {#1}{\urlprefix }}%
\providecommand \urlprefix  [0]{URL }%
\providecommand \Eprint [0]{\href }%
\providecommand \doibase [0]{http://dx.doi.org/}%
\providecommand \selectlanguage [0]{\@gobble}%
\providecommand \bibinfo  [0]{\@secondoftwo}%
\providecommand \bibfield  [0]{\@secondoftwo}%
\providecommand \translation [1]{[#1]}%
\providecommand \BibitemOpen [0]{}%
\providecommand \bibitemStop [0]{}%
\providecommand \bibitemNoStop [0]{.\EOS\space}%
\providecommand \EOS [0]{\spacefactor3000\relax}%
\providecommand \BibitemShut  [1]{\csname bibitem#1\endcsname}%
\let\auto@bib@innerbib\@empty
\bibitem [{\citenamefont {Giddings}\ and\ \citenamefont {Thomas}(2002)}]{Giddings:2001bu}%
  \BibitemOpen
  \bibfield  {author} {\bibinfo {author} {\bibfnamefont {S.~B.}\ \bibnamefont {Giddings}}\ and\ \bibinfo {author} {\bibfnamefont {S.~D.}\ \bibnamefont {Thomas}},\ }\href {\doibase 10.1103/PhysRevD.65.056010} {\bibfield  {journal} {\bibinfo  {journal} {Phys. Rev. D}\ }\textbf {\bibinfo {volume} {65}},\ \bibinfo {pages} {056010} (\bibinfo {year} {2002})},\ \Eprint {http://arxiv.org/abs/hep-ph/0106219} {arXiv:hep-ph/0106219} \BibitemShut {NoStop}%
\bibitem [{\citenamefont {Giddings}\ and\ \citenamefont {Mangano}(2008)}]{Giddings:2008gr}%
  \BibitemOpen
  \bibfield  {author} {\bibinfo {author} {\bibfnamefont {S.~B.}\ \bibnamefont {Giddings}}\ and\ \bibinfo {author} {\bibfnamefont {M.~L.}\ \bibnamefont {Mangano}},\ }\href {\doibase 10.1103/PhysRevD.78.035009} {\bibfield  {journal} {\bibinfo  {journal} {Phys. Rev. D}\ }\textbf {\bibinfo {volume} {78}},\ \bibinfo {pages} {035009} (\bibinfo {year} {2008})},\ \Eprint {http://arxiv.org/abs/0806.3381} {arXiv:0806.3381 [hep-ph]} \BibitemShut {NoStop}%
\bibitem [{\citenamefont {Emparan}\ and\ \citenamefont {Reall}(2008)}]{Emparan:2008eg}%
  \BibitemOpen
  \bibfield  {author} {\bibinfo {author} {\bibfnamefont {R.}~\bibnamefont {Emparan}}\ and\ \bibinfo {author} {\bibfnamefont {H.~S.}\ \bibnamefont {Reall}},\ }\href {\doibase 10.12942/lrr-2008-6} {\bibfield  {journal} {\bibinfo  {journal} {Living Rev. Rel.}\ }\textbf {\bibinfo {volume} {11}},\ \bibinfo {pages} {6} (\bibinfo {year} {2008})},\ \Eprint {http://arxiv.org/abs/0801.3471} {arXiv:0801.3471 [hep-th]} \BibitemShut {NoStop}%
\bibitem [{\citenamefont {Dimopoulos}\ and\ \citenamefont {Landsberg}(2001)}]{Dimopoulos:2001hw}%
  \BibitemOpen
  \bibfield  {author} {\bibinfo {author} {\bibfnamefont {S.}~\bibnamefont {Dimopoulos}}\ and\ \bibinfo {author} {\bibfnamefont {G.~L.}\ \bibnamefont {Landsberg}},\ }\href {\doibase 10.1103/PhysRevLett.87.161602} {\bibfield  {journal} {\bibinfo  {journal} {Phys. Rev. Lett.}\ }\textbf {\bibinfo {volume} {87}},\ \bibinfo {pages} {161602} (\bibinfo {year} {2001})},\ \Eprint {http://arxiv.org/abs/hep-ph/0106295} {arXiv:hep-ph/0106295} \BibitemShut {NoStop}%
\bibitem [{\citenamefont {Kanti}(2004)}]{Kanti:2004nr}%
  \BibitemOpen
  \bibfield  {author} {\bibinfo {author} {\bibfnamefont {P.}~\bibnamefont {Kanti}},\ }\href {\doibase 10.1142/S0217751X04018324} {\bibfield  {journal} {\bibinfo  {journal} {Int. J. Mod. Phys. A}\ }\textbf {\bibinfo {volume} {19}},\ \bibinfo {pages} {4899} (\bibinfo {year} {2004})},\ \Eprint {http://arxiv.org/abs/hep-ph/0402168} {arXiv:hep-ph/0402168} \BibitemShut {NoStop}%
\bibitem [{\citenamefont {Antoniadis}\ \emph {et~al.}(1998)\citenamefont {Antoniadis}, \citenamefont {Arkani-Hamed}, \citenamefont {Dimopoulos},\ and\ \citenamefont {Dvali}}]{Antoniadis:1998ig}%
  \BibitemOpen
  \bibfield  {author} {\bibinfo {author} {\bibfnamefont {I.}~\bibnamefont {Antoniadis}}, \bibinfo {author} {\bibfnamefont {N.}~\bibnamefont {Arkani-Hamed}}, \bibinfo {author} {\bibfnamefont {S.}~\bibnamefont {Dimopoulos}}, \ and\ \bibinfo {author} {\bibfnamefont {G.~R.}\ \bibnamefont {Dvali}},\ }\href {\doibase 10.1016/S0370-2693(98)00860-0} {\bibfield  {journal} {\bibinfo  {journal} {Phys. Lett. B}\ }\textbf {\bibinfo {volume} {436}},\ \bibinfo {pages} {257} (\bibinfo {year} {1998})},\ \Eprint {http://arxiv.org/abs/hep-ph/9804398} {arXiv:hep-ph/9804398} \BibitemShut {NoStop}%
\bibitem [{\citenamefont {Randall}\ and\ \citenamefont {Sundrum}(1999)}]{Randall:1999ee}%
  \BibitemOpen
  \bibfield  {author} {\bibinfo {author} {\bibfnamefont {L.}~\bibnamefont {Randall}}\ and\ \bibinfo {author} {\bibfnamefont {R.}~\bibnamefont {Sundrum}},\ }\href {\doibase 10.1103/PhysRevLett.83.3370} {\bibfield  {journal} {\bibinfo  {journal} {Phys. Rev. Lett.}\ }\textbf {\bibinfo {volume} {83}},\ \bibinfo {pages} {3370} (\bibinfo {year} {1999})},\ \Eprint {http://arxiv.org/abs/hep-ph/9905221} {arXiv:hep-ph/9905221} \BibitemShut {NoStop}%
\bibitem [{\citenamefont {Arkani-Hamed}\ \emph {et~al.}(1999)\citenamefont {Arkani-Hamed}, \citenamefont {Dimopoulos},\ and\ \citenamefont {Dvali}}]{Arkani-Hamed:1998sfv}%
  \BibitemOpen
  \bibfield  {author} {\bibinfo {author} {\bibfnamefont {N.}~\bibnamefont {Arkani-Hamed}}, \bibinfo {author} {\bibfnamefont {S.}~\bibnamefont {Dimopoulos}}, \ and\ \bibinfo {author} {\bibfnamefont {G.~R.}\ \bibnamefont {Dvali}},\ }\href {\doibase 10.1103/PhysRevD.59.086004} {\bibfield  {journal} {\bibinfo  {journal} {Phys. Rev. D}\ }\textbf {\bibinfo {volume} {59}},\ \bibinfo {pages} {086004} (\bibinfo {year} {1999})},\ \Eprint {http://arxiv.org/abs/hep-ph/9807344} {arXiv:hep-ph/9807344} \BibitemShut {NoStop}%
\bibitem [{\citenamefont {Arkani-Hamed}\ \emph {et~al.}(1998)\citenamefont {Arkani-Hamed}, \citenamefont {Dimopoulos},\ and\ \citenamefont {Dvali}}]{Arkani-Hamed:1998jmv}%
  \BibitemOpen
  \bibfield  {author} {\bibinfo {author} {\bibfnamefont {N.}~\bibnamefont {Arkani-Hamed}}, \bibinfo {author} {\bibfnamefont {S.}~\bibnamefont {Dimopoulos}}, \ and\ \bibinfo {author} {\bibfnamefont {G.~R.}\ \bibnamefont {Dvali}},\ }\href {\doibase 10.1016/S0370-2693(98)00466-3} {\bibfield  {journal} {\bibinfo  {journal} {Phys. Lett. B}\ }\textbf {\bibinfo {volume} {429}},\ \bibinfo {pages} {263} (\bibinfo {year} {1998})},\ \Eprint {http://arxiv.org/abs/hep-ph/9803315} {arXiv:hep-ph/9803315} \BibitemShut {NoStop}%
\bibitem [{\citenamefont {Charmousis}(2009)}]{Charmousis:2008kc}%
  \BibitemOpen
  \bibfield  {author} {\bibinfo {author} {\bibfnamefont {C.}~\bibnamefont {Charmousis}},\ }\href {\doibase 10.1007/978-3-540-88460-6_8} {\bibfield  {journal} {\bibinfo  {journal} {Lect. Notes Phys.}\ }\textbf {\bibinfo {volume} {769}},\ \bibinfo {pages} {299} (\bibinfo {year} {2009})},\ \Eprint {http://arxiv.org/abs/0805.0568} {arXiv:0805.0568 [gr-qc]} \BibitemShut {NoStop}%
\bibitem [{\citenamefont {Clifton}\ \emph {et~al.}(2012)\citenamefont {Clifton}, \citenamefont {Ferreira}, \citenamefont {Padilla},\ and\ \citenamefont {Skordis}}]{Clifton:2011jh}%
  \BibitemOpen
  \bibfield  {author} {\bibinfo {author} {\bibfnamefont {T.}~\bibnamefont {Clifton}}, \bibinfo {author} {\bibfnamefont {P.~G.}\ \bibnamefont {Ferreira}}, \bibinfo {author} {\bibfnamefont {A.}~\bibnamefont {Padilla}}, \ and\ \bibinfo {author} {\bibfnamefont {C.}~\bibnamefont {Skordis}},\ }\href {\doibase 10.1016/j.physrep.2012.01.001} {\bibfield  {journal} {\bibinfo  {journal} {Phys. Rept.}\ }\textbf {\bibinfo {volume} {513}},\ \bibinfo {pages} {1} (\bibinfo {year} {2012})},\ \Eprint {http://arxiv.org/abs/1106.2476} {arXiv:1106.2476 [astro-ph.CO]} \BibitemShut {NoStop}%
\bibitem [{\citenamefont {Lovelock}(1971)}]{Lovelock:1971yv}%
  \BibitemOpen
  \bibfield  {author} {\bibinfo {author} {\bibfnamefont {D.}~\bibnamefont {Lovelock}},\ }\href {\doibase 10.1063/1.1665613} {\bibfield  {journal} {\bibinfo  {journal} {J. Math. Phys.}\ }\textbf {\bibinfo {volume} {12}},\ \bibinfo {pages} {498} (\bibinfo {year} {1971})}\BibitemShut {NoStop}%
\bibitem [{\citenamefont {Lovelock}(1972)}]{Lovelock:1972vz}%
  \BibitemOpen
  \bibfield  {author} {\bibinfo {author} {\bibfnamefont {D.}~\bibnamefont {Lovelock}},\ }\href {\doibase 10.1063/1.1666069} {\bibfield  {journal} {\bibinfo  {journal} {J. Math. Phys.}\ }\textbf {\bibinfo {volume} {13}},\ \bibinfo {pages} {874} (\bibinfo {year} {1972})}\BibitemShut {NoStop}%
\bibitem [{\citenamefont {Zwiebach}(1985)}]{Zwiebach:1985uq}%
  \BibitemOpen
  \bibfield  {author} {\bibinfo {author} {\bibfnamefont {B.}~\bibnamefont {Zwiebach}},\ }\href {\doibase 10.1016/0370-2693(85)91616-8} {\bibfield  {journal} {\bibinfo  {journal} {Phys. Lett. B}\ }\textbf {\bibinfo {volume} {156}},\ \bibinfo {pages} {315} (\bibinfo {year} {1985})}\BibitemShut {NoStop}%
\bibitem [{\citenamefont {Gross}\ and\ \citenamefont {Witten}(1986)}]{Gross:1986iv}%
  \BibitemOpen
  \bibfield  {author} {\bibinfo {author} {\bibfnamefont {D.~J.}\ \bibnamefont {Gross}}\ and\ \bibinfo {author} {\bibfnamefont {E.}~\bibnamefont {Witten}},\ }\href {\doibase 10.1016/0550-3213(86)90429-3} {\bibfield  {journal} {\bibinfo  {journal} {Nucl. Phys. B}\ }\textbf {\bibinfo {volume} {277}},\ \bibinfo {pages} {1} (\bibinfo {year} {1986})}\BibitemShut {NoStop}%
\bibitem [{\citenamefont {Boulware}\ and\ \citenamefont {Deser}(1985)}]{Boulware:1985wk}%
  \BibitemOpen
  \bibfield  {author} {\bibinfo {author} {\bibfnamefont {D.~G.}\ \bibnamefont {Boulware}}\ and\ \bibinfo {author} {\bibfnamefont {S.}~\bibnamefont {Deser}},\ }\href {\doibase 10.1103/PhysRevLett.55.2656} {\bibfield  {journal} {\bibinfo  {journal} {Phys. Rev. Lett.}\ }\textbf {\bibinfo {volume} {55}},\ \bibinfo {pages} {2656} (\bibinfo {year} {1985})}\BibitemShut {NoStop}%
\bibitem [{\citenamefont {Wheeler}(1986)}]{Wheeler:1985nh}%
  \BibitemOpen
  \bibfield  {author} {\bibinfo {author} {\bibfnamefont {J.~T.}\ \bibnamefont {Wheeler}},\ }\href {\doibase 10.1016/0550-3213(86)90268-3} {\bibfield  {journal} {\bibinfo  {journal} {Nucl. Phys. B}\ }\textbf {\bibinfo {volume} {268}},\ \bibinfo {pages} {737} (\bibinfo {year} {1986})}\BibitemShut {NoStop}%
\bibitem [{\citenamefont {Cai}(2002)}]{Cai:2001dz}%
  \BibitemOpen
  \bibfield  {author} {\bibinfo {author} {\bibfnamefont {R.-G.}\ \bibnamefont {Cai}},\ }\href {\doibase 10.1103/PhysRevD.65.084014} {\bibfield  {journal} {\bibinfo  {journal} {Phys. Rev. D}\ }\textbf {\bibinfo {volume} {65}},\ \bibinfo {pages} {084014} (\bibinfo {year} {2002})},\ \Eprint {http://arxiv.org/abs/hep-th/0109133} {arXiv:hep-th/0109133} \BibitemShut {NoStop}%
\bibitem [{\citenamefont {Cai}\ and\ \citenamefont {Guo}(2004)}]{Cai:2003gr}%
  \BibitemOpen
  \bibfield  {author} {\bibinfo {author} {\bibfnamefont {R.-G.}\ \bibnamefont {Cai}}\ and\ \bibinfo {author} {\bibfnamefont {Q.}~\bibnamefont {Guo}},\ }\href {\doibase 10.1103/PhysRevD.69.104025} {\bibfield  {journal} {\bibinfo  {journal} {Phys. Rev. D}\ }\textbf {\bibinfo {volume} {69}},\ \bibinfo {pages} {104025} (\bibinfo {year} {2004})},\ \Eprint {http://arxiv.org/abs/hep-th/0311020} {arXiv:hep-th/0311020} \BibitemShut {NoStop}%
\bibitem [{\citenamefont {Brigante}\ \emph {et~al.}(2008{\natexlab{a}})\citenamefont {Brigante}, \citenamefont {Liu}, \citenamefont {Myers}, \citenamefont {Shenker},\ and\ \citenamefont {Yaida}}]{Brigante:2008gz}%
  \BibitemOpen
  \bibfield  {author} {\bibinfo {author} {\bibfnamefont {M.}~\bibnamefont {Brigante}}, \bibinfo {author} {\bibfnamefont {H.}~\bibnamefont {Liu}}, \bibinfo {author} {\bibfnamefont {R.~C.}\ \bibnamefont {Myers}}, \bibinfo {author} {\bibfnamefont {S.}~\bibnamefont {Shenker}}, \ and\ \bibinfo {author} {\bibfnamefont {S.}~\bibnamefont {Yaida}},\ }\href {\doibase 10.1103/PhysRevLett.100.191601} {\bibfield  {journal} {\bibinfo  {journal} {Phys. Rev. Lett.}\ }\textbf {\bibinfo {volume} {100}},\ \bibinfo {pages} {191601} (\bibinfo {year} {2008}{\natexlab{a}})},\ \Eprint {http://arxiv.org/abs/0802.3318} {arXiv:0802.3318 [hep-th]} \BibitemShut {NoStop}%
\bibitem [{\citenamefont {Brigante}\ \emph {et~al.}(2008{\natexlab{b}})\citenamefont {Brigante}, \citenamefont {Liu}, \citenamefont {Myers}, \citenamefont {Shenker},\ and\ \citenamefont {Yaida}}]{Brigante:2007nu}%
  \BibitemOpen
  \bibfield  {author} {\bibinfo {author} {\bibfnamefont {M.}~\bibnamefont {Brigante}}, \bibinfo {author} {\bibfnamefont {H.}~\bibnamefont {Liu}}, \bibinfo {author} {\bibfnamefont {R.~C.}\ \bibnamefont {Myers}}, \bibinfo {author} {\bibfnamefont {S.}~\bibnamefont {Shenker}}, \ and\ \bibinfo {author} {\bibfnamefont {S.}~\bibnamefont {Yaida}},\ }\href {\doibase 10.1103/PhysRevD.77.126006} {\bibfield  {journal} {\bibinfo  {journal} {Phys. Rev. D}\ }\textbf {\bibinfo {volume} {77}},\ \bibinfo {pages} {126006} (\bibinfo {year} {2008}{\natexlab{b}})},\ \Eprint {http://arxiv.org/abs/0712.0805} {arXiv:0712.0805 [hep-th]} \BibitemShut {NoStop}%
\bibitem [{\citenamefont {Cvetic}\ \emph {et~al.}(2002)\citenamefont {Cvetic}, \citenamefont {Nojiri},\ and\ \citenamefont {Odintsov}}]{Cvetic:2001bk}%
  \BibitemOpen
  \bibfield  {author} {\bibinfo {author} {\bibfnamefont {M.}~\bibnamefont {Cvetic}}, \bibinfo {author} {\bibfnamefont {S.}~\bibnamefont {Nojiri}}, \ and\ \bibinfo {author} {\bibfnamefont {S.~D.}\ \bibnamefont {Odintsov}},\ }\href {\doibase 10.1016/S0550-3213(02)00075-5} {\bibfield  {journal} {\bibinfo  {journal} {Nucl. Phys. B}\ }\textbf {\bibinfo {volume} {628}},\ \bibinfo {pages} {295} (\bibinfo {year} {2002})},\ \Eprint {http://arxiv.org/abs/hep-th/0112045} {arXiv:hep-th/0112045} \BibitemShut {NoStop}%
\bibitem [{\citenamefont {Maeda}\ and\ \citenamefont {Nozawa}(2008)}]{Maeda:2008nz}%
  \BibitemOpen
  \bibfield  {author} {\bibinfo {author} {\bibfnamefont {H.}~\bibnamefont {Maeda}}\ and\ \bibinfo {author} {\bibfnamefont {M.}~\bibnamefont {Nozawa}},\ }\href {\doibase 10.1103/PhysRevD.78.024005} {\bibfield  {journal} {\bibinfo  {journal} {Phys. Rev. D}\ }\textbf {\bibinfo {volume} {78}},\ \bibinfo {pages} {024005} (\bibinfo {year} {2008})},\ \Eprint {http://arxiv.org/abs/0803.1704} {arXiv:0803.1704 [gr-qc]} \BibitemShut {NoStop}%
\bibitem [{\citenamefont {Cao}\ \emph {et~al.}(2023)\citenamefont {Cao}, \citenamefont {Wu}, \citenamefont {Zhao},\ and\ \citenamefont {Zhou}}]{Cao:2023zhy}%
  \BibitemOpen
  \bibfield  {author} {\bibinfo {author} {\bibfnamefont {L.-M.}\ \bibnamefont {Cao}}, \bibinfo {author} {\bibfnamefont {L.-B.}\ \bibnamefont {Wu}}, \bibinfo {author} {\bibfnamefont {Y.}~\bibnamefont {Zhao}}, \ and\ \bibinfo {author} {\bibfnamefont {Y.-S.}\ \bibnamefont {Zhou}},\ }\href {\doibase 10.1103/PhysRevD.108.124023} {\bibfield  {journal} {\bibinfo  {journal} {Phys. Rev. D}\ }\textbf {\bibinfo {volume} {108}},\ \bibinfo {pages} {124023} (\bibinfo {year} {2023})}\BibitemShut {NoStop}%
\bibitem [{\citenamefont {Chandrasekhar}(1998)}]{chandrasekhar1998mathematical}%
  \BibitemOpen
  \bibfield  {author} {\bibinfo {author} {\bibfnamefont {S.}~\bibnamefont {Chandrasekhar}},\ }\href {https://books.google.co.jp/books?id=LBOVcrzFfhsC} {\emph {\bibinfo {title} {The Mathematical Theory of Black Holes}}},\ International series of monographs on physics\ (\bibinfo  {publisher} {Clarendon Press},\ \bibinfo {year} {1998})\BibitemShut {NoStop}%
\bibitem [{\citenamefont {Nollert}(1999)}]{Nollert:1999ji}%
  \BibitemOpen
  \bibfield  {author} {\bibinfo {author} {\bibfnamefont {H.-P.}\ \bibnamefont {Nollert}},\ }\href {\doibase 10.1088/0264-9381/16/12/201} {\bibfield  {journal} {\bibinfo  {journal} {Class. Quant. Grav.}\ }\textbf {\bibinfo {volume} {16}},\ \bibinfo {pages} {R159} (\bibinfo {year} {1999})}\BibitemShut {NoStop}%
\bibitem [{\citenamefont {Kokkotas}\ and\ \citenamefont {Schmidt}(1999)}]{Kokkotas:1999bd}%
  \BibitemOpen
  \bibfield  {author} {\bibinfo {author} {\bibfnamefont {K.~D.}\ \bibnamefont {Kokkotas}}\ and\ \bibinfo {author} {\bibfnamefont {B.~G.}\ \bibnamefont {Schmidt}},\ }\href {\doibase 10.12942/lrr-1999-2} {\bibfield  {journal} {\bibinfo  {journal} {Living Rev. Rel.}\ }\textbf {\bibinfo {volume} {2}},\ \bibinfo {pages} {2} (\bibinfo {year} {1999})},\ \Eprint {http://arxiv.org/abs/gr-qc/9909058} {arXiv:gr-qc/9909058} \BibitemShut {NoStop}%
\bibitem [{\citenamefont {Berti}\ \emph {et~al.}(2009)\citenamefont {Berti}, \citenamefont {Cardoso},\ and\ \citenamefont {Starinets}}]{Berti:2009kk}%
  \BibitemOpen
  \bibfield  {author} {\bibinfo {author} {\bibfnamefont {E.}~\bibnamefont {Berti}}, \bibinfo {author} {\bibfnamefont {V.}~\bibnamefont {Cardoso}}, \ and\ \bibinfo {author} {\bibfnamefont {A.~O.}\ \bibnamefont {Starinets}},\ }\href {\doibase 10.1088/0264-9381/26/16/163001} {\bibfield  {journal} {\bibinfo  {journal} {Class. Quant. Grav.}\ }\textbf {\bibinfo {volume} {26}},\ \bibinfo {pages} {163001} (\bibinfo {year} {2009})},\ \Eprint {http://arxiv.org/abs/0905.2975} {arXiv:0905.2975 [gr-qc]} \BibitemShut {NoStop}%
\bibitem [{\citenamefont {Konoplya}\ and\ \citenamefont {Zhidenko}(2011)}]{Konoplya:2011qq}%
  \BibitemOpen
  \bibfield  {author} {\bibinfo {author} {\bibfnamefont {R.~A.}\ \bibnamefont {Konoplya}}\ and\ \bibinfo {author} {\bibfnamefont {A.}~\bibnamefont {Zhidenko}},\ }\href {\doibase 10.1103/RevModPhys.83.793} {\bibfield  {journal} {\bibinfo  {journal} {Rev. Mod. Phys.}\ }\textbf {\bibinfo {volume} {83}},\ \bibinfo {pages} {793} (\bibinfo {year} {2011})},\ \Eprint {http://arxiv.org/abs/1102.4014} {arXiv:1102.4014 [gr-qc]} \BibitemShut {NoStop}%
\bibitem [{\citenamefont {Leaver}(1986)}]{Leaver:1986gd}%
  \BibitemOpen
  \bibfield  {author} {\bibinfo {author} {\bibfnamefont {E.~W.}\ \bibnamefont {Leaver}},\ }\href {\doibase 10.1103/PhysRevD.34.384} {\bibfield  {journal} {\bibinfo  {journal} {Phys. Rev. D}\ }\textbf {\bibinfo {volume} {34}},\ \bibinfo {pages} {384} (\bibinfo {year} {1986})}\BibitemShut {NoStop}%
\bibitem [{\citenamefont {Price}(1972{\natexlab{a}})}]{Price:1971fb}%
  \BibitemOpen
  \bibfield  {author} {\bibinfo {author} {\bibfnamefont {R.~H.}\ \bibnamefont {Price}},\ }\href {\doibase 10.1103/PhysRevD.5.2419} {\bibfield  {journal} {\bibinfo  {journal} {Phys. Rev. D}\ }\textbf {\bibinfo {volume} {5}},\ \bibinfo {pages} {2419} (\bibinfo {year} {1972}{\natexlab{a}})}\BibitemShut {NoStop}%
\bibitem [{\citenamefont {Price}(1972{\natexlab{b}})}]{Price:1972pw}%
  \BibitemOpen
  \bibfield  {author} {\bibinfo {author} {\bibfnamefont {R.~H.}\ \bibnamefont {Price}},\ }\href {\doibase 10.1103/PhysRevD.5.2439} {\bibfield  {journal} {\bibinfo  {journal} {Phys. Rev. D}\ }\textbf {\bibinfo {volume} {5}},\ \bibinfo {pages} {2439} (\bibinfo {year} {1972}{\natexlab{b}})}\BibitemShut {NoStop}%
\bibitem [{\citenamefont {Baibhav}\ \emph {et~al.}(2023)\citenamefont {Baibhav}, \citenamefont {Cheung}, \citenamefont {Berti}, \citenamefont {Cardoso}, \citenamefont {Carullo}, \citenamefont {Cotesta}, \citenamefont {Del~Pozzo},\ and\ \citenamefont {Duque}}]{Baibhav:2023clw}%
  \BibitemOpen
  \bibfield  {author} {\bibinfo {author} {\bibfnamefont {V.}~\bibnamefont {Baibhav}}, \bibinfo {author} {\bibfnamefont {M.~H.-Y.}\ \bibnamefont {Cheung}}, \bibinfo {author} {\bibfnamefont {E.}~\bibnamefont {Berti}}, \bibinfo {author} {\bibfnamefont {V.}~\bibnamefont {Cardoso}}, \bibinfo {author} {\bibfnamefont {G.}~\bibnamefont {Carullo}}, \bibinfo {author} {\bibfnamefont {R.}~\bibnamefont {Cotesta}}, \bibinfo {author} {\bibfnamefont {W.}~\bibnamefont {Del~Pozzo}}, \ and\ \bibinfo {author} {\bibfnamefont {F.}~\bibnamefont {Duque}},\ }\href {\doibase 10.1103/PhysRevD.108.104020} {\bibfield  {journal} {\bibinfo  {journal} {Phys. Rev. D}\ }\textbf {\bibinfo {volume} {108}},\ \bibinfo {pages} {104020} (\bibinfo {year} {2023})},\ \Eprint {http://arxiv.org/abs/2302.03050} {arXiv:2302.03050 [gr-qc]} \BibitemShut {NoStop}%
\bibitem [{\citenamefont {Giesler}\ \emph {et~al.}(2019)\citenamefont {Giesler}, \citenamefont {Isi}, \citenamefont {Scheel},\ and\ \citenamefont {Teukolsky}}]{Giesler:2019uxc}%
  \BibitemOpen
  \bibfield  {author} {\bibinfo {author} {\bibfnamefont {M.}~\bibnamefont {Giesler}}, \bibinfo {author} {\bibfnamefont {M.}~\bibnamefont {Isi}}, \bibinfo {author} {\bibfnamefont {M.~A.}\ \bibnamefont {Scheel}}, \ and\ \bibinfo {author} {\bibfnamefont {S.}~\bibnamefont {Teukolsky}},\ }\href {\doibase 10.1103/PhysRevX.9.041060} {\bibfield  {journal} {\bibinfo  {journal} {Phys. Rev. X}\ }\textbf {\bibinfo {volume} {9}},\ \bibinfo {pages} {041060} (\bibinfo {year} {2019})},\ \Eprint {http://arxiv.org/abs/1903.08284} {arXiv:1903.08284 [gr-qc]} \BibitemShut {NoStop}%
\bibitem [{\citenamefont {Abbott}\ \emph {et~al.}(2016)\citenamefont {Abbott} \emph {et~al.}}]{LIGOScientific:2016aoc}%
  \BibitemOpen
  \bibfield  {author} {\bibinfo {author} {\bibfnamefont {B.~P.}\ \bibnamefont {Abbott}} \emph {et~al.} (\bibinfo {collaboration} {LIGO Scientific, Virgo}),\ }\href {\doibase 10.1103/PhysRevLett.116.061102} {\bibfield  {journal} {\bibinfo  {journal} {Phys. Rev. Lett.}\ }\textbf {\bibinfo {volume} {116}},\ \bibinfo {pages} {061102} (\bibinfo {year} {2016})},\ \Eprint {http://arxiv.org/abs/1602.03837} {arXiv:1602.03837 [gr-qc]} \BibitemShut {NoStop}%
\bibitem [{\citenamefont {Berti}\ \emph {et~al.}(2018)\citenamefont {Berti}, \citenamefont {Yagi}, \citenamefont {Yang},\ and\ \citenamefont {Yunes}}]{Berti:2018vdi}%
  \BibitemOpen
  \bibfield  {author} {\bibinfo {author} {\bibfnamefont {E.}~\bibnamefont {Berti}}, \bibinfo {author} {\bibfnamefont {K.}~\bibnamefont {Yagi}}, \bibinfo {author} {\bibfnamefont {H.}~\bibnamefont {Yang}}, \ and\ \bibinfo {author} {\bibfnamefont {N.}~\bibnamefont {Yunes}},\ }\href {\doibase 10.1007/s10714-018-2372-6} {\bibfield  {journal} {\bibinfo  {journal} {Gen. Rel. Grav.}\ }\textbf {\bibinfo {volume} {50}},\ \bibinfo {pages} {49} (\bibinfo {year} {2018})},\ \Eprint {http://arxiv.org/abs/1801.03587} {arXiv:1801.03587 [gr-qc]} \BibitemShut {NoStop}%
\bibitem [{\citenamefont {Maggiore}(2018)}]{Maggiore}%
  \BibitemOpen
  \bibfield  {author} {\bibinfo {author} {\bibfnamefont {M.}~\bibnamefont {Maggiore}},\ }\href {\doibase 10.1093/oso/9780198570899.001.0001} {\emph {\bibinfo {title} {Gravitational Waves: Volume 2: Astrophysics and Cosmology}}}\ (\bibinfo  {publisher} {Oxford University Press},\ \bibinfo {year} {2018})\BibitemShut {NoStop}%
\bibitem [{\citenamefont {Dotti}\ and\ \citenamefont {Gleiser}(2005{\natexlab{a}})}]{Dotti:2004sh}%
  \BibitemOpen
  \bibfield  {author} {\bibinfo {author} {\bibfnamefont {G.}~\bibnamefont {Dotti}}\ and\ \bibinfo {author} {\bibfnamefont {R.~J.}\ \bibnamefont {Gleiser}},\ }\href {\doibase 10.1088/0264-9381/22/1/L01} {\bibfield  {journal} {\bibinfo  {journal} {Class. Quant. Grav.}\ }\textbf {\bibinfo {volume} {22}},\ \bibinfo {pages} {L1} (\bibinfo {year} {2005}{\natexlab{a}})},\ \Eprint {http://arxiv.org/abs/gr-qc/0409005} {arXiv:gr-qc/0409005} \BibitemShut {NoStop}%
\bibitem [{\citenamefont {Dotti}\ and\ \citenamefont {Gleiser}(2005{\natexlab{b}})}]{Dotti:2005sq}%
  \BibitemOpen
  \bibfield  {author} {\bibinfo {author} {\bibfnamefont {G.}~\bibnamefont {Dotti}}\ and\ \bibinfo {author} {\bibfnamefont {R.~J.}\ \bibnamefont {Gleiser}},\ }\href {\doibase 10.1103/PhysRevD.72.044018} {\bibfield  {journal} {\bibinfo  {journal} {Phys. Rev. D}\ }\textbf {\bibinfo {volume} {72}},\ \bibinfo {pages} {044018} (\bibinfo {year} {2005}{\natexlab{b}})},\ \Eprint {http://arxiv.org/abs/gr-qc/0503117} {arXiv:gr-qc/0503117} \BibitemShut {NoStop}%
\bibitem [{\citenamefont {Gleiser}\ and\ \citenamefont {Dotti}(2005)}]{Gleiser:2005ra}%
  \BibitemOpen
  \bibfield  {author} {\bibinfo {author} {\bibfnamefont {R.~J.}\ \bibnamefont {Gleiser}}\ and\ \bibinfo {author} {\bibfnamefont {G.}~\bibnamefont {Dotti}},\ }\href {\doibase 10.1103/PhysRevD.72.124002} {\bibfield  {journal} {\bibinfo  {journal} {Phys. Rev. D}\ }\textbf {\bibinfo {volume} {72}},\ \bibinfo {pages} {124002} (\bibinfo {year} {2005})},\ \Eprint {http://arxiv.org/abs/gr-qc/0510069} {arXiv:gr-qc/0510069} \BibitemShut {NoStop}%
\bibitem [{\citenamefont {Beroiz}\ \emph {et~al.}(2007)\citenamefont {Beroiz}, \citenamefont {Dotti},\ and\ \citenamefont {Gleiser}}]{Beroiz:2007gp}%
  \BibitemOpen
  \bibfield  {author} {\bibinfo {author} {\bibfnamefont {M.}~\bibnamefont {Beroiz}}, \bibinfo {author} {\bibfnamefont {G.}~\bibnamefont {Dotti}}, \ and\ \bibinfo {author} {\bibfnamefont {R.~J.}\ \bibnamefont {Gleiser}},\ }\href {\doibase 10.1103/PhysRevD.76.024012} {\bibfield  {journal} {\bibinfo  {journal} {Phys. Rev. D}\ }\textbf {\bibinfo {volume} {76}},\ \bibinfo {pages} {024012} (\bibinfo {year} {2007})},\ \Eprint {http://arxiv.org/abs/hep-th/0703074} {arXiv:hep-th/0703074} \BibitemShut {NoStop}%
\bibitem [{\citenamefont {Konoplya}\ and\ \citenamefont {Zhidenko}(2008)}]{Konoplya:2008ix}%
  \BibitemOpen
  \bibfield  {author} {\bibinfo {author} {\bibfnamefont {R.~A.}\ \bibnamefont {Konoplya}}\ and\ \bibinfo {author} {\bibfnamefont {A.}~\bibnamefont {Zhidenko}},\ }\href {\doibase 10.1103/PhysRevD.77.104004} {\bibfield  {journal} {\bibinfo  {journal} {Phys. Rev. D}\ }\textbf {\bibinfo {volume} {77}},\ \bibinfo {pages} {104004} (\bibinfo {year} {2008})},\ \Eprint {http://arxiv.org/abs/0802.0267} {arXiv:0802.0267 [hep-th]} \BibitemShut {NoStop}%
\bibitem [{\citenamefont {Takahashi}\ and\ \citenamefont {Soda}(2009{\natexlab{a}})}]{Takahashi:2009dz}%
  \BibitemOpen
  \bibfield  {author} {\bibinfo {author} {\bibfnamefont {T.}~\bibnamefont {Takahashi}}\ and\ \bibinfo {author} {\bibfnamefont {J.}~\bibnamefont {Soda}},\ }\href {\doibase 10.1103/PhysRevD.79.104025} {\bibfield  {journal} {\bibinfo  {journal} {Phys. Rev. D}\ }\textbf {\bibinfo {volume} {79}},\ \bibinfo {pages} {104025} (\bibinfo {year} {2009}{\natexlab{a}})},\ \Eprint {http://arxiv.org/abs/0902.2921} {arXiv:0902.2921 [gr-qc]} \BibitemShut {NoStop}%
\bibitem [{\citenamefont {Takahashi}\ and\ \citenamefont {Soda}(2009{\natexlab{b}})}]{Takahashi:2009xh}%
  \BibitemOpen
  \bibfield  {author} {\bibinfo {author} {\bibfnamefont {T.}~\bibnamefont {Takahashi}}\ and\ \bibinfo {author} {\bibfnamefont {J.}~\bibnamefont {Soda}},\ }\href {\doibase 10.1103/PhysRevD.80.104021} {\bibfield  {journal} {\bibinfo  {journal} {Phys. Rev. D}\ }\textbf {\bibinfo {volume} {80}},\ \bibinfo {pages} {104021} (\bibinfo {year} {2009}{\natexlab{b}})},\ \Eprint {http://arxiv.org/abs/0907.0556} {arXiv:0907.0556 [gr-qc]} \BibitemShut {NoStop}%
\bibitem [{\citenamefont {Takahashi}\ and\ \citenamefont {Soda}(2010{\natexlab{a}})}]{Takahashi:2010ye}%
  \BibitemOpen
  \bibfield  {author} {\bibinfo {author} {\bibfnamefont {T.}~\bibnamefont {Takahashi}}\ and\ \bibinfo {author} {\bibfnamefont {J.}~\bibnamefont {Soda}},\ }\href {\doibase 10.1143/PTP.124.911} {\bibfield  {journal} {\bibinfo  {journal} {Prog. Theor. Phys.}\ }\textbf {\bibinfo {volume} {124}},\ \bibinfo {pages} {911} (\bibinfo {year} {2010}{\natexlab{a}})},\ \Eprint {http://arxiv.org/abs/1008.1385} {arXiv:1008.1385 [gr-qc]} \BibitemShut {NoStop}%
\bibitem [{\citenamefont {Takahashi}\ and\ \citenamefont {Soda}(2010{\natexlab{b}})}]{Takahashi:2010gz}%
  \BibitemOpen
  \bibfield  {author} {\bibinfo {author} {\bibfnamefont {T.}~\bibnamefont {Takahashi}}\ and\ \bibinfo {author} {\bibfnamefont {J.}~\bibnamefont {Soda}},\ }\href {\doibase 10.1143/PTP.124.711} {\bibfield  {journal} {\bibinfo  {journal} {Prog. Theor. Phys.}\ }\textbf {\bibinfo {volume} {124}},\ \bibinfo {pages} {711} (\bibinfo {year} {2010}{\natexlab{b}})},\ \Eprint {http://arxiv.org/abs/1008.1618} {arXiv:1008.1618 [gr-qc]} \BibitemShut {NoStop}%
\bibitem [{\citenamefont {Takahashi}\ and\ \citenamefont {Soda}(2011)}]{Takahashi:2011cgy}%
  \BibitemOpen
  \bibfield  {author} {\bibinfo {author} {\bibfnamefont {T.}~\bibnamefont {Takahashi}}\ and\ \bibinfo {author} {\bibfnamefont {J.}~\bibnamefont {Soda}},\ }in\ \href@noop {} {\emph {\bibinfo {booktitle} {{20th Workshop on General Relativity and Gravitation in Japan}}}}\ (\bibinfo {year} {2011})\ pp.\ \bibinfo {pages} {387--390}\BibitemShut {NoStop}%
\bibitem [{\citenamefont {Yoshida}\ and\ \citenamefont {Soda}(2016)}]{Yoshida:2015vua}%
  \BibitemOpen
  \bibfield  {author} {\bibinfo {author} {\bibfnamefont {D.}~\bibnamefont {Yoshida}}\ and\ \bibinfo {author} {\bibfnamefont {J.}~\bibnamefont {Soda}},\ }\href {\doibase 10.1103/PhysRevD.93.044024} {\bibfield  {journal} {\bibinfo  {journal} {Phys. Rev. D}\ }\textbf {\bibinfo {volume} {93}},\ \bibinfo {pages} {044024} (\bibinfo {year} {2016})},\ \Eprint {http://arxiv.org/abs/1512.05865} {arXiv:1512.05865 [gr-qc]} \BibitemShut {NoStop}%
\bibitem [{\citenamefont {Abdalla}\ \emph {et~al.}(2005)\citenamefont {Abdalla}, \citenamefont {Konoplya},\ and\ \citenamefont {Molina}}]{Abdalla:2005hu}%
  \BibitemOpen
  \bibfield  {author} {\bibinfo {author} {\bibfnamefont {E.}~\bibnamefont {Abdalla}}, \bibinfo {author} {\bibfnamefont {R.~A.}\ \bibnamefont {Konoplya}}, \ and\ \bibinfo {author} {\bibfnamefont {C.}~\bibnamefont {Molina}},\ }\href {\doibase 10.1103/PhysRevD.72.084006} {\bibfield  {journal} {\bibinfo  {journal} {Phys. Rev. D}\ }\textbf {\bibinfo {volume} {72}},\ \bibinfo {pages} {084006} (\bibinfo {year} {2005})},\ \Eprint {http://arxiv.org/abs/hep-th/0507100} {arXiv:hep-th/0507100} \BibitemShut {NoStop}%
\bibitem [{\citenamefont {Zenginoglu}(2008)}]{Zenginoglu:2007jw}%
  \BibitemOpen
  \bibfield  {author} {\bibinfo {author} {\bibfnamefont {A.}~\bibnamefont {Zenginoglu}},\ }\href {\doibase 10.1088/0264-9381/25/14/145002} {\bibfield  {journal} {\bibinfo  {journal} {Class. Quant. Grav.}\ }\textbf {\bibinfo {volume} {25}},\ \bibinfo {pages} {145002} (\bibinfo {year} {2008})},\ \Eprint {http://arxiv.org/abs/0712.4333} {arXiv:0712.4333 [gr-qc]} \BibitemShut {NoStop}%
\bibitem [{\citenamefont {Ansorg}\ and\ \citenamefont {Panosso~Macedo}(2016)}]{Ansorg:2016ztf}%
  \BibitemOpen
  \bibfield  {author} {\bibinfo {author} {\bibfnamefont {M.}~\bibnamefont {Ansorg}}\ and\ \bibinfo {author} {\bibfnamefont {R.}~\bibnamefont {Panosso~Macedo}},\ }\href {\doibase 10.1103/PhysRevD.93.124016} {\bibfield  {journal} {\bibinfo  {journal} {Phys. Rev. D}\ }\textbf {\bibinfo {volume} {93}},\ \bibinfo {pages} {124016} (\bibinfo {year} {2016})},\ \Eprint {http://arxiv.org/abs/1604.02261} {arXiv:1604.02261 [gr-qc]} \BibitemShut {NoStop}%
\bibitem [{\citenamefont {Panosso~Macedo}\ \emph {et~al.}(2018)\citenamefont {Panosso~Macedo}, \citenamefont {Jaramillo},\ and\ \citenamefont {Ansorg}}]{PanossoMacedo:2018hab}%
  \BibitemOpen
  \bibfield  {author} {\bibinfo {author} {\bibfnamefont {R.}~\bibnamefont {Panosso~Macedo}}, \bibinfo {author} {\bibfnamefont {J.~L.}\ \bibnamefont {Jaramillo}}, \ and\ \bibinfo {author} {\bibfnamefont {M.}~\bibnamefont {Ansorg}},\ }\href {\doibase 10.1103/PhysRevD.98.124005} {\bibfield  {journal} {\bibinfo  {journal} {Phys. Rev. D}\ }\textbf {\bibinfo {volume} {98}},\ \bibinfo {pages} {124005} (\bibinfo {year} {2018})},\ \Eprint {http://arxiv.org/abs/1809.02837} {arXiv:1809.02837 [gr-qc]} \BibitemShut {NoStop}%
\bibitem [{\citenamefont {Zenginoglu}(2011)}]{Zenginoglu:2011jz}%
  \BibitemOpen
  \bibfield  {author} {\bibinfo {author} {\bibfnamefont {A.}~\bibnamefont {Zenginoglu}},\ }\href {\doibase 10.1103/PhysRevD.83.127502} {\bibfield  {journal} {\bibinfo  {journal} {Phys. Rev. D}\ }\textbf {\bibinfo {volume} {83}},\ \bibinfo {pages} {127502} (\bibinfo {year} {2011})},\ \Eprint {http://arxiv.org/abs/1102.2451} {arXiv:1102.2451 [gr-qc]} \BibitemShut {NoStop}%
\bibitem [{\citenamefont {Panosso~Macedo}(2024)}]{PanossoMacedo:2023qzp}%
  \BibitemOpen
  \bibfield  {author} {\bibinfo {author} {\bibfnamefont {R.}~\bibnamefont {Panosso~Macedo}},\ }\href {\doibase 10.1098/rsta.2023.0046} {\bibfield  {journal} {\bibinfo  {journal} {Phil. Trans. Roy. Soc. Lond. A}\ }\textbf {\bibinfo {volume} {382}},\ \bibinfo {pages} {20230046} (\bibinfo {year} {2024})},\ \Eprint {http://arxiv.org/abs/2307.15735} {arXiv:2307.15735 [gr-qc]} \BibitemShut {NoStop}%
\bibitem [{\citenamefont {Zengino\u{g}lu}(2024)}]{Zenginoglu:2024bzs}%
  \BibitemOpen
  \bibfield  {author} {\bibinfo {author} {\bibfnamefont {A.}~\bibnamefont {Zengino\u{g}lu}},\ }\href {\doibase 10.1119/5.0214271} {\bibfield  {journal} {\bibinfo  {journal} {Am. J. Phys.}\ }\textbf {\bibinfo {volume} {92}},\ \bibinfo {pages} {965} (\bibinfo {year} {2024})},\ \Eprint {http://arxiv.org/abs/2404.01528} {arXiv:2404.01528 [gr-qc]} \BibitemShut {NoStop}%
\bibitem [{\citenamefont {Panosso~Macedo}\ and\ \citenamefont {Zenginoglu}(2024)}]{PanossoMacedo:2024nkw}%
  \BibitemOpen
  \bibfield  {author} {\bibinfo {author} {\bibfnamefont {R.}~\bibnamefont {Panosso~Macedo}}\ and\ \bibinfo {author} {\bibfnamefont {A.}~\bibnamefont {Zenginoglu}},\ }\href@noop {} {\  (\bibinfo {year} {2024})},\ \Eprint {http://arxiv.org/abs/2409.11478} {arXiv:2409.11478 [gr-qc]} \BibitemShut {NoStop}%
\bibitem [{\citenamefont {Panosso~Macedo}(2020)}]{PanossoMacedo:2019npm}%
  \BibitemOpen
  \bibfield  {author} {\bibinfo {author} {\bibfnamefont {R.}~\bibnamefont {Panosso~Macedo}},\ }\href {\doibase 10.1088/1361-6382/ab6e3e} {\bibfield  {journal} {\bibinfo  {journal} {Class. Quant. Grav.}\ }\textbf {\bibinfo {volume} {37}},\ \bibinfo {pages} {065019} (\bibinfo {year} {2020})},\ \Eprint {http://arxiv.org/abs/1910.13452} {arXiv:1910.13452 [gr-qc]} \BibitemShut {NoStop}%
\bibitem [{\citenamefont {Barausse}\ \emph {et~al.}(2014)\citenamefont {Barausse}, \citenamefont {Cardoso},\ and\ \citenamefont {Pani}}]{Barausse:2014tra}%
  \BibitemOpen
  \bibfield  {author} {\bibinfo {author} {\bibfnamefont {E.}~\bibnamefont {Barausse}}, \bibinfo {author} {\bibfnamefont {V.}~\bibnamefont {Cardoso}}, \ and\ \bibinfo {author} {\bibfnamefont {P.}~\bibnamefont {Pani}},\ }\href {\doibase 10.1103/PhysRevD.89.104059} {\bibfield  {journal} {\bibinfo  {journal} {Phys. Rev. D}\ }\textbf {\bibinfo {volume} {89}},\ \bibinfo {pages} {104059} (\bibinfo {year} {2014})},\ \Eprint {http://arxiv.org/abs/1404.7149} {arXiv:1404.7149 [gr-qc]} \BibitemShut {NoStop}%
\bibitem [{\citenamefont {Nollert}(1996)}]{Nollert:1996rf}%
  \BibitemOpen
  \bibfield  {author} {\bibinfo {author} {\bibfnamefont {H.-P.}\ \bibnamefont {Nollert}},\ }\href {\doibase 10.1103/PhysRevD.53.4397} {\bibfield  {journal} {\bibinfo  {journal} {Phys. Rev. D}\ }\textbf {\bibinfo {volume} {53}},\ \bibinfo {pages} {4397} (\bibinfo {year} {1996})},\ \Eprint {http://arxiv.org/abs/gr-qc/9602032} {arXiv:gr-qc/9602032} \BibitemShut {NoStop}%
\bibitem [{\citenamefont {Nollert}\ and\ \citenamefont {Price}(1999)}]{Nollert:1998ys}%
  \BibitemOpen
  \bibfield  {author} {\bibinfo {author} {\bibfnamefont {H.-P.}\ \bibnamefont {Nollert}}\ and\ \bibinfo {author} {\bibfnamefont {R.~H.}\ \bibnamefont {Price}},\ }\href {\doibase 10.1063/1.532698} {\bibfield  {journal} {\bibinfo  {journal} {J. Math. Phys.}\ }\textbf {\bibinfo {volume} {40}},\ \bibinfo {pages} {980} (\bibinfo {year} {1999})},\ \Eprint {http://arxiv.org/abs/gr-qc/9810074} {arXiv:gr-qc/9810074} \BibitemShut {NoStop}%
\bibitem [{\citenamefont {Daghigh}\ \emph {et~al.}(2020)\citenamefont {Daghigh}, \citenamefont {Green},\ and\ \citenamefont {Morey}}]{Daghigh:2020jyk}%
  \BibitemOpen
  \bibfield  {author} {\bibinfo {author} {\bibfnamefont {R.~G.}\ \bibnamefont {Daghigh}}, \bibinfo {author} {\bibfnamefont {M.~D.}\ \bibnamefont {Green}}, \ and\ \bibinfo {author} {\bibfnamefont {J.~C.}\ \bibnamefont {Morey}},\ }\href {\doibase 10.1103/PhysRevD.101.104009} {\bibfield  {journal} {\bibinfo  {journal} {Phys. Rev. D}\ }\textbf {\bibinfo {volume} {101}},\ \bibinfo {pages} {104009} (\bibinfo {year} {2020})},\ \Eprint {http://arxiv.org/abs/2002.07251} {arXiv:2002.07251 [gr-qc]} \BibitemShut {NoStop}%
\bibitem [{\citenamefont {Qian}\ \emph {et~al.}(2021)\citenamefont {Qian}, \citenamefont {Lin}, \citenamefont {Shao}, \citenamefont {Wang},\ and\ \citenamefont {Yue}}]{Qian:2020cnz}%
  \BibitemOpen
  \bibfield  {author} {\bibinfo {author} {\bibfnamefont {W.-L.}\ \bibnamefont {Qian}}, \bibinfo {author} {\bibfnamefont {K.}~\bibnamefont {Lin}}, \bibinfo {author} {\bibfnamefont {C.-Y.}\ \bibnamefont {Shao}}, \bibinfo {author} {\bibfnamefont {B.}~\bibnamefont {Wang}}, \ and\ \bibinfo {author} {\bibfnamefont {R.-H.}\ \bibnamefont {Yue}},\ }\href {\doibase 10.1103/PhysRevD.103.024019} {\bibfield  {journal} {\bibinfo  {journal} {Phys. Rev. D}\ }\textbf {\bibinfo {volume} {103}},\ \bibinfo {pages} {024019} (\bibinfo {year} {2021})},\ \Eprint {http://arxiv.org/abs/2009.11627} {arXiv:2009.11627 [gr-qc]} \BibitemShut {NoStop}%
\bibitem [{\citenamefont {Destounis}\ and\ \citenamefont {Duque}(2023)}]{Destounis:2023ruj}%
  \BibitemOpen
  \bibfield  {author} {\bibinfo {author} {\bibfnamefont {K.}~\bibnamefont {Destounis}}\ and\ \bibinfo {author} {\bibfnamefont {F.}~\bibnamefont {Duque}}\ }(\bibinfo {year} {2023})\ \Eprint {http://arxiv.org/abs/2308.16227} {arXiv:2308.16227 [gr-qc]} \BibitemShut {NoStop}%
\bibitem [{\citenamefont {Jaramillo}\ \emph {et~al.}(2021)\citenamefont {Jaramillo}, \citenamefont {Panosso~Macedo},\ and\ \citenamefont {Al~Sheikh}}]{Jaramillo:2020tuu}%
  \BibitemOpen
  \bibfield  {author} {\bibinfo {author} {\bibfnamefont {J.~L.}\ \bibnamefont {Jaramillo}}, \bibinfo {author} {\bibfnamefont {R.}~\bibnamefont {Panosso~Macedo}}, \ and\ \bibinfo {author} {\bibfnamefont {L.}~\bibnamefont {Al~Sheikh}},\ }\href {\doibase 10.1103/PhysRevX.11.031003} {\bibfield  {journal} {\bibinfo  {journal} {Phys. Rev. X}\ }\textbf {\bibinfo {volume} {11}},\ \bibinfo {pages} {031003} (\bibinfo {year} {2021})},\ \Eprint {http://arxiv.org/abs/2004.06434} {arXiv:2004.06434 [gr-qc]} \BibitemShut {NoStop}%
\bibitem [{\citenamefont {Trefethen}\ and\ \citenamefont {Embree}(2005)}]{trefethen2020spectra}%
  \BibitemOpen
  \bibfield  {author} {\bibinfo {author} {\bibfnamefont {L.}~\bibnamefont {Trefethen}}\ and\ \bibinfo {author} {\bibfnamefont {M.}~\bibnamefont {Embree}},\ }\href {\doibase 10.2307/j.ctvzxx9kj} {\emph {\bibinfo {title} {Spectra and Pseudospectra: The Behavior of Nonnormal Matrices and Operators}}}\ (\bibinfo  {publisher} {Princeton university press},\ \bibinfo {year} {2005})\BibitemShut {NoStop}%
\bibitem [{\citenamefont {Destounis}\ \emph {et~al.}(2021)\citenamefont {Destounis}, \citenamefont {Macedo}, \citenamefont {Berti}, \citenamefont {Cardoso},\ and\ \citenamefont {Jaramillo}}]{Destounis:2021lum}%
  \BibitemOpen
  \bibfield  {author} {\bibinfo {author} {\bibfnamefont {K.}~\bibnamefont {Destounis}}, \bibinfo {author} {\bibfnamefont {R.~P.}\ \bibnamefont {Macedo}}, \bibinfo {author} {\bibfnamefont {E.}~\bibnamefont {Berti}}, \bibinfo {author} {\bibfnamefont {V.}~\bibnamefont {Cardoso}}, \ and\ \bibinfo {author} {\bibfnamefont {J.~L.}\ \bibnamefont {Jaramillo}},\ }\href {\doibase 10.1103/PhysRevD.104.084091} {\bibfield  {journal} {\bibinfo  {journal} {Phys. Rev. D}\ }\textbf {\bibinfo {volume} {104}},\ \bibinfo {pages} {084091} (\bibinfo {year} {2021})},\ \Eprint {http://arxiv.org/abs/2107.09673} {arXiv:2107.09673 [gr-qc]} \BibitemShut {NoStop}%
\bibitem [{\citenamefont {Cao}\ \emph {et~al.}(2024)\citenamefont {Cao}, \citenamefont {Chen}, \citenamefont {Wu}, \citenamefont {Xie},\ and\ \citenamefont {Zhou}}]{Cao:2024oud}%
  \BibitemOpen
  \bibfield  {author} {\bibinfo {author} {\bibfnamefont {L.-M.}\ \bibnamefont {Cao}}, \bibinfo {author} {\bibfnamefont {J.-N.}\ \bibnamefont {Chen}}, \bibinfo {author} {\bibfnamefont {L.-B.}\ \bibnamefont {Wu}}, \bibinfo {author} {\bibfnamefont {L.}~\bibnamefont {Xie}}, \ and\ \bibinfo {author} {\bibfnamefont {Y.-S.}\ \bibnamefont {Zhou}},\ }\href {\doibase 10.1007/s11433-024-2435-5} {\bibfield  {journal} {\bibinfo  {journal} {Sci. China Phys. Mech. Astron.}\ }\textbf {\bibinfo {volume} {67}},\ \bibinfo {pages} {100412} (\bibinfo {year} {2024})},\ \Eprint {http://arxiv.org/abs/2401.09907} {arXiv:2401.09907 [gr-qc]} \BibitemShut {NoStop}%
\bibitem [{\citenamefont {Sarkar}\ \emph {et~al.}(2023)\citenamefont {Sarkar}, \citenamefont {Rahman},\ and\ \citenamefont {Chakraborty}}]{Sarkar:2023rhp}%
  \BibitemOpen
  \bibfield  {author} {\bibinfo {author} {\bibfnamefont {S.}~\bibnamefont {Sarkar}}, \bibinfo {author} {\bibfnamefont {M.}~\bibnamefont {Rahman}}, \ and\ \bibinfo {author} {\bibfnamefont {S.}~\bibnamefont {Chakraborty}},\ }\href {\doibase 10.1103/PhysRevD.108.104002} {\bibfield  {journal} {\bibinfo  {journal} {Phys. Rev. D}\ }\textbf {\bibinfo {volume} {108}},\ \bibinfo {pages} {104002} (\bibinfo {year} {2023})},\ \Eprint {http://arxiv.org/abs/2304.06829} {arXiv:2304.06829 [gr-qc]} \BibitemShut {NoStop}%
\bibitem [{\citenamefont {Destounis}\ \emph {et~al.}(2024)\citenamefont {Destounis}, \citenamefont {Boyanov},\ and\ \citenamefont {Panosso~Macedo}}]{Destounis:2023nmb}%
  \BibitemOpen
  \bibfield  {author} {\bibinfo {author} {\bibfnamefont {K.}~\bibnamefont {Destounis}}, \bibinfo {author} {\bibfnamefont {V.}~\bibnamefont {Boyanov}}, \ and\ \bibinfo {author} {\bibfnamefont {R.}~\bibnamefont {Panosso~Macedo}},\ }\href {\doibase 10.1103/PhysRevD.109.044023} {\bibfield  {journal} {\bibinfo  {journal} {Phys. Rev. D}\ }\textbf {\bibinfo {volume} {109}},\ \bibinfo {pages} {044023} (\bibinfo {year} {2024})},\ \Eprint {http://arxiv.org/abs/2312.11630} {arXiv:2312.11630 [gr-qc]} \BibitemShut {NoStop}%
\bibitem [{\citenamefont {Luo}(2024)}]{Luo:2024dxl}%
  \BibitemOpen
  \bibfield  {author} {\bibinfo {author} {\bibfnamefont {S.}~\bibnamefont {Luo}},\ }\href {\doibase 10.1103/PhysRevD.110.084071} {\bibfield  {journal} {\bibinfo  {journal} {Phys. Rev. D}\ }\textbf {\bibinfo {volume} {110}},\ \bibinfo {pages} {084071} (\bibinfo {year} {2024})},\ \Eprint {http://arxiv.org/abs/2408.08139} {arXiv:2408.08139 [gr-qc]} \BibitemShut {NoStop}%
\bibitem [{\citenamefont {Warnick}(2024)}]{Warnick:2024usx}%
  \BibitemOpen
  \bibfield  {author} {\bibinfo {author} {\bibfnamefont {C.}~\bibnamefont {Warnick}},\ }\href@noop {} {\  (\bibinfo {year} {2024})},\ \Eprint {http://arxiv.org/abs/2407.19850} {arXiv:2407.19850 [gr-qc]} \BibitemShut {NoStop}%
\bibitem [{\citenamefont {Arean}\ \emph {et~al.}(2024)\citenamefont {Arean}, \citenamefont {Garcia-Fari\~na},\ and\ \citenamefont {Landsteiner}}]{Arean:2024afl}%
  \BibitemOpen
  \bibfield  {author} {\bibinfo {author} {\bibfnamefont {D.}~\bibnamefont {Arean}}, \bibinfo {author} {\bibfnamefont {D.}~\bibnamefont {Garcia-Fari\~na}}, \ and\ \bibinfo {author} {\bibfnamefont {K.}~\bibnamefont {Landsteiner}},\ }\href {\doibase 10.3389/fphy.2024.1460268} {\bibfield  {journal} {\bibinfo  {journal} {Front. in Phys.}\ }\textbf {\bibinfo {volume} {12}},\ \bibinfo {pages} {1460268} (\bibinfo {year} {2024})},\ \Eprint {http://arxiv.org/abs/2407.04372} {arXiv:2407.04372 [hep-th]} \BibitemShut {NoStop}%
\bibitem [{\citenamefont {Cownden}\ \emph {et~al.}(2024)\citenamefont {Cownden}, \citenamefont {Pantelidou},\ and\ \citenamefont {Zilh\~ao}}]{Cownden:2023dam}%
  \BibitemOpen
  \bibfield  {author} {\bibinfo {author} {\bibfnamefont {B.}~\bibnamefont {Cownden}}, \bibinfo {author} {\bibfnamefont {C.}~\bibnamefont {Pantelidou}}, \ and\ \bibinfo {author} {\bibfnamefont {M.}~\bibnamefont {Zilh\~ao}},\ }\href {\doibase 10.1007/JHEP05(2024)202} {\bibfield  {journal} {\bibinfo  {journal} {JHEP}\ }\textbf {\bibinfo {volume} {05}},\ \bibinfo {pages} {202} (\bibinfo {year} {2024})},\ \Eprint {http://arxiv.org/abs/2312.08352} {arXiv:2312.08352 [gr-qc]} \BibitemShut {NoStop}%
\bibitem [{\citenamefont {Boyanov}\ \emph {et~al.}(2024)\citenamefont {Boyanov}, \citenamefont {Cardoso}, \citenamefont {Destounis}, \citenamefont {Jaramillo},\ and\ \citenamefont {Panosso~Macedo}}]{Boyanov:2023qqf}%
  \BibitemOpen
  \bibfield  {author} {\bibinfo {author} {\bibfnamefont {V.}~\bibnamefont {Boyanov}}, \bibinfo {author} {\bibfnamefont {V.}~\bibnamefont {Cardoso}}, \bibinfo {author} {\bibfnamefont {K.}~\bibnamefont {Destounis}}, \bibinfo {author} {\bibfnamefont {J.~L.}\ \bibnamefont {Jaramillo}}, \ and\ \bibinfo {author} {\bibfnamefont {R.}~\bibnamefont {Panosso~Macedo}},\ }\href {\doibase 10.1103/PhysRevD.109.064068} {\bibfield  {journal} {\bibinfo  {journal} {Phys. Rev. D}\ }\textbf {\bibinfo {volume} {109}},\ \bibinfo {pages} {064068} (\bibinfo {year} {2024})},\ \Eprint {http://arxiv.org/abs/2312.11998} {arXiv:2312.11998 [gr-qc]} \BibitemShut {NoStop}%
\bibitem [{\citenamefont {Garcia-Fari\~na}\ \emph {et~al.}(2025)\citenamefont {Garcia-Fari\~na}, \citenamefont {Landsteiner}, \citenamefont {Romeu},\ and\ \citenamefont {Saura-Bastida}}]{Garcia-Farina:2024pdd}%
  \BibitemOpen
  \bibfield  {author} {\bibinfo {author} {\bibfnamefont {D.}~\bibnamefont {Garcia-Fari\~na}}, \bibinfo {author} {\bibfnamefont {K.}~\bibnamefont {Landsteiner}}, \bibinfo {author} {\bibfnamefont {P.~G.}\ \bibnamefont {Romeu}}, \ and\ \bibinfo {author} {\bibfnamefont {P.}~\bibnamefont {Saura-Bastida}},\ }\href {\doibase 10.1007/JHEP01(2025)185} {\bibfield  {journal} {\bibinfo  {journal} {JHEP}\ }\textbf {\bibinfo {volume} {01}},\ \bibinfo {pages} {185} (\bibinfo {year} {2025})},\ \Eprint {http://arxiv.org/abs/2407.06104} {arXiv:2407.06104 [hep-th]} \BibitemShut {NoStop}%
\bibitem [{\citenamefont {Are\'an}\ \emph {et~al.}(2023)\citenamefont {Are\'an}, \citenamefont {Fari\~na},\ and\ \citenamefont {Landsteiner}}]{Arean:2023ejh}%
  \BibitemOpen
  \bibfield  {author} {\bibinfo {author} {\bibfnamefont {D.}~\bibnamefont {Are\'an}}, \bibinfo {author} {\bibfnamefont {D.~G.}\ \bibnamefont {Fari\~na}}, \ and\ \bibinfo {author} {\bibfnamefont {K.}~\bibnamefont {Landsteiner}},\ }\href {\doibase 10.1007/JHEP12(2023)187} {\bibfield  {journal} {\bibinfo  {journal} {JHEP}\ }\textbf {\bibinfo {volume} {12}},\ \bibinfo {pages} {187} (\bibinfo {year} {2023})},\ \Eprint {http://arxiv.org/abs/2307.08751} {arXiv:2307.08751 [hep-th]} \BibitemShut {NoStop}%
\bibitem [{\citenamefont {Boyanov}\ \emph {et~al.}(2023)\citenamefont {Boyanov}, \citenamefont {Destounis}, \citenamefont {Panosso~Macedo}, \citenamefont {Cardoso},\ and\ \citenamefont {Jaramillo}}]{Boyanov:2022ark}%
  \BibitemOpen
  \bibfield  {author} {\bibinfo {author} {\bibfnamefont {V.}~\bibnamefont {Boyanov}}, \bibinfo {author} {\bibfnamefont {K.}~\bibnamefont {Destounis}}, \bibinfo {author} {\bibfnamefont {R.}~\bibnamefont {Panosso~Macedo}}, \bibinfo {author} {\bibfnamefont {V.}~\bibnamefont {Cardoso}}, \ and\ \bibinfo {author} {\bibfnamefont {J.~L.}\ \bibnamefont {Jaramillo}},\ }\href {\doibase 10.1103/PhysRevD.107.064012} {\bibfield  {journal} {\bibinfo  {journal} {Phys. Rev. D}\ }\textbf {\bibinfo {volume} {107}},\ \bibinfo {pages} {064012} (\bibinfo {year} {2023})},\ \Eprint {http://arxiv.org/abs/2209.12950} {arXiv:2209.12950 [gr-qc]} \BibitemShut {NoStop}%
\bibitem [{\citenamefont {Yang}\ \emph {et~al.}(2024)\citenamefont {Yang}, \citenamefont {Mai}, \citenamefont {Yang}, \citenamefont {Shao},\ and\ \citenamefont {Berti}}]{Yang:2024vor}%
  \BibitemOpen
  \bibfield  {author} {\bibinfo {author} {\bibfnamefont {Y.}~\bibnamefont {Yang}}, \bibinfo {author} {\bibfnamefont {Z.-F.}\ \bibnamefont {Mai}}, \bibinfo {author} {\bibfnamefont {R.-Q.}\ \bibnamefont {Yang}}, \bibinfo {author} {\bibfnamefont {L.}~\bibnamefont {Shao}}, \ and\ \bibinfo {author} {\bibfnamefont {E.}~\bibnamefont {Berti}},\ }\href {\doibase 10.1103/PhysRevD.110.084018} {\bibfield  {journal} {\bibinfo  {journal} {Phys. Rev. D}\ }\textbf {\bibinfo {volume} {110}},\ \bibinfo {pages} {084018} (\bibinfo {year} {2024})},\ \Eprint {http://arxiv.org/abs/2407.20131} {arXiv:2407.20131 [gr-qc]} \BibitemShut {NoStop}%
\bibitem [{\citenamefont {Ianniccari}\ \emph {et~al.}(2024)\citenamefont {Ianniccari}, \citenamefont {Iovino}, \citenamefont {Kehagias}, \citenamefont {Pani}, \citenamefont {Perna}, \citenamefont {Perrone},\ and\ \citenamefont {Riotto}}]{Ianniccari:2024ysv}%
  \BibitemOpen
  \bibfield  {author} {\bibinfo {author} {\bibfnamefont {A.}~\bibnamefont {Ianniccari}}, \bibinfo {author} {\bibfnamefont {A.~J.}\ \bibnamefont {Iovino}}, \bibinfo {author} {\bibfnamefont {A.}~\bibnamefont {Kehagias}}, \bibinfo {author} {\bibfnamefont {P.}~\bibnamefont {Pani}}, \bibinfo {author} {\bibfnamefont {G.}~\bibnamefont {Perna}}, \bibinfo {author} {\bibfnamefont {D.}~\bibnamefont {Perrone}}, \ and\ \bibinfo {author} {\bibfnamefont {A.}~\bibnamefont {Riotto}},\ }\href {\doibase 10.1103/PhysRevLett.133.211401} {\bibfield  {journal} {\bibinfo  {journal} {Phys. Rev. Lett.}\ }\textbf {\bibinfo {volume} {133}},\ \bibinfo {pages} {211401} (\bibinfo {year} {2024})},\ \Eprint {http://arxiv.org/abs/2407.20144} {arXiv:2407.20144 [gr-qc]} \BibitemShut {NoStop}%
\bibitem [{\citenamefont {Berti}\ \emph {et~al.}(2022)\citenamefont {Berti}, \citenamefont {Cardoso}, \citenamefont {Cheung}, \citenamefont {Di~Filippo}, \citenamefont {Duque}, \citenamefont {Martens},\ and\ \citenamefont {Mukohyama}}]{Berti:2022xfj}%
  \BibitemOpen
  \bibfield  {author} {\bibinfo {author} {\bibfnamefont {E.}~\bibnamefont {Berti}}, \bibinfo {author} {\bibfnamefont {V.}~\bibnamefont {Cardoso}}, \bibinfo {author} {\bibfnamefont {M.~H.-Y.}\ \bibnamefont {Cheung}}, \bibinfo {author} {\bibfnamefont {F.}~\bibnamefont {Di~Filippo}}, \bibinfo {author} {\bibfnamefont {F.}~\bibnamefont {Duque}}, \bibinfo {author} {\bibfnamefont {P.}~\bibnamefont {Martens}}, \ and\ \bibinfo {author} {\bibfnamefont {S.}~\bibnamefont {Mukohyama}},\ }\href {\doibase 10.1103/PhysRevD.106.084011} {\bibfield  {journal} {\bibinfo  {journal} {Phys. Rev. D}\ }\textbf {\bibinfo {volume} {106}},\ \bibinfo {pages} {084011} (\bibinfo {year} {2022})},\ \Eprint {http://arxiv.org/abs/2205.08547} {arXiv:2205.08547 [gr-qc]} \BibitemShut {NoStop}%
\bibitem [{\citenamefont {Kyutoku}\ \emph {et~al.}(2023)\citenamefont {Kyutoku}, \citenamefont {Motohashi},\ and\ \citenamefont {Tanaka}}]{Kyutoku:2022gbr}%
  \BibitemOpen
  \bibfield  {author} {\bibinfo {author} {\bibfnamefont {K.}~\bibnamefont {Kyutoku}}, \bibinfo {author} {\bibfnamefont {H.}~\bibnamefont {Motohashi}}, \ and\ \bibinfo {author} {\bibfnamefont {T.}~\bibnamefont {Tanaka}},\ }\href {\doibase 10.1103/PhysRevD.107.044012} {\bibfield  {journal} {\bibinfo  {journal} {Phys. Rev. D}\ }\textbf {\bibinfo {volume} {107}},\ \bibinfo {pages} {044012} (\bibinfo {year} {2023})},\ \Eprint {http://arxiv.org/abs/2206.00671} {arXiv:2206.00671 [gr-qc]} \BibitemShut {NoStop}%
\bibitem [{\citenamefont {Torres}(2023)}]{Torres:2023nqg}%
  \BibitemOpen
  \bibfield  {author} {\bibinfo {author} {\bibfnamefont {T.}~\bibnamefont {Torres}},\ }\href {\doibase 10.1103/PhysRevLett.131.111401} {\bibfield  {journal} {\bibinfo  {journal} {Phys. Rev. Lett.}\ }\textbf {\bibinfo {volume} {131}},\ \bibinfo {pages} {111401} (\bibinfo {year} {2023})},\ \Eprint {http://arxiv.org/abs/2304.10252} {arXiv:2304.10252 [gr-qc]} \BibitemShut {NoStop}%
\bibitem [{\citenamefont {Rosato}\ \emph {et~al.}(2024)\citenamefont {Rosato}, \citenamefont {Destounis},\ and\ \citenamefont {Pani}}]{Rosato:2024arw}%
  \BibitemOpen
  \bibfield  {author} {\bibinfo {author} {\bibfnamefont {R.~F.}\ \bibnamefont {Rosato}}, \bibinfo {author} {\bibfnamefont {K.}~\bibnamefont {Destounis}}, \ and\ \bibinfo {author} {\bibfnamefont {P.}~\bibnamefont {Pani}},\ }\href {\doibase 10.1103/PhysRevD.110.L121501} {\bibfield  {journal} {\bibinfo  {journal} {Phys. Rev. D}\ }\textbf {\bibinfo {volume} {110}},\ \bibinfo {pages} {L121501} (\bibinfo {year} {2024})},\ \Eprint {http://arxiv.org/abs/2406.01692} {arXiv:2406.01692 [gr-qc]} \BibitemShut {NoStop}%
\bibitem [{\citenamefont {Oshita}\ \emph {et~al.}(2024)\citenamefont {Oshita}, \citenamefont {Takahashi},\ and\ \citenamefont {Mukohyama}}]{Oshita:2024fzf}%
  \BibitemOpen
  \bibfield  {author} {\bibinfo {author} {\bibfnamefont {N.}~\bibnamefont {Oshita}}, \bibinfo {author} {\bibfnamefont {K.}~\bibnamefont {Takahashi}}, \ and\ \bibinfo {author} {\bibfnamefont {S.}~\bibnamefont {Mukohyama}},\ }\href {\doibase 10.1103/PhysRevD.110.084070} {\bibfield  {journal} {\bibinfo  {journal} {Phys. Rev. D}\ }\textbf {\bibinfo {volume} {110}},\ \bibinfo {pages} {084070} (\bibinfo {year} {2024})},\ \Eprint {http://arxiv.org/abs/2406.04525} {arXiv:2406.04525 [gr-qc]} \BibitemShut {NoStop}%
\bibitem [{\citenamefont {Wu}\ \emph {et~al.}(2025)\citenamefont {Wu}, \citenamefont {Cai},\ and\ \citenamefont {Xie}}]{Wu:2024ldo}%
  \BibitemOpen
  \bibfield  {author} {\bibinfo {author} {\bibfnamefont {L.-B.}\ \bibnamefont {Wu}}, \bibinfo {author} {\bibfnamefont {R.-G.}\ \bibnamefont {Cai}}, \ and\ \bibinfo {author} {\bibfnamefont {L.}~\bibnamefont {Xie}},\ }\href {\doibase 10.1103/PhysRevD.111.044066} {\bibfield  {journal} {\bibinfo  {journal} {Phys. Rev. D}\ }\textbf {\bibinfo {volume} {111}},\ \bibinfo {pages} {044066} (\bibinfo {year} {2025})},\ \Eprint {http://arxiv.org/abs/2411.07734} {arXiv:2411.07734 [gr-qc]} \BibitemShut {NoStop}%
\bibitem [{\citenamefont {Spieksma}\ \emph {et~al.}(2025)\citenamefont {Spieksma}, \citenamefont {Cardoso}, \citenamefont {Carullo}, \citenamefont {Della~Rocca},\ and\ \citenamefont {Duque}}]{Spieksma:2024voy}%
  \BibitemOpen
  \bibfield  {author} {\bibinfo {author} {\bibfnamefont {T.~F.~M.}\ \bibnamefont {Spieksma}}, \bibinfo {author} {\bibfnamefont {V.}~\bibnamefont {Cardoso}}, \bibinfo {author} {\bibfnamefont {G.}~\bibnamefont {Carullo}}, \bibinfo {author} {\bibfnamefont {M.}~\bibnamefont {Della~Rocca}}, \ and\ \bibinfo {author} {\bibfnamefont {F.}~\bibnamefont {Duque}},\ }\href {\doibase 10.1103/PhysRevLett.134.081402} {\bibfield  {journal} {\bibinfo  {journal} {Phys. Rev. Lett.}\ }\textbf {\bibinfo {volume} {134}},\ \bibinfo {pages} {081402} (\bibinfo {year} {2025})},\ \Eprint {http://arxiv.org/abs/2409.05950} {arXiv:2409.05950 [gr-qc]} \BibitemShut {NoStop}%
\bibitem [{\citenamefont {Cao}\ and\ \citenamefont {Wu}(2021)}]{Cao:2021sty}%
  \BibitemOpen
  \bibfield  {author} {\bibinfo {author} {\bibfnamefont {L.-M.}\ \bibnamefont {Cao}}\ and\ \bibinfo {author} {\bibfnamefont {L.-B.}\ \bibnamefont {Wu}},\ }\href {\doibase 10.1103/PhysRevD.103.064054} {\bibfield  {journal} {\bibinfo  {journal} {Phys. Rev. D}\ }\textbf {\bibinfo {volume} {103}},\ \bibinfo {pages} {064054} (\bibinfo {year} {2021})},\ \Eprint {http://arxiv.org/abs/2101.02461} {arXiv:2101.02461 [gr-qc]} \BibitemShut {NoStop}%
\bibitem [{\citenamefont {Konoplya}\ and\ \citenamefont {Zhidenko}(2010)}]{Konoplya:2010vz}%
  \BibitemOpen
  \bibfield  {author} {\bibinfo {author} {\bibfnamefont {R.~A.}\ \bibnamefont {Konoplya}}\ and\ \bibinfo {author} {\bibfnamefont {A.}~\bibnamefont {Zhidenko}},\ }\href {\doibase 10.1103/PhysRevD.82.084003} {\bibfield  {journal} {\bibinfo  {journal} {Phys. Rev. D}\ }\textbf {\bibinfo {volume} {82}},\ \bibinfo {pages} {084003} (\bibinfo {year} {2010})},\ \Eprint {http://arxiv.org/abs/1004.3772} {arXiv:1004.3772 [hep-th]} \BibitemShut {NoStop}%
\bibitem [{\citenamefont {Konoplya}\ and\ \citenamefont {Zhidenko}(2017)}]{Konoplya:2017ymp}%
  \BibitemOpen
  \bibfield  {author} {\bibinfo {author} {\bibfnamefont {R.~A.}\ \bibnamefont {Konoplya}}\ and\ \bibinfo {author} {\bibfnamefont {A.}~\bibnamefont {Zhidenko}},\ }\href {\doibase 10.1103/PhysRevD.95.104005} {\bibfield  {journal} {\bibinfo  {journal} {Phys. Rev. D}\ }\textbf {\bibinfo {volume} {95}},\ \bibinfo {pages} {104005} (\bibinfo {year} {2017})},\ \Eprint {http://arxiv.org/abs/1701.01652} {arXiv:1701.01652 [hep-th]} \BibitemShut {NoStop}%
\bibitem [{\citenamefont {Boyd}(2001)}]{boyd2001chebyshev}%
  \BibitemOpen
  \bibfield  {author} {\bibinfo {author} {\bibfnamefont {J.~P.}\ \bibnamefont {Boyd}},\ }\href@noop {} {\emph {\bibinfo {title} {Chebyshev and Fourier spectral methods}}}\ (\bibinfo  {publisher} {Courier Corporation},\ \bibinfo {year} {2001})\BibitemShut {NoStop}%
\bibitem [{\citenamefont {Chen}\ \emph {et~al.}(2024)\citenamefont {Chen}, \citenamefont {Wu},\ and\ \citenamefont {Guo}}]{Chen:2024mon}%
  \BibitemOpen
  \bibfield  {author} {\bibinfo {author} {\bibfnamefont {J.-N.}\ \bibnamefont {Chen}}, \bibinfo {author} {\bibfnamefont {L.-B.}\ \bibnamefont {Wu}}, \ and\ \bibinfo {author} {\bibfnamefont {Z.-K.}\ \bibnamefont {Guo}},\ }\href {\doibase 10.1088/1361-6382/ad89a1} {\bibfield  {journal} {\bibinfo  {journal} {Class. Quant. Grav.}\ }\textbf {\bibinfo {volume} {41}},\ \bibinfo {pages} {235015} (\bibinfo {year} {2024})},\ \Eprint {http://arxiv.org/abs/2407.03907} {arXiv:2407.03907 [gr-qc]} \BibitemShut {NoStop}%
\bibitem [{\citenamefont {Boyd}(1996)}]{BOYD199611}%
  \BibitemOpen
  \bibfield  {author} {\bibinfo {author} {\bibfnamefont {J.~P.}\ \bibnamefont {Boyd}},\ }\href {\doibase https://doi.org/10.1006/jcph.1996.0116} {\bibfield  {journal} {\bibinfo  {journal} {Journal of Computational Physics}\ }\textbf {\bibinfo {volume} {126}},\ \bibinfo {pages} {11} (\bibinfo {year} {1996})}\BibitemShut {NoStop}%
\bibitem [{\citenamefont {Markakis}\ \emph {et~al.}(2019)\citenamefont {Markakis}, \citenamefont {O'Boyle}, \citenamefont {Glennon}, \citenamefont {Tran}, \citenamefont {Brubeck}, \citenamefont {Haas}, \citenamefont {Schive},\ and\ \citenamefont {Uryū}}]{markakis2019timesymmetry}%
  \BibitemOpen
  \bibfield  {author} {\bibinfo {author} {\bibfnamefont {C.~M.}\ \bibnamefont {Markakis}}, \bibinfo {author} {\bibfnamefont {M.~F.}\ \bibnamefont {O'Boyle}}, \bibinfo {author} {\bibfnamefont {D.}~\bibnamefont {Glennon}}, \bibinfo {author} {\bibfnamefont {K.}~\bibnamefont {Tran}}, \bibinfo {author} {\bibfnamefont {P.}~\bibnamefont {Brubeck}}, \bibinfo {author} {\bibfnamefont {R.}~\bibnamefont {Haas}}, \bibinfo {author} {\bibfnamefont {H.-Y.}\ \bibnamefont {Schive}}, \ and\ \bibinfo {author} {\bibfnamefont {K.}~\bibnamefont {Uryū}},\ }\href@noop {} {\  (\bibinfo {year} {2019})},\ \Eprint {http://arxiv.org/abs/1901.09967} {arXiv:1901.09967 [math.NA]} \BibitemShut {NoStop}%
\bibitem [{\citenamefont {Markakis}\ \emph {et~al.}(2023)\citenamefont {Markakis}, \citenamefont {Bray},\ and\ \citenamefont {Zengino\u{g}lu}}]{Markakis:2023pfh}%
  \BibitemOpen
  \bibfield  {author} {\bibinfo {author} {\bibfnamefont {C.}~\bibnamefont {Markakis}}, \bibinfo {author} {\bibfnamefont {S.}~\bibnamefont {Bray}}, \ and\ \bibinfo {author} {\bibfnamefont {A.}~\bibnamefont {Zengino\u{g}lu}},\ }\href@noop {} {\  (\bibinfo {year} {2023})},\ \Eprint {http://arxiv.org/abs/2303.08153} {arXiv:2303.08153 [gr-qc]} \BibitemShut {NoStop}%
\bibitem [{\citenamefont {Da~Silva}(2024)}]{DaSilva:2024yea}%
  \BibitemOpen
  \bibfield  {author} {\bibinfo {author} {\bibfnamefont {L.~J.~G.}\ \bibnamefont {Da~Silva}},\ }\href@noop {} {\  (\bibinfo {year} {2024})},\ \Eprint {http://arxiv.org/abs/2401.08758} {arXiv:2401.08758 [gr-qc]} \BibitemShut {NoStop}%
\bibitem [{\citenamefont {O'Boyle}(2022)}]{OBoyle:2022lek}%
  \BibitemOpen
  \bibfield  {author} {\bibinfo {author} {\bibfnamefont {M.~F.}\ \bibnamefont {O'Boyle}},\ }\emph {\bibinfo {title} {{Time-symmetric integration of partial differential equations with applications to black hole physics}}},\ \href@noop {} {Ph.D. thesis},\ \bibinfo  {school} {Illinois U., Urbana (main)} (\bibinfo {year} {2022})\BibitemShut {NoStop}%
\bibitem [{\citenamefont {O'Boyle}\ \emph {et~al.}(2022)\citenamefont {O'Boyle}, \citenamefont {Markakis}, \citenamefont {Da~Silva}, \citenamefont {Panosso~Macedo},\ and\ \citenamefont {Kroon}}]{OBoyle:2022yhp}%
  \BibitemOpen
  \bibfield  {author} {\bibinfo {author} {\bibfnamefont {M.~F.}\ \bibnamefont {O'Boyle}}, \bibinfo {author} {\bibfnamefont {C.}~\bibnamefont {Markakis}}, \bibinfo {author} {\bibfnamefont {L.~J.~G.}\ \bibnamefont {Da~Silva}}, \bibinfo {author} {\bibfnamefont {R.}~\bibnamefont {Panosso~Macedo}}, \ and\ \bibinfo {author} {\bibfnamefont {J.~A.~V.}\ \bibnamefont {Kroon}},\ }\href@noop {} {\  (\bibinfo {year} {2022})},\ \Eprint {http://arxiv.org/abs/2210.02550} {arXiv:2210.02550 [gr-qc]} \BibitemShut {NoStop}%
\bibitem [{\citenamefont {Besson}\ and\ \citenamefont {Jaramillo}(2024)}]{Besson:2024adi}%
  \BibitemOpen
  \bibfield  {author} {\bibinfo {author} {\bibfnamefont {J.}~\bibnamefont {Besson}}\ and\ \bibinfo {author} {\bibfnamefont {J.~L.}\ \bibnamefont {Jaramillo}},\ }\href@noop {} {\  (\bibinfo {year} {2024})},\ \Eprint {http://arxiv.org/abs/2412.02793} {arXiv:2412.02793 [gr-qc]} \BibitemShut {NoStop}%
\bibitem [{\citenamefont {Cardoso}\ \emph {et~al.}(2024)\citenamefont {Cardoso}, \citenamefont {Carullo}, \citenamefont {De~Amicis}, \citenamefont {Duque}, \citenamefont {Katagiri}, \citenamefont {Pereniguez}, \citenamefont {Redondo-Yuste}, \citenamefont {Spieksma},\ and\ \citenamefont {Zhong}}]{Cardoso:2024jme}%
  \BibitemOpen
  \bibfield  {author} {\bibinfo {author} {\bibfnamefont {V.}~\bibnamefont {Cardoso}}, \bibinfo {author} {\bibfnamefont {G.}~\bibnamefont {Carullo}}, \bibinfo {author} {\bibfnamefont {M.}~\bibnamefont {De~Amicis}}, \bibinfo {author} {\bibfnamefont {F.}~\bibnamefont {Duque}}, \bibinfo {author} {\bibfnamefont {T.}~\bibnamefont {Katagiri}}, \bibinfo {author} {\bibfnamefont {D.}~\bibnamefont {Pereniguez}}, \bibinfo {author} {\bibfnamefont {J.}~\bibnamefont {Redondo-Yuste}}, \bibinfo {author} {\bibfnamefont {T.~F.~M.}\ \bibnamefont {Spieksma}}, \ and\ \bibinfo {author} {\bibfnamefont {Z.}~\bibnamefont {Zhong}},\ }\href {\doibase 10.1103/PhysRevD.109.L121502} {\bibfield  {journal} {\bibinfo  {journal} {Phys. Rev. D}\ }\textbf {\bibinfo {volume} {109}},\ \bibinfo {pages} {L121502} (\bibinfo {year} {2024})},\ \Eprint {http://arxiv.org/abs/2405.12290} {arXiv:2405.12290 [gr-qc]} \BibitemShut {NoStop}%
\bibitem [{\citenamefont {Ching}\ \emph {et~al.}(1995{\natexlab{a}})\citenamefont {Ching}, \citenamefont {Leung}, \citenamefont {Suen},\ and\ \citenamefont {Young}}]{Ching:1994bd}%
  \BibitemOpen
  \bibfield  {author} {\bibinfo {author} {\bibfnamefont {E.~S.~C.}\ \bibnamefont {Ching}}, \bibinfo {author} {\bibfnamefont {P.~T.}\ \bibnamefont {Leung}}, \bibinfo {author} {\bibfnamefont {W.~M.}\ \bibnamefont {Suen}}, \ and\ \bibinfo {author} {\bibfnamefont {K.}~\bibnamefont {Young}},\ }\href {\doibase 10.1103/PhysRevLett.74.2414} {\bibfield  {journal} {\bibinfo  {journal} {Phys. Rev. Lett.}\ }\textbf {\bibinfo {volume} {74}},\ \bibinfo {pages} {2414} (\bibinfo {year} {1995}{\natexlab{a}})},\ \Eprint {http://arxiv.org/abs/gr-qc/9410044} {arXiv:gr-qc/9410044} \BibitemShut {NoStop}%
\bibitem [{\citenamefont {Ching}\ \emph {et~al.}(1995{\natexlab{b}})\citenamefont {Ching}, \citenamefont {Leung}, \citenamefont {Suen},\ and\ \citenamefont {Young}}]{Ching:1995tj}%
  \BibitemOpen
  \bibfield  {author} {\bibinfo {author} {\bibfnamefont {E.~S.~C.}\ \bibnamefont {Ching}}, \bibinfo {author} {\bibfnamefont {P.~T.}\ \bibnamefont {Leung}}, \bibinfo {author} {\bibfnamefont {W.~M.}\ \bibnamefont {Suen}}, \ and\ \bibinfo {author} {\bibfnamefont {K.}~\bibnamefont {Young}},\ }\href {\doibase 10.1103/PhysRevD.52.2118} {\bibfield  {journal} {\bibinfo  {journal} {Phys. Rev. D}\ }\textbf {\bibinfo {volume} {52}},\ \bibinfo {pages} {2118} (\bibinfo {year} {1995}{\natexlab{b}})},\ \Eprint {http://arxiv.org/abs/gr-qc/9507035} {arXiv:gr-qc/9507035} \BibitemShut {NoStop}%
\bibitem [{\citenamefont {Da~Silva}\ \emph {et~al.}(2023)\citenamefont {Da~Silva}, \citenamefont {Panosso~Macedo}, \citenamefont {Thompson}, \citenamefont {Kroon}, \citenamefont {Durkan},\ and\ \citenamefont {Long}}]{DaSilva:2023xif}%
  \BibitemOpen
  \bibfield  {author} {\bibinfo {author} {\bibfnamefont {L.~J.~G.}\ \bibnamefont {Da~Silva}}, \bibinfo {author} {\bibfnamefont {R.}~\bibnamefont {Panosso~Macedo}}, \bibinfo {author} {\bibfnamefont {J.~E.}\ \bibnamefont {Thompson}}, \bibinfo {author} {\bibfnamefont {J.~A.~V.}\ \bibnamefont {Kroon}}, \bibinfo {author} {\bibfnamefont {L.}~\bibnamefont {Durkan}}, \ and\ \bibinfo {author} {\bibfnamefont {O.}~\bibnamefont {Long}},\ }\href@noop {} {\  (\bibinfo {year} {2023})},\ \Eprint {http://arxiv.org/abs/2306.13153} {arXiv:2306.13153 [gr-qc]} \BibitemShut {NoStop}%
\bibitem [{\citenamefont {Cardoso}\ \emph {et~al.}(2003)\citenamefont {Cardoso}, \citenamefont {Yoshida}, \citenamefont {Dias},\ and\ \citenamefont {Lemos}}]{cardoso:2003jf}%
  \BibitemOpen
  \bibfield  {author} {\bibinfo {author} {\bibfnamefont {V.}~\bibnamefont {Cardoso}}, \bibinfo {author} {\bibfnamefont {S.}~\bibnamefont {Yoshida}}, \bibinfo {author} {\bibfnamefont {O.~J.~C.}\ \bibnamefont {Dias}}, \ and\ \bibinfo {author} {\bibfnamefont {J.~P.~S.}\ \bibnamefont {Lemos}},\ }\href {\doibase 10.1103/PhysRevD.68.061503} {\bibfield  {journal} {\bibinfo  {journal} {Phys. Rev. D}\ }\textbf {\bibinfo {volume} {68}},\ \bibinfo {pages} {061503} (\bibinfo {year} {2003})},\ \Eprint {http://arxiv.org/abs/hep-th/0307122} {arXiv:hep-th/0307122} \BibitemShut {NoStop}%
\bibitem [{\citenamefont {Ashida}\ \emph {et~al.}(2021)\citenamefont {Ashida}, \citenamefont {Gong},\ and\ \citenamefont {Ueda}}]{Ashida:2020dkc}%
  \BibitemOpen
  \bibfield  {author} {\bibinfo {author} {\bibfnamefont {Y.}~\bibnamefont {Ashida}}, \bibinfo {author} {\bibfnamefont {Z.}~\bibnamefont {Gong}}, \ and\ \bibinfo {author} {\bibfnamefont {M.}~\bibnamefont {Ueda}},\ }\href {\doibase 10.1080/00018732.2021.1876991} {\bibfield  {journal} {\bibinfo  {journal} {Adv. Phys.}\ }\textbf {\bibinfo {volume} {69}},\ \bibinfo {pages} {249} (\bibinfo {year} {2021})},\ \Eprint {http://arxiv.org/abs/2006.01837} {arXiv:2006.01837 [cond-mat.mes-hall]} \BibitemShut {NoStop}%
\bibitem [{\citenamefont {Gasperin}\ and\ \citenamefont {Jaramillo}(2022)}]{Gasperin:2021kfv}%
  \BibitemOpen
  \bibfield  {author} {\bibinfo {author} {\bibfnamefont {E.}~\bibnamefont {Gasperin}}\ and\ \bibinfo {author} {\bibfnamefont {J.~L.}\ \bibnamefont {Jaramillo}},\ }\href {\doibase 10.1088/1361-6382/ac5054} {\bibfield  {journal} {\bibinfo  {journal} {Class. Quant. Grav.}\ }\textbf {\bibinfo {volume} {39}},\ \bibinfo {pages} {115010} (\bibinfo {year} {2022})},\ \Eprint {http://arxiv.org/abs/2107.12865} {arXiv:2107.12865 [gr-qc]} \BibitemShut {NoStop}%
\bibitem [{\citenamefont {Cheung}\ \emph {et~al.}(2022)\citenamefont {Cheung}, \citenamefont {Destounis}, \citenamefont {Macedo}, \citenamefont {Berti},\ and\ \citenamefont {Cardoso}}]{Cheung:2021bol}%
  \BibitemOpen
  \bibfield  {author} {\bibinfo {author} {\bibfnamefont {M.~H.-Y.}\ \bibnamefont {Cheung}}, \bibinfo {author} {\bibfnamefont {K.}~\bibnamefont {Destounis}}, \bibinfo {author} {\bibfnamefont {R.~P.}\ \bibnamefont {Macedo}}, \bibinfo {author} {\bibfnamefont {E.}~\bibnamefont {Berti}}, \ and\ \bibinfo {author} {\bibfnamefont {V.}~\bibnamefont {Cardoso}},\ }\href {\doibase 10.1103/PhysRevLett.128.111103} {\bibfield  {journal} {\bibinfo  {journal} {Phys. Rev. Lett.}\ }\textbf {\bibinfo {volume} {128}},\ \bibinfo {pages} {111103} (\bibinfo {year} {2022})},\ \Eprint {http://arxiv.org/abs/2111.05415} {arXiv:2111.05415 [gr-qc]} \BibitemShut {NoStop}%
\bibitem [{\citenamefont {Boyanov}(2024)}]{Boyanov:2024fgc}%
  \BibitemOpen
  \bibfield  {author} {\bibinfo {author} {\bibfnamefont {V.}~\bibnamefont {Boyanov}},\ }\href {\doibase 10.3389/fphy.2024.1511757} {\bibfield  {journal} {\bibinfo  {journal} {Front. in Phys.}\ }\textbf {\bibinfo {volume} {12}},\ \bibinfo {pages} {1511757} (\bibinfo {year} {2024})},\ \Eprint {http://arxiv.org/abs/2410.11547} {arXiv:2410.11547 [gr-qc]} \BibitemShut {NoStop}%
\end{thebibliography}%
\bibliographystyle{apsrev4-1}

\end{document}